%% file: arxiv_stokes_fenecr_hinf_2013Feb18.tex
\documentclass[10pt]{jfm1}
\usepackage{graphicx}

\usepackage{natbib}            
\usepackage{setspace}
\usepackage{amsfonts,amssymb,amsmath,dsfont}
\usepackage{setspace,subfigure}
\usepackage{latexsym}
\usepackage{pifont}
\usepackage{mathtools}
\usepackage{mathdots}
\usepackage{multirow}
\usepackage{multicol}
\usepackage{color}
\usepackage{url}

\usepackage{rotating}
\usepackage{array}

\input{commands}
\newcommand{\cO}{{\cal O}}
\newcommand{\bA}{\mathbf{A}}
\newcommand{\bB}{\mathbf{B}}
\newcommand{\bC}{\mathbf{C}}
\newcommand{\bE}{\mathbf{E}}
\newcommand{\bG}{\mathbf{G}}
\newcommand{\bH}{\mathbf{H}}

\newcommand{\bR}{\mathbf{R}}

\newcommand{\bbR}{\mathbb{R}}
\newcommand{\bbC}{\mathbb{C}}

\newcommand{\non}{\nonumber}
\newcommand{\ds}{\displaystyle}

\newcommand{\mrd}{\mathrm{d}}
\newcommand{\mre}{\mathrm{e}}
\newcommand{\mri}{\mathrm{i}}

\newcommand{\bd}{\mathbf{d}}
\newcommand{\bD}{{\bf D}}

\newcommand{\bF}{{\bf F}}

\newcommand{\bphi}{\mbox{\boldmath$\phi$}}

\newcommand{\bvarphi}{\mbox{\boldmath$\varphi$}}

\newcommand{\bnabla}{\mbox{\boldmath$\nabla$}}

\newcommand{\px}{\partial_x}
\newcommand{\py}{\partial_y}
\newcommand{\pz}{\partial_z}

\newcommand{\pyy}{\partial_{yy}}
\newcommand{\pzz}{\partial_{zz}}
\newcommand{\pyz}{\partial_{yz}}

\newcommand{\pyyyy}{\partial_{yyyy}}

\newcommand{\bV}{{\bf V}}
\newcommand{\bT}{{\bf T}}
\newcommand{\bI}{{\bf I}}

\newcommand{\btau}{\mbox{\boldmath$\tau$}}
\newcommand{\bkappa}{\mbox{\boldmath$\kappa$}}
\newcommand{\We}{W\!e}

\newcommand{\Rbar}{\bar{R}}

\newcommand{\brr}{\mathbf{r}}

\newcommand{\bGamma}{\mbox{\boldmath$\Gamma$}}

\newcommand{\tc}{\textcolor}
\definecolor{dred}{rgb}{0.4,0.2,0}

\title{Worst-case amplification of disturbances in inertialess Couette flow of viscoelastic fluids}

\shorttitle{Worst-case amplification in inertialess Couette flow of viscoelastic fluids}

\author{Binh K. Lieu$^{1}$, Mihailo R.\ Jovanovi\'c$^{1}$, \and Satish Kumar$^{2}$}

\affiliation{$^{1}$Department of Electrical and Computer Engineering,
University of Minnesota, \\ Minneapolis, MN 55455, USA
\\[\affilskip]
$^{2}$Department of Chemical Engineering and Materials Science, University of Minnesota, \\ Minneapolis, MN 55455 USA}

\shortauthor{B. K. Lieu, M. R. Jovanovi\'c \& S. Kumar}

\begin{document}

\maketitle

\begin{abstract}
	Amplification of deterministic disturbances in inertialess shear-driven channel flows of viscoelastic fluids is examined by analyzing the frequency responses from spatio-temporal body forces to the velocity and polymer stress fluctuations. In strongly elastic flows, we show that disturbances with large streamwise length scales may be significantly amplified even in the absence of inertia. For fluctuations without streamwise variations, we derive explicit analytical expressions for the dependence of the worst-case amplification (from different forcing to different velocity and polymer stress components) on the Weissenberg number ($\We$), the maximum extensibility of the polymer chains ($L$), the viscosity ratio, and the spanwise wavenumber. For the Oldroyd-B model, the amplification of the most energetic components of velocity and polymer stress fields scales as $\We^{2}$ and $\We^{4}$. On the other hand, finite extensibility of polymer molecules limits the largest achievable amplification even in flows with infinitely large Weissenberg numbers: in the presence of wall-normal and spanwise forces the amplification of the streamwise velocity and polymer stress fluctuations is bounded by quadratic and quartic functions of $L$. This high amplification signals low robustness to modeling imperfections of inertialess channel flows of viscoelastic fluids. The underlying physical mechanism involves interactions of polymer stress fluctuations with a base shear, and it represents a close analog of the lift-up mechanism that initiates a bypass transition in inertial flows of Newtonian fluids.
    \end{abstract}

      \vspace*{-2ex}
\section{Introduction}

	Newtonian fluids transition to turbulence under the influence of inertia. In stark contrast, recent experiments have shown that flows of viscoelastic fluids may undergo a transition to a time-dependent disordered flow state and become turbulent even when inertial forces are considerably weaker than viscous forces~\citep*{lar00,groste00,groste04,arrthodiogol06}.  Since viscoelastic fluid flows are often encountered in commercially important settings, understanding transition to {\em elastic turbulence\/} in such flows is important from both fundamental and technological standpoints.  In polymer processing, for example, elastic turbulence is not desirable because it compromises quality of the final product~\citep{lar92}. But in microfluidic devices elastic turbulence can help promote transport, thereby improving the quality of mixing~\citep{groste01,ottwig04}.
	
	Transition in the experiments of~\citet{groste00,groste04} is thought to be initiated by the occurrence of a linear instability that arises from the presence of curved streamlines~\citep*{larshamul90,lar92}. However, the question of whether and how transition can occur in channel flows of viscoelastic fluids with straight streamlines remains wide open. Standard modal stability analysis of the upper convected Maxwell and Oldroyd-B constitutive equations shows that these flows are linearly stable when inertial effects are negligible; yet, they exhibit complex dynamical responses in strongly elastic regimes~\citep{yes02,yes09,boningamakel11,panmorarr11}. Thus, if an inertialess transition can indeed be described using such basic constitutive models, it would likely involve finite-amplitude disturbances that would trigger nonlinear effects~\citep{meustomorsar04,morvan05}. However, the lack of a modal instability does not rule out the possibility that the early stages of transition can be described by the linearized equations. If non-modal growth is present, initially small-amplitude disturbances could grow to a finite amplitude at intermediate times before decaying at long times. For sufficiently large disturbance amplitudes the flow could enter a regime where nonlinear interactions are no longer negligible. This can induce secondary amplification and instability of the flow structures that are selected by the linearized dynamics and promote eventual transition to elastic turbulence.

	\citet*{hodjovkumJFM08,hodjovkumJFM09} recently employed tools from linear systems theory to study the amplification of stochastic spatio-temporal body forces in plane Couette and Poiseuille flows of viscoelastic fluids with nonzero Reynolds numbers. In strongly elastic flows, the results of~\citet{hodjovkumJFM08,hodjovkumJFM09} indicate that significant amplification of streamwise-constant velocity fluctuations can occur even when inertial forces are weak. As in Newtonian fluids, this amplification is caused by non-normality of the underlying operators and it cannot be predicted via standard linear stability analysis. Furthermore, recent work of~\citet{jovkumPOF10,jovkumJNNFM11} shows that this large amplification arises from the interactions between the polymer stress fluctuations in the wall-normal/spanwise plane with the base shear. Through these interactions weak streamwise vortices induce a viscoelastic analogue of the lift-up mechanism which is responsible for the creation of alternating regions of high and low streamwise velocities (relative to the mean flow).~\citet{jovkumPOF10,jovkumJNNFM11} demonstrated significant conceptual similarities between this purely elastic mechanism and the well-known inertial vortex tilting mechanism that initiates a bypass transition \mbox{in shear flows of Newtonian fluids.}
	
	Despite this recent progress, analytical results that quantify influence of finite extensibility of polymer molecules on the amplification of disturbances in channel flows of viscoelastic fluids without inertia are still lacking. Such results may provide physical insight into the early stages of transition and help benchmark direct numerical simulations. Analogous results have been extremely helpful in understanding the early stages of transition to turbulence in wall-bounded shear flows of Newtonian fluids~\citep{farioa93,tretrereddri93,jovbamJFM05,sch07}.

It is worth noting that the problem of determining the amplification of white-in-time stochastic forcing is ill-posed when inertia is completely absent. This restricts the results of~\citet{hodjovkumJFM08,hodjovkumJFM09} to cases where the flow has finite inertia.~\citet{jovkumJNNFM11} used singular perturbation methods to identify the spatial structure of velocity and polymer stress fluctuations that exhibit the highest amplification in stochastically forced weakly inertial channel flows of viscoelastic fluids. As the influence of inertial forces vanishes, it was demonstrated that the velocity fluctuations become white-in-time, thereby exhibiting infinite variance. Furthermore,~\cite{hodjovkumJFM08,hodjovkumJFM09} and \cite{jovkumPOF10,jovkumJNNFM11} employed the Oldroyd-B constitutive model to investigate energy amplification of velocity and polymer stress fluctuations. However, since the Oldroyd-B model allows the polymers to stretch indefinitely, examining the role of finite extensibility of polymer molecules on the amplification of disturbances remains an open question.

In the present work, we address these issues by examining the worst-case amplification of deterministic disturbances in inertialess (i.e., creeping) shear-driven channel flows of viscoelastic fluids. We consider spatially distributed and temporally varying forcing that is purely harmonic in the horizontal directions and time, and deterministic in the wall-normal direction. The motivation for studying creeping flows arises from the observation that viscoelastic fluids can become turbulent even in low inertial regimes, i.e., at small Reynolds number~\citep{lar92,groste00}. Furthermore, the present analysis uses the finitely extensible nonlinear elastic Chilcott-Rallison (FENE-CR) model~\citep{chiral88}, which captures the finite extensibility of the polymer molecules. It is well-known that, for infinitely extensible polymers, the FENE-CR model simplifies to the Oldroyd-B model.

With our approach, we show that velocity and polymer stress fluctuations can exhibit significant amplification and that the most energetic flow structures have large streamwise length scales. In the absence of streamwise variations, we derive explicit expressions for the worst-case amplification from different components of the forcing field to different components of velocity and polymer stress fluctuations. For the Oldroyd-B model, the wall-normal and spanwise forces induce amplification of the streamwise components of velocity and polymer stress fields that scales quadratically and quarticly with the Weissenberg number. On the other hand, we demonstrate that finite extensibility of the polymer molecules saturates the largest achievable amplification even for flows with infinitely large Weissenberg numbers. The functions that bound the worst-case amplification of the velocity and polymer stress fluctuations scale quadratically and quarticly with the largest extensibility of the polymer molecules. We also identify the viscoelastic analogue of the well-known inertial lift-up mechanism as the primary driving force for high flow sensitivity; the underlying mechanism arises from interactions of polymer stress fluctuations with a base shear and it is facilitated by spanwise variations in flow fluctuations~\citep{jovkumJNNFM11}. This non-modal amplification may provide a route by which infinitesimal disturbances can grow to finite amplitude and perhaps trigger a transition to elastic turbulence in channel flows of viscoelastic fluids.

To facilitate development of explicit analytical expressions for worst-case amplification of velocity and polymer stress fluctuations, we restrict our study to an inertialess shear-driven channel (Couette) flow of FENE-CR fluids. Even though the current analysis can be readily applied to the FENE-P model and to a pressure-driven channel (Poiseuille) flow, more complicated base state removes algebraic convenience encountered in Couette flow of FENE-CR fluids. We note that all physical mechanisms identified in this paper remain at play in inertialess Poiseuille flow of both FENE-CR and FENE-P fluids.

	The rest of this paper is organized as follows: in~\S~\ref{sec.prelim}, we use a simple example to illustrate how techniques from control theory can be used to quantify amplification of disturbances and robustness to modeling imperfections. In~\S~\ref{sec.prob-form}, we describe the governing equations for inertialess channel flows of FENE-CR fluids, provide the evolution model, and briefly discuss the essential features of the frequency response analysis. In~\S~\ref{sec.3d}, we examine the frequency responses of three-dimensional velocity fluctuations in inertialess Couette of FENE-CR fluids. In~\S~\ref{sec.kx0}, we provide analytical expressions for the worst-case amplification from the forcing to velocity fluctuations using a streamwise-constant linearized model. We also identify the spatial structures of the dominant forcing and velocity components and demonstrate the importance of the viscoelastic lift-up mechanism. In~\S~\ref{sec.polymer-stress-2d3c}, we study the dynamics of streamwise-constant polymer stress fluctuations. Finally, in~\S~\ref{sec.conclusion}, we summarize the major contributions of this work and discuss future research directions.

    \vspace*{-2ex}
\section{The role of uncertainty: an illustrative example}
    \label{sec.prelim}

	In the course of addressing the issue of disturbance amplification, the present paper provides insight into the robustness of viscoelastic flow models. In this section, we briefly summarize how the tools from control theory facilitate quantification of sensitivity and robustness of a system to modeling imperfections via frequency response analysis. The approach taken in the present paper is closely related to classical frequency response studies of systems arising in physics and engineering. For example, in the design of operational amplifiers it is well-known that caution must be exercised with models that show a low degree of robustness because small modeling errors could cause otherwise stable dynamics to become unstable. Similar ideas have found use in fluid mechanics, including the analysis of the early stages of transition in shear flows of \mbox{Newtonian fluids~\citep{sch07}.}

\begin{figure}
    	\centering
    	\begin{tabular}{ccc}
    		\begin{tabular}{c}
    			{
			\setlength{\unitlength}{0.8cm}
    			\input{simpleNEW}
    			}
			\\
			\subfigure[]
			{
			\hspace{0.5cm}
			\label{fig.block}
			}
    		\end{tabular}
    		&
    		\hspace*{-0.5cm}
    		&
    		\begin{tabular}{c}
    			{
			\setlength{\unitlength}{0.8cm}
    			\input{simpleGnew}
    			}
			\\
			\subfigure[]
			{
			\hspace{0.5cm}
			\label{fig.sg}
			}
    		\end{tabular}
    	\end{tabular}
    	\caption{Block diagrams of (a) system~(\ref{EX}); and (b) system~(\ref{EX}) connected in feedback with norm-bounded unstructured uncertainty $\Gamma$. Here, $s \in \bbC$ denotes the temporal Laplace transform variable, and the transfer function, from $d$ to $\varphi$, is determined by $H (s) = R/\big( (s + \lambda_1) (s + \lambda_2) \big)$.}
   \end{figure}
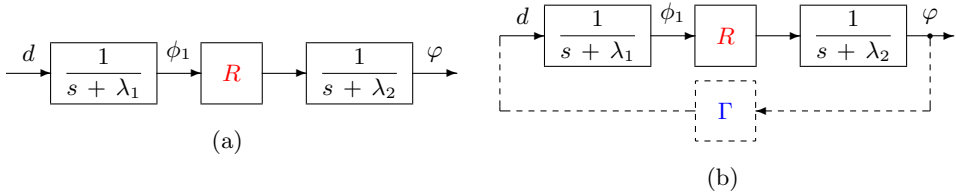

	To fix ideas, let us begin with a simple system of two coupled first-order differential equations
	    \beq
	    \ba{rcl}
	    \tbo{\dot{\phi}_1}{\dot{\phi}_2}
	    & \!\! = \!\! &
	    \tbt{-\lambda_1}{0}{R}{-\lambda_2}
	    \tbo{\phi_1}{\phi_2}
	    \; + \;
	    \tbo{1}{0}
	    d,
	    \\[0.35cm]
	    \varphi
	    & \!\! = \!\! &
	    \obt{0}{1}
	    \tbo{\phi_1}{\phi_2},
	    \ea
	    \label{EX}
	    \eeq
	where $\phi_1$ and $\phi_2$ are the states, $d$ is the disturbance, and $\varphi$ is the output. We assume positivity of scalars $\lambda_1$ and $\lambda_2$, which guarantees modal stability of~(\ref{EX}). Equivalently, this system can be represented via its block-diagram in figure~\ref{fig.block}. Clearly, we have a cascade connection of two stable first-order systems with parameter $R$ determining the strength of coupling between the two subsystems. For $\lambda_1 \neq \lambda_2$ the solution of the unforced problem, i.e.\ with $d \equiv 0$, is determined by
	\begin{equation}
		\begin{array}{rcl}
			\phi_1 (t)
			& \!\! = \!\! &
			\mre^{-\lambda_1 t}
			\phi_1 (0),
			\\[0.1cm]
			\phi_2 (t)
			& \!\! = \!\! &
			\mre^{-\lambda_2 t}
			\phi_2 (0)
			\; + \;
			\dfrac{R}{\lambda_2 \, - \, \lambda_1}
			\left( \mre^{-\lambda_1 t} \, - \, \mre^{-\lambda_2 t} \right)
			\phi_1 (0).
		\end{array}
	\end{equation}
	Thus, the initial conditions in $\phi_i$ create monotonically decaying responses of $\phi_i$, with a rate of decay determined by $\lambda_i$. In contrast, the response of $\phi_2$ arising from the initial condition in $\phi_1$ is characterized by two competing exponentials, and it vanishes both for $t = 0$ and for asymptotically large times. For finite times, however, transient growth, directly proportional to the coupling coefficient $R$, is exhibited with the largest value of transient response taking place at $t = (1/(\lambda_1 - \lambda_2)) \log \, (\lambda_1/\lambda_2)$. This transient growth does not require the presence of near resonances (i.e., $\lambda_1 \approx \lambda_2$) or modes with algebraic growth (i.e., $\lambda_1 = \lambda_2$); it is instead caused by the non-normality of the dynamical generator in~(\ref{EX}).
	
	We note that transient growth represents one particular manifestation of the non-normality of the dynamical generator in the above example (see~\cite{gro00} for a comprehensive treatment). Additional features can be observed by analyzing~(\ref{EX}) in the frequency domain. The frequency response is obtained by evaluating the transfer function $H (s)$ (from input $d$ to output $\varphi$, $\varphi (s) = H(s) d(s)$) on the $\mri \omega$ axis, where $\omega \in \bbR$ is the temporal frequency and $\mri = \sqrt{-1}$. The largest value of $| H (\mri \omega) |$ determines the so-called $H_\infty$ norm. This measure of input-output amplification has an appealing physical interpretation: it quantifies the worst-case amplification of finite-energy disturbances~\citep{zhodoyglo96}. In the above example, $\| H \|_\infty = | R |/(\lambda_1 \lambda_2)$ indicates the existence of a unit energy disturbance that generates output whose energy is given by $| R |^2/(\lambda_1 \lambda_2)^2$.
	
	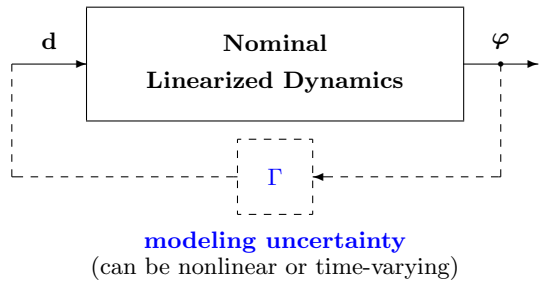
\begin{figure}
	\centering
		\begin{tabular}{c}
			{
			\setlength{\unitlength}{1.0cm}
    			\input{fbkGamma}
    			}
			\\[1.0cm]
    		\end{tabular}
	\caption{Block diagram of a system connected in feedback with norm-bounded unstructured uncertainty $\Gamma$.}
	\label{fig.fbkgamma}
	\end{figure}

	Furthermore, the $H_\infty$ norm has an interesting robustness interpretation which is closely related to the analysis of pseudospectra of linear operators~\citep{treemb05}. Namely, $\| H \|_\infty$ determines the size of modeling uncertainty, $d (s) = \Gamma (s) \varphi (s)$, that can destabilize the system; see figure~\ref{fig.fbkgamma} for an illustration. This uncertainty may arise from the inevitable imperfections in the laboratory environment or from the approximate nature of the governing equations (caused by, e.g., high-frequency unmodeled dynamics,  parametric variations, neglected nonlinearities, or crude physical assumptions made in modeling). In particular, system~(\ref{EX}) with $d (s) = \Gamma (s) \varphi (s)$ can be represented by a feedback interconnection in figure~\ref{fig.sg}. If, apart from being norm-bounded, there are {\em no structural restrictions\/} on uncertainty $\Gamma$, then the {\em necessary and sufficient condition\/} for stability of a feedback interconnection in figure~\ref{fig.sg} is given by the so-called small-gain theorem, $\| \Gamma \|_\infty < 1/\| H \|_\infty$. In the above example, this condition simplifies to $\| \Gamma \|_\infty < \lambda_1 \lambda_2/|R|$. In particular, it is easy to establish the existence of a constant gain uncertainty, $\Gamma (s) = \gamma = \mbox{\rm const.}$, of magnitude larger than $\lambda_1 \lambda_2/|R|$ that makes the system in figure~\ref{fig.sg} unstable.

	The above example illustrates that, in systems with non-normal generators, the eigenvalues may represent misleading measures of both transient growth and input-output amplification. While they successfully predict system's behavior for asymptotically large times, they may fail to capture important aspects of short-time behavior, disturbance propagation, and robustness. In particular, the coupling $R$ between subsystems in~(\ref{EX}) plays a crucial role in determining transient and input-output features of the system's response: large values of $R$ signal large transient responses, poor stability margins, and large amplification of disturbances. In the absence of the coupling, i.e.,\ for $R = 0$, the stability margins of subsystems in~(\ref{EX}) are determined by $\lambda_1$ and $\lambda_2$. In contrast, for non-zero $R$ this margin is determined by $\lambda_1 \lambda_2/|R|$, thereby indicating that, even for subsystems with large stability margins, small uncertainties can have a destabilizing effect on the overall system if the coupling between the subsystems is large enough.

	The main ideas from the above example extend to multivariable and infinite dimensional systems. For these problems, the {\em singular values\/} of the frequency response operator can be used to determine input-output amplification in the presence of disturbances. Furthermore, the analysis of spatio-temporal frequency responses for spatially distributed systems can be used to identify prevalent spatial length scales and spatio-temporal patterns that are most amplified by the system's dynamics. For example, if instead of being constant scalars, parameters $\lambda_i$ and $R$ in~(\ref{EX}) are given by $\{ \lambda_i = a_i + b_i \kappa^2$, $R = c \, \mri \kappa \}$, then~(\ref{EX}) can be interpreted as an equivalent of the following system
	\begin{equation*}
	\begin{array}{rcl}
		\phi_{1t} (x,t)
		& \!\! = \!\! &
		b_1 \, \phi_{1xx} (x,t)
		\, - \,
		a_1 \, \phi_1 (x,t)
		\, + \,
		d (x,t),
		\\[0.1cm]
		\phi_{2t} (x,t)
		& \!\! = \!\! &
		c \, \phi_{1x} (x,t)
		\, + \,
		b_2 \, \phi_{2xx} (x,t)
		\, - \,
		a_2 \, \phi_2 (x,t),
		\\[0.1cm]
		\varphi (x,t)
		& \!\! = \!\! &
		\phi_2 (x,t),
		~~~
		x
		\, \in \,
		\bbR,
	\end{array}
	\end{equation*}
	in the spatial frequency domain after applying the spatial Fourier transform to the above system. Here, $\kappa \in \bbR$ denotes the spatial wavenumber, and $(a_i,b_i,c)$ denote positive reaction, diffusion, and convection coefficients. The $\kappa$-parameterized $H_\infty$ norm,
	    $
	    c
	    | \kappa |
	    /
	    ( (a_1 + b_1 \kappa^2) (a_2 + b_2 \kappa^2) ),
	    $
	disappears for $\kappa = 0$ and as $\kappa \rightarrow \infty$, thereby achieving its peak for non-zero $\kappa$, $\bar{\kappa}$. This value of $\kappa$ identifies the spatial length scale, $2 \pi/\bar{\kappa}$, that has the smallest stability margin and that is most amplified by deterministic disturbances. Thus, convective coupling in reaction-diffusion systems can provide dynamical responses that cannot be inferred by analyzing subsystems in isolation.
	
	We finally note that this simple example captures the essential features of nonmodal amplification in wall-bounded shear flows of both Newtonian and viscoelastic fluids. In Newtonian fluids the subsystems in figure~\ref{fig.block} would correspond to the Orr-Sommerfeld and Squire equations and the coupling between them would represent the vortex tilting term whose strength is directly proportional to the Reynolds number~\citep{jovbamJFM05}. In a study focusing on transient growth in inertialess channel flows of viscoelastic fluids,~\citet{jovkumPOF10} showed that polymer stretching and the Weissenberg number effectively take the role that vortex tilting and the Reynolds number play in inertial flows of Newtonian fluids.

    \vspace*{-2ex}
\section{Problem formulation}
\label{sec.prob-form}
	
In this section, we present the governing equations for inertialess shear-driven channel flow of viscoelastic fluids. We show how the linearized equations can be cast into an evolution form that is amenable to both analytical and computational developments. We then provide a brief description of frequency responses and input-output norms, along with the numerical tools for computing them.
	
\subsection{Governing equations}
	\label{sec.governing-eqs}
	
\begin{figure}
		\begin{center}
			{
				\includegraphics[width=0.5\columnwidth]
				{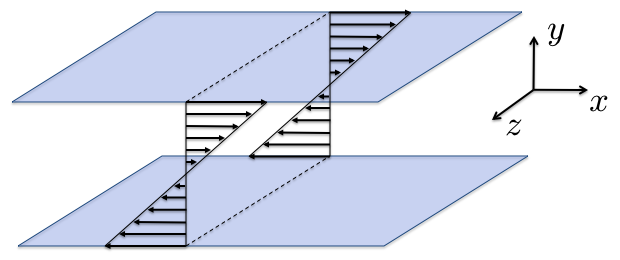}
			}
		\end{center}
		\caption{Geometry of a three-dimensional shear-driven channel flow.}
		\label{fig.channel}
	\end{figure}

	The non-dimensional momentum, continuity, and constitutive equations for an incompressible shear-driven channel flow of viscoelastic fluids, with geometry shown in figure~\ref{fig.channel}, are given by~\citep{bircurarmhas87v2,lar99}
	\begin{subequations}
	\label{eq.governing}
		\begin{align}
			\label{eq.momentum}
			Re \, \dot{\bV}
			\;\; = \;\; &
			\We
			\left(
			\left(1 \, - \, \beta \right) \bnabla \cdot \bT
			\, + \,
			\beta \, \bnabla^2 \bV
			\, - \,
			\bnabla P
			\, - \,
			Re \, \bV \cdot \bnabla \bV
			\right),
			\\[0.15cm]
			\label{eq.continuity}
			0
			\;\; = \;\; &
			\nabla \cdot \bV,
			\\[0.1cm]
			\label{eq.constitutive}
			\dot{\bR}
			\;\; = \;\; &
			\We
			\left(
			\bR \cdot \bnabla \bV
			\, + \,
			\left(
			\bR \cdot \bnabla \bV
			\right)^{T}
			\, - \,
			\bV \cdot \bnabla \bR
			\, - \,
			\bT
			\right).
		\end{align}
	\end{subequations}
	Here, the overdot denotes a partial derivative with respect to time $t$, $\bV$ is the velocity vector, $P$ is the pressure, $\bT$ is the polymer stress tensor, $\bR$ is the conformation tensor, $\bnabla$ is the gradient, and $\bnabla^2$ is the Laplacian. System~(\ref{eq.governing}) governs the behavior of dilute polymer solutions with fluid density $\rho$, and it has been obtained by scaling length with the channel half-height $h$, time with the fluid relaxation time $\lambda$, velocity with the largest base flow velocity $U_0$, polymer stresses with $\eta_{p} U_{0} / h$, and pressure with $\left( \eta_{s} + \eta_{p} \right) U_0 / h$, where $\eta_{s}$ and $\eta_{p}$ are the solvent and polymer viscosities. This scaling leads to three parameters that characterize the properties of~(\ref{eq.governing}): the viscosity ratio, $\beta = \eta_{s}/\left( \eta_{s} + \eta_{p} \right)$; the Weissenberg number, $\We = \lambda U_{0} / h$; and the Reynolds number, $Re = \rho U_{0} h / (\eta_{s} + \eta_{p})$. While the Reynolds number quantifies the ratio of inertial to viscous forces, the Weissenberg number determines the ratio of the fluid relaxation time $\lambda$ to the characteristic flow time $h / U_{0}$.
	
	The momentum~(\ref{eq.momentum}) and continuity~(\ref{eq.continuity}) equations describe the motion of an incompressible viscoelastic fluid. For given $\bT$, the pressure adjusts itself so that the velocity satisfies the continuity equation~(\ref{eq.continuity}). In our previous work~\citep{hodjovkumJFM08,hodjovkumJFM09,jovkumPOF10,jovkumJNNFM11}, we used the Oldroyd-B model, which is based on a linear bead-spring dumbbell, to relate the polymeric stress tensor to the conformation tensor. However, it is well-known that the Oldroyd-B model does not account for the finite extensibility of the polymer chains. In this work, we address this issue by using the FENE-CR model, which utilizes a nonlinear relationship between the polymeric stress tensor $\bT$ and the conformation tensor $\bR$~\citep{bircurarmhas87v2},
	\begin{equation}
	\label{eq.TC}
		\bT \; = \; \cfrac{f}{\We} \; \left( \bR \, - \, \bI \right).
	\end{equation}
Here, $\bI$ is the unit tensor, and the function $f$ (that quantifies the influence of the nonlinear spring) is determined by the trace of the conformation tensor, $\trace \left( \bR \right)$, and the square of the maximum extensibility of polymer chains,
	\begin{equation}
	\label{eq.f}
		f
        \; = \;
        \dfrac{L^2 \, - \, 3}{L^2 \, - \, \trace \left( \bR \right)}.
	\end{equation}
Note that $\bR$ and $L^2$ are made dimensionless with respect to $k T / c$, where $k$, $T$, and $c$ denote the Boltzmann constant, the absolute temperature, and the spring constant of the Hookean dumbbell, respectively. In the limiting case $L \rightarrow \infty$, we have $f \rightarrow 1$; consequently, the nonlinear spring becomes linear and the FENE-CR model simplifies to the Oldroyd-B model.

	In a shear-driven channel flow, system of equations~(\ref{eq.governing})~--~\eqref{eq.TC} exhibits the following steady-state solution
	\begin{equation*}
		\begin{array}{rcl}
			\bar{\bv}
			& = &
			\left[
			\begin{array}{ccc}
				U(y) & 0 & 	0
			\end{array}\right]^{T},
			\\[0.3cm]
			\bar{\bR}
			& = &
			\left[
			\begin{array}{ccc}
				\Rbar_{11} & \Rbar_{21} & \Rbar_{13} \\[0.1cm]
				\Rbar_{12} & \Rbar_{22} & \Rbar_{23} \\[0.1cm]
				\Rbar_{13} & \Rbar_{23} & \Rbar_{33}
			\end{array}
			\right]
			\, = \,
			\left[
			\begin{array}{ccc}
				1 + 2 \We^2 / \bar{f}^2 & \We / \bar{f} & 0 \\[0.1cm]
				\We / \bar{f} & 1 & 0 \\[0.1cm]
				0 & 0 & 1
			\end{array}
			\right],
		\end{array}
	\end{equation*}
	where
	\begin{equation*}
		U(y)
		\; = \;
		y,
		\;\;\;\;
		\bar{f}
		\; = \;
		\cfrac{1}{2}
		\left( 1
		\, + \,
		\sqrt{1 \, + \, \cfrac{8 \, \We^2}{\bar{L}^2}} \right),
        \;\;\;\;
        \bar{L}^2
        \; = \;
        L^2
        \, - \,
        3.
	\end{equation*}
	We also note that the first normal stress difference in Couette flow is determined by
    \beq
    \bar{N}_1
    \; = \;
    \Rbar_{11}
    \, - \,
    \Rbar_{22}
    \; = \;
    2 \left( \We / \bar{f} \right)^2.
    \non
    \eeq
In~\S~\ref{sec.kx0} and~\S~\ref{sec.polymer-stress-2d3c} we will show that this parameter, that can take values between $0$ and $\bar{L}^2$, plays a key role in the dynamics of velocity and polymer stress fluctuations.
	
	In the absence of inertia, i.e.\ in flows with $Re = 0$, the dynamics of infinitesimal velocity, $\bv$, pressure, $p$, and conformation tensor, ${\brr}$, fluctuations around the base flow $\left( \bar{\bv}, \bar{\bR} \right)$ are governed by
	\begin{subequations}
	\label{eq.stokes-governing}
		\begin{align}
			\label{eq.stokes}
			0
			\;\; = \;\; &
			-\bnabla p
			\, + \,
			\left(1 \, - \, \beta \right) \bnabla \cdot \btau
			\, + \,
			\beta \, \bnabla^2 \bv
			\, + \,
			\bd,
			\\[0.1cm]
			\label{eq.vcontinuity}
			0
			\;\; = \;\; &
			\bnabla \cdot \bv,
			\\[0.1cm]
			\nonumber
			\dot{\brr}
			\;\; = \;\; &
			\We
			\,
			\Big(
			{\brr} \cdot \bnabla \bar{\bv}
			\, + \,
			\bar{\bR} \cdot \bnabla \bv
            		\, + \,
			\left( {\brr} \cdot \bnabla \bar{\bv} \right)^{T}
			\, + \,
			\\[0.1cm]
			\label{eq.tconstitutive}
			&
			\left( \bar{\bR} \cdot \bnabla \bv \right)^{T}
			\, - \,
			\bv \cdot \bnabla \bar{\bR}
			\, - \,
			\bar{\bv} \cdot \bnabla {\brr}
			\, - \,
			\btau
			\Big),
			\\[0.1cm]
            		\label{eq.tau-r}
			\btau
			\;\; = \;\; &
			\cfrac{\bar{f}}{\We} \;
			\left(
			{\brr}
			\; + \;
			\cfrac{\bar{f} \left( \bar{\bR} \, - \, \bI \right)}{\bar{L}^2}
            		\;
			\trace({\brr})
			\right).
		\end{align}
	\end{subequations}
	Here,~(\ref{eq.tau-r}) establishes a relation between polymer stress and conformation tensor fluctuations. Furthermore, $u$, $v$, and $w$ are the components of the velocity fluctuation vector $\bv = \left[ \begin{array}{ccc} u & v & w \end{array} \right]^{T}$ in the streamwise ($x$), wall-normal ($y$), and spanwise ($z$) directions, respectively. The momentum equation~(\ref{eq.stokes}) is driven by a spatially distributed and temporally varying body forcing, $\bd = \left[ \begin{array}{ccc} d_1 & d_2 & d_3 \end{array} \right]^{T}$, where $d_1$, $d_2$, and $d_3$ are the forcing fluctuations in the streamwise, wall-normal, and spanwise directions, respectively. In prior work using the Oldroyd-B constitutive equations~\citep{hodjovkumJFM08,hodjovkumJFM09,jovkumJNNFM11}, the three-dimensional body forcing varies harmonically in the horizontal directions and stochastically in the wall-normal direction and in time. However, given the static-in-time momentum equation~(\ref{eq.stokes}), white-in-time stochastic disturbances $\bd$ induce white-in-time velocity fluctuations $\bv$, and the problem of variance amplification (in the absence of inertia) becomes ill-posed~\citep{jovkumJNNFM11}. Hence, in this work, we consider the body forcing $\bd$ to be purely harmonic in the horizontal directions and time, and deterministic in the wall-normal direction, and we study the worst-case amplification of deterministic disturbances.

	\subsection{Model in the evolution form}
	
	We note that~(\ref{eq.stokes}) and~(\ref{eq.vcontinuity}) can be simplified by expressing the velocity fields in terms of the wall-normal velocity ($v$) and vorticity ($\eta = \pz u - \px w$) fluctuations. This is done by first taking the divergence of~(\ref{eq.stokes}) and using~(\ref{eq.vcontinuity}) to get an expression for $p$. The equation for $v$ is then obtained by eliminating $p$ from~(\ref{eq.stokes}). The equation for $\eta$ can be obtained by taking the curl of~(\ref{eq.stokes}). Finally, by rearranging the components of the conformation tensor into the vector
	\[
	\bphi
	\; = \;
	\left[
	\begin{array}{cc}
	\bphi_{1}^T & \bphi_{2}^T
	\end{array}
	\right]^{T},
	\]
	with
	\[
	\bphi_1
	\; = \;
	\left[
	\begin{array}{ccc}
	r_{22} & r_{23} & r_{33}
	\end{array}
	\right]^{T},
	\;\;\;
	\bphi_2
	\; = \;
	\left[
	\begin{array}{ccc}
	r_{13} & r_{12} & r_{11}
	\end{array}
	\right]^{T},
	\]
	and by applying the Fourier transform in the $x$- and $z$-directions, we arrive at the following static-in-time expressions for $v$ and $\eta$ in terms of the conformation tensor and body-forcing fluctuations
	\begin{equation}
	\label{eq.veta}
		\begin{array}{rcl}
			v
			& = &
			\bC_{v} \, \bphi
			\, + \,
			\bD_{v} \, \bd,
			\\[0.1cm]
			\eta
			& = &
			\bC_{\eta} \, \bphi
			\, + \,
			\bD_{\eta} \, \bd.
		\end{array}
	\end{equation}
	In addition, equation~(\ref{eq.tconstitutive}) can be brought to the following form
	\begin{equation}
	\label{eq.psi}
		\begin{array}{rcl}
			\dot{\bphi}_{1}
			& = &
			\bF_{1 1} \, \bphi_{1} \, + \, \bF_{1 v} \, v \, + \, \bF_{1 \eta} \, \eta,
			\\[0.1cm]
			\dot{\bphi}_{2}
			& = &
			\bF_{2 1} \, \bphi_{1} \, + \, \bF_{2 2} \, \bphi_{2} \, + \, \bF_{2 v} \, v \, + \, \bF_{2 \eta} \, \eta.
		\end{array}
	\end{equation}
	The operators in~(\ref{eq.veta}) and~(\ref{eq.psi}) are defined in Appendix~\ref{sec.app-operators}. For notational convenience, we have suppressed the dependence of $\{ v, \eta, \bphi_{i}, \bd \}$ on $( \bkappa, y, t; \beta, \We, L )$, where $\bkappa = (k_x, k_z)$ with $k_x$ and $k_z$ denoting the horizontal wavenumbers.
	
	The boundary conditions on the wall-normal velocity and vorticity are dictated by the no-slip and no-penetration requirements
	\begin{equation*}
		v \left( \bkappa, y = \pm 1, t \right)
        \, = \,
        \py v \left( \bkappa, y = \pm 1, t \right)
        \, = \,
        \eta \left( \bkappa, y = \pm 1, t \right)
        \, = \,
        0.
	\end{equation*}
	We note that there are no boundary conditions on the components of $\bR$.
		
	An evolution model for~(\ref{eq.stokes-governing}) can be obtained by substituting~\eqref{eq.veta} into~\eqref{eq.psi} which yields
	\begin{subequations}
	\label{eq.ss}
		\begin{align}
			\dot{\bphi} \left( \bkappa, y, t \right)
			\; = & \;\;
			\bA \left( \bkappa \right) \, \bphi \left( \bkappa, y, t \right)
			\, + \,
			\bB \left( \bkappa \right) \, \bd \left( \bkappa, y, t \right),
			\\[0.1cm]
			\bv \left( \bkappa, y, t \right)
			\; = & \;\;
			\bC \left( \bkappa \right) \, \bphi \left( \bkappa, y, t \right)
			\, + \,
			\bD \left( \bkappa \right) \, \bd \left( \bkappa, y, t \right),
		\end{align}
	\end{subequations}
	where the operators $\bA$, $\bB$, $\bC$, and $\bD$ are defined in Appendix~\ref{sec.app-operators}.
	
	\subsection{Spatio-temporal frequency responses}
	
	Application of the temporal Fourier transform yields the frequency response operator for system~(\ref{eq.ss})
	\begin{equation}
		\bH \left( \bkappa, \omega; \beta, \We, L \right) \; = \; \bC \left( \bkappa \right) \left( \mri \, \omega \, \bI \, - \, \bA \left( \bkappa \right) \right)^{-1} \, \bB \left( \bkappa \right) \, + \, \bD \left( \bkappa \right),
		\label{eq.freq-response}
	\end{equation}
	where $\omega$ is the temporal frequency, and $\bI$ is the identity operator. For a stable system~(\ref{eq.ss}),~(\ref{eq.freq-response}) can be used to characterize the steady-state response to harmonic input signals across spatial wavenumbers $\bkappa$ and temporal frequency $\omega$. Namely, if the input
$\bd$ is harmonic in $x$, $z$, and $t$, i.e.,
    \[
    \bd(x, y, z, t)
    \, = \,
    \bar{\bd}(y)
    \,
    {\mre}^{ \mri \left(\bar{k}_{x} \, x \, + \ \bar{k}_{z} \, z \, + \, \bar{\omega} \, t \right)},
    \]
with $\bar{\bd}(y)$ denoting some spatial distribution in the wall-normal direction, then the output $\bv$ is also harmonic in $x$, $z$, and $t$ with the same frequencies but with a modified amplitude and phase
	\begin{equation*}
	\begin{array}{rcl}
    		\bv(x,y,z,t)
		& \!\! = \!\! &
		\left(
		\left[
		\bH \left( \bar{k}_{x}, \bar{k}_{z}, \bar{\omega} \right) \bar{\bd}
		\right] (y)
		\right)
		\mre^{ \mri \left( \bar{k}_x \, x \, + \, \bar{k}_{z} \, z \, + \, \bar{\omega} t \right)},
    		\non
	\end{array}
	\end{equation*}
	where the amplitude and phase are precisely determined by the frequency response at the input frequencies ($\bar{k}_{x}, \bar{k}_{z},\bar{\omega}$). Note that we have dropped the dependence of the frequency response operator on $\We$, $\beta$, and $L$ for notational convenience.

	The $n$th singular value of the frequency response operator $\bH$ is determined by
	\beq
		\sigma_{n}^2
		\left( \bH \right)
		~=~
		\lambda_{n}
		\left(
		\bH^{\star}
        \,
        \bH
		\right),
	\non
	\eeq
	where $\lambda_{n} (\cdot)$ denotes the $n$th eigenvalue of a given self-adjoint operator and $\bH^{\star}$ is the adjoint of $\bH$. For any $(k_x,k_z,\omega)$, $\sigma_{\max} (\bH) = \max_n \sigma_n ( \bH)$ determines the largest amplification from $\bd$ to $\bv$. Furthermore, the temporal supremum of the maximal singular value of $\bH$ determines the $H_\infty$ norm of system~(\ref{eq.ss})~\citep{zhodoyglo96}
	\beq
	\ba{rcl}
		G \left( \bkappa; \beta, \We, L \right)
		& = &
		{\displaystyle \sup_{\omega} \, \sigma_{\max}^{2} \left( \bH \left( \bkappa, \omega; \beta, \We, L \right) \right).}
	\ea
	\non
	\eeq
	This measure of input-output amplification has several appealing interpretations:
	\begin{itemize}
		\item[(a)] for any ($k_x,k_z$), the $H_\infty$ norm represents the worst-case amplification of purely harmonic (in $x$, $z$, and $t$) deterministic (in $y$) disturbances. This worst-case input-output gain is obtained by maximizing over input temporal frequencies ($\sup$ over $\omega$) and wall-normal shapes (maximal singular value of $\bH$);
		
		\item[(b)] in the temporal domain, the $H_\infty$ norm represents the energy gain from forcing to velocity fluctuations
	    \beq
	    G(\bkappa)
	    ~=~
	    \sup_{E_\bd (\bkappa) ~\leq~ 1}
	    \dfrac{E_\bv (\bkappa)}{E_\bd (\bkappa)},
	    \non
	    \eeq
where $E_\bv (\bkappa)$ denotes the \bkappa-parameterized energy of velocity fluctuations, i.e.,
	    \beq
	    E_\bv (\bkappa)
	    ~=~
	    \int_{0}^{\infty}
        \int_{-1}^{1}
        \bv^*(\bkappa,y,t)
        \,
        \bv(\bkappa,y,t)
        \, \mrd y
        \, \mrd t.
	    \non
	    \eeq
In other words, for a unit-energy forcing, $G (\bkappa)$ captures the largest possible energy of velocity fluctuations across wavenumbers $\bkappa$; and

		\item[(c)] at any ($k_x,k_z$), the inverse of the $H_\infty$ norm quantifies the size of an additive unstructured modeling uncertainty $\bGamma$ that can destabilize generator $\bA$ in~(\ref{eq.ss}). As described in \S~\ref{sec.prelim}, large $H_\infty$ norm indicates small stability margins (i.e., low robustness to modeling imperfections). For systems with poor robustness properties, even small modeling uncertainties (captured by operator $\bGamma$) can lead to instability of operator $\bA + \bGamma$.
	\end{itemize}
	
	We also note that the frequency response of system~(\ref{eq.ss}) can be further decomposed into $3 \times 3$ block-operator form
	\begin{equation}
		\begin{array}{rcl}
			\left[
			\begin{array}{c}
				u \\
				v \\
				w
			\end{array}
			\right]
			& \!\! = \!\! &
			\left[
			\begin{array}{ccc}
				\bH_{u 1} & \bH_{u 2} & \bH_{u 3}
				\\
				\bH_{v 1} & \bH_{v 2} & \bH_{v 3}
				\\
				\bH_{w 1} & \bH_{w 2} & \bH_{w 3}
			\end{array}
			\right]
			\left[
			\begin{array}{c}
				d_1 \\
				d_2 \\
				d_3
			\end{array}
			\right].
		\end{array}
		\label{eq.uHd}	
	\end{equation}
	This form is suitable for identifying forcing components that introduce the largest amplification of velocity fluctuations. In~(\ref{eq.uHd}), $\bH_{s j}$ maps $d_{j}$ to $s$, and
    \[
    G_{s j} (\bkappa;\beta, \We, L)
    \; = \;
    \sup_{\omega}
    \,
    \sigma_{\max}^{2} \left( \bH_{sj} \left(\bkappa, \omega; \beta, \We, L \right) \right),
    ~~
    s
    \, = \,
    \{ u, v, w\},
    ~~
    j \, = \, \{ 1, 2, 3\}.
    \]

	The finite-dimensional approximations of the underlying operators are obtained using {\sc Matlab} Differentiation Matrix Suite~\citep*{weired00}, which utilizes pseudospectral methods to approximate differential operators. After discretization in the wall-normal direction, each component in~(\ref{eq.uHd}) becomes an $N \times N$ matrix, where $N$ denotes the number of Chebyshev collocation points in $y$. All computations are performed in {\sc Matlab} and grid-point convergence is confirmed by running additional computations with larger number of grid points in $y$.

	After discretization in $y$, the $H_\infty$ norm of the frequency response matrix can, in principle, be computed by determining $\sigma_{\max}(\bH(\omega))$ for many values of $\omega$ and by choosing the resulting maximum value. However, there are two obvious problems associated with such a method: difficulty in determining the range and spacing of the temporal frequencies, and the large number of computations. To avoid these issues,~\citet*{boybalkab89} devised a bisection method that can efficiently compute the $H_\infty$ norm. Furthermore,~\cite{bruste90} introduced a fast algorithm that utilizes an efficient method of choosing the temporal frequency for computing the $H_{\infty}$ norm. This fast algorithm is utilized in our computations and it is based on the relation between the singular values of the frequency response matrix and the eigenvalues of a related Hamiltonian matrix.
	
	All of our results are confirmed by additional frequency response computations that utilize the integral formulation of~(\ref{eq.ss}). This is accomplished by rewriting the evolution equations~(\ref{eq.ss}) into an equivalent two-point boundary value problem and then reformulating it into a system of integral equations. The procedure for achieving this along with easy-to-use {\sc Matlab} source codes is provided in~\citet{liejovJCP11}. This new paradigm for computing frequency responses utilizes the {\sc Chebfun} computing environment~\citep{chebfunv4} and it exhibits superior numerical accuracy compared to conventional numerical schemes.
	
	\begin{figure}
   	\begin{center}
       		\begin{tabular}{m{5cm}m{5cm}m{5cm}}
			\begin{sideways}
				\hspace{0.2cm}
				{\large $k_x$}
			\end{sideways}
			\hspace*{-0.4cm}
			\begin{tabular}{c}
				$G(\bkappa;0.5,10,10)$
				\\
				{
				\includegraphics[width=0.33\columnwidth]
				{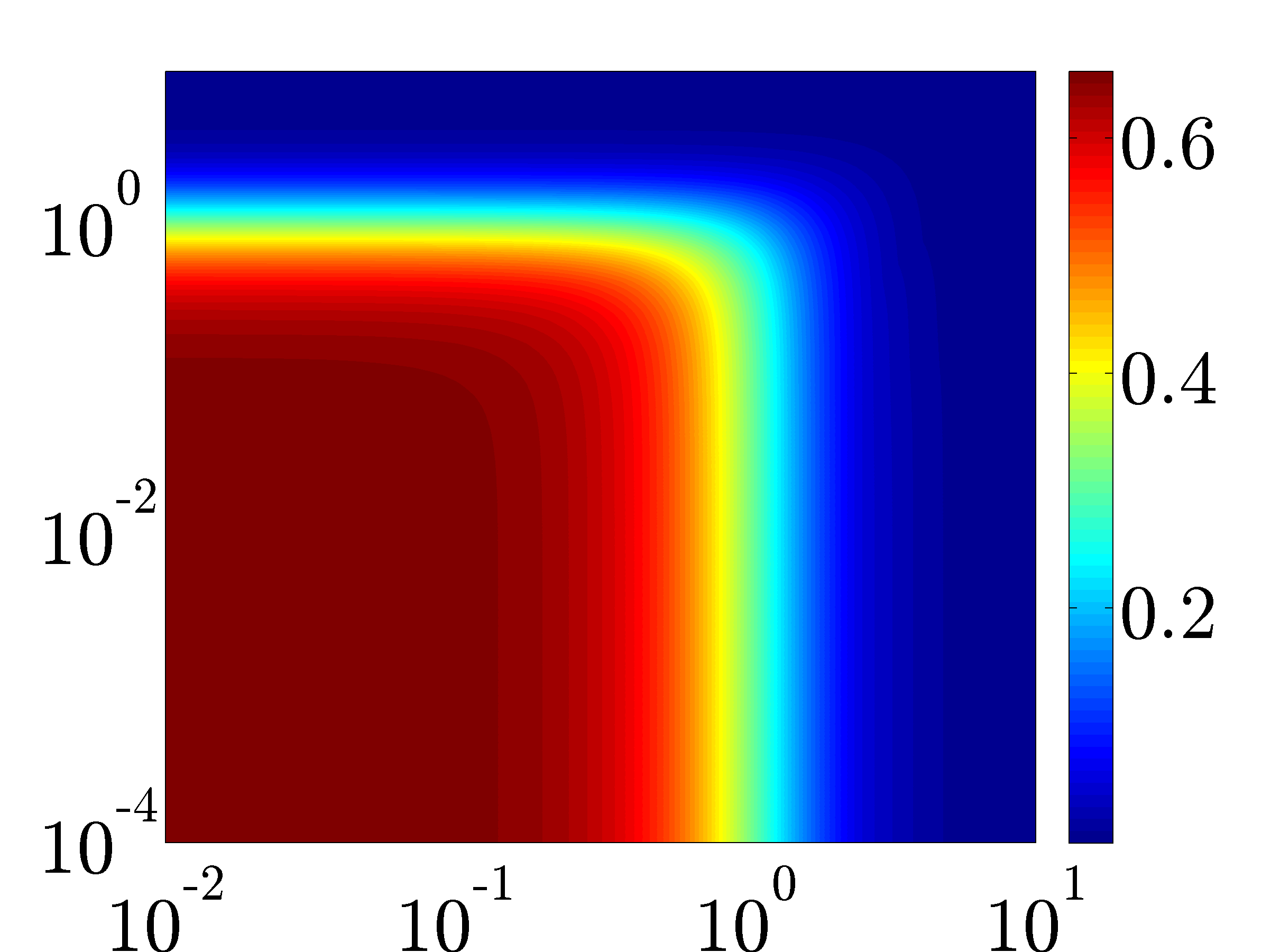}
				}
				\\
				{\large $k_z$}
				\\
				\subfigure[]{
				\label{fig.hinf-ud-We10-L10}
				}
			\end{tabular}
			&
			\hspace{-0.6cm}
			\begin{tabular}{c}
				$G(\bkappa;0.5,10,50)$
				\\
				{
				\includegraphics[width=0.33\columnwidth]
				{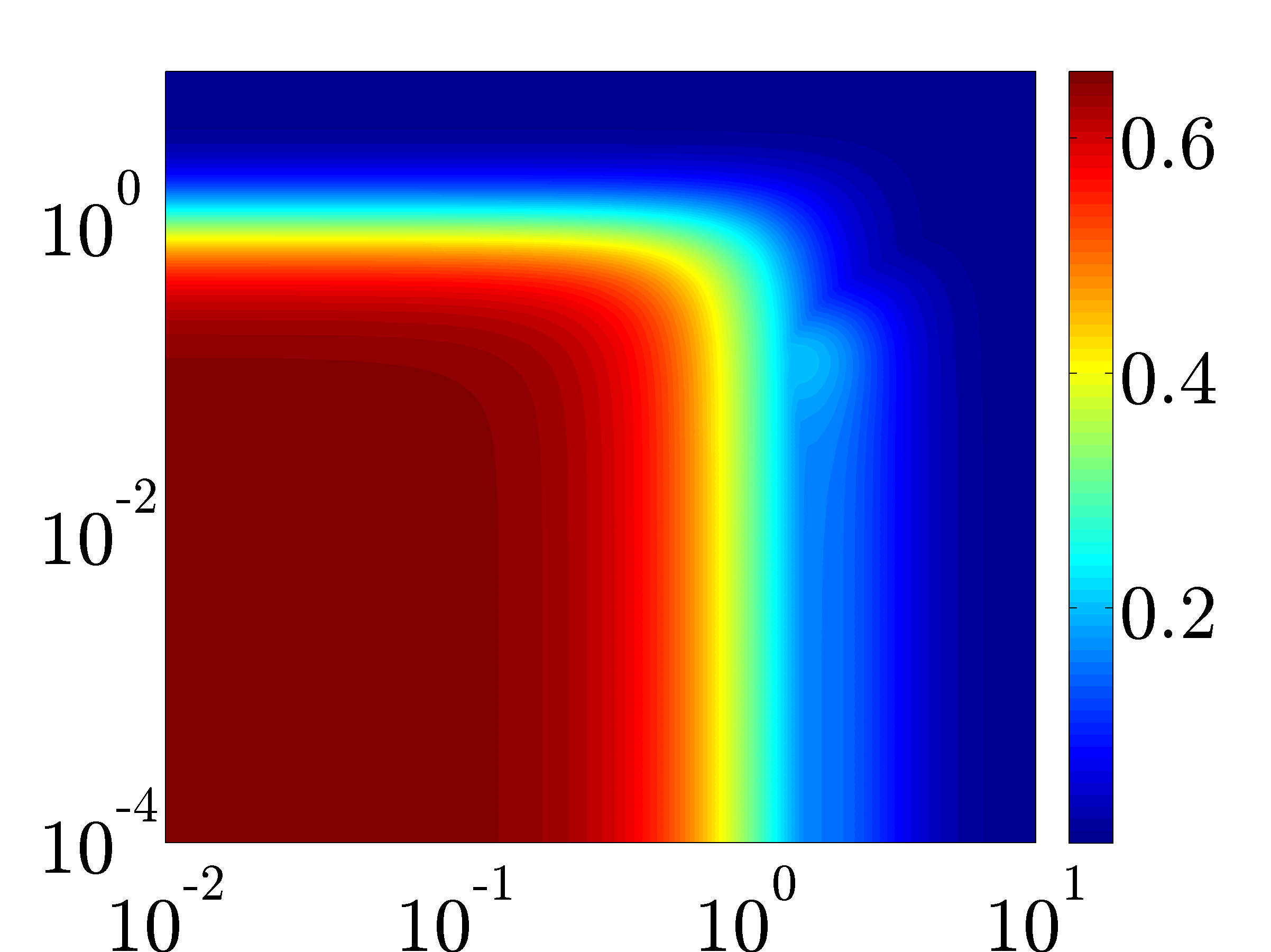}
				}
				\\
				{\large $k_z$}
				\\
				\subfigure[]{
				\label{fig.hinf-ud-We10-L50}
				}
			\end{tabular}
			&
			\hspace{-1.3cm}
			\begin{tabular}{c}
				$G(\bkappa;0.5,10,100)$
				\\
				{
				\includegraphics[width=0.33\columnwidth]
				{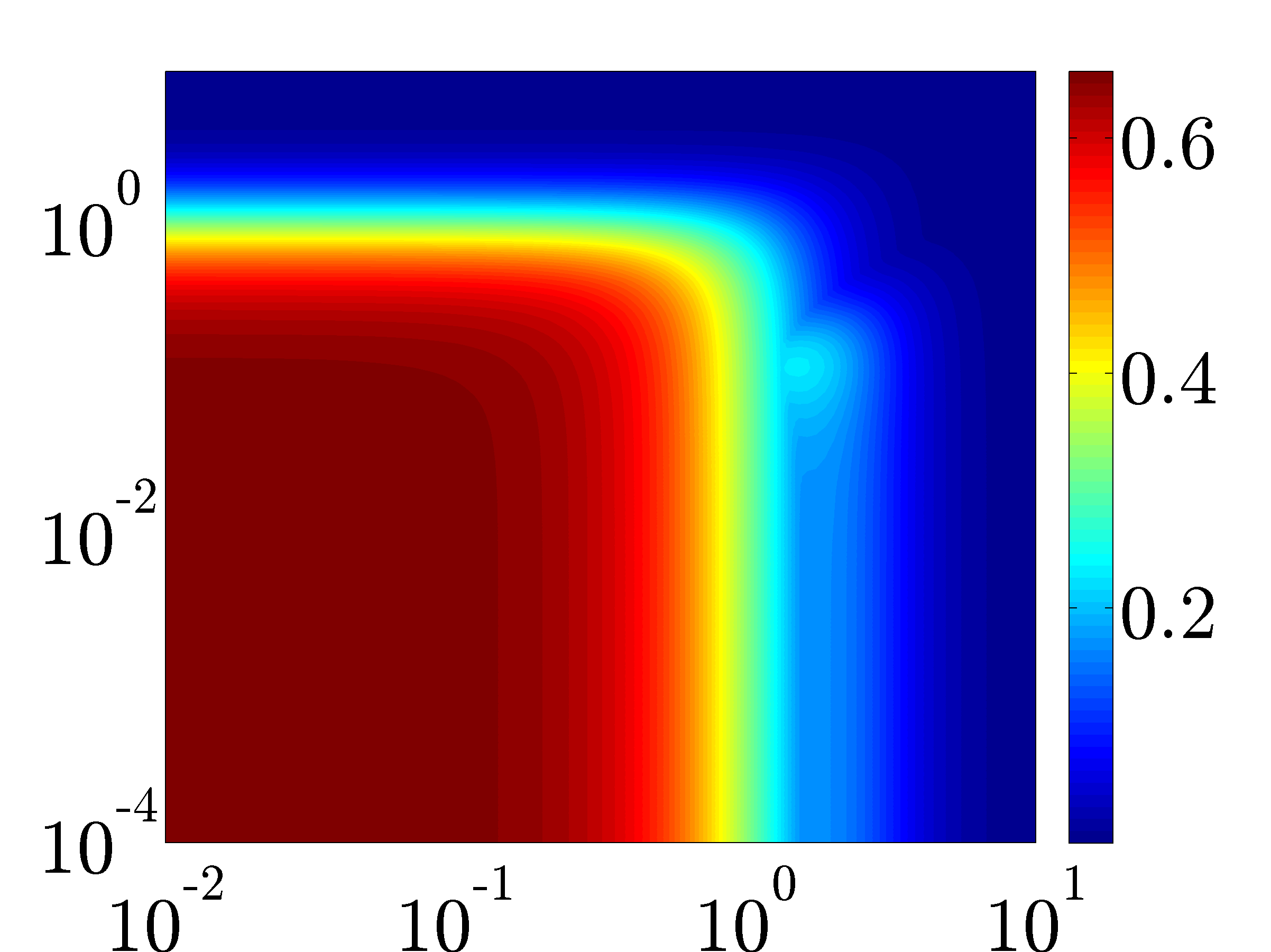}
				}
				\\
				{\large $k_z$}
				\\
				\subfigure[]{
				\label{fig.hinf-ud-We10-L100}
				}
			\end{tabular}
			\\
			\begin{sideways}
				\hspace{0.2cm}
				{\large $k_x$}
			\end{sideways}
			\hspace*{-0.4cm}
			\begin{tabular}{c}
				$G(\bkappa;0.5,50,10)$
				\\
				{
				\includegraphics[width=0.33\columnwidth]
				{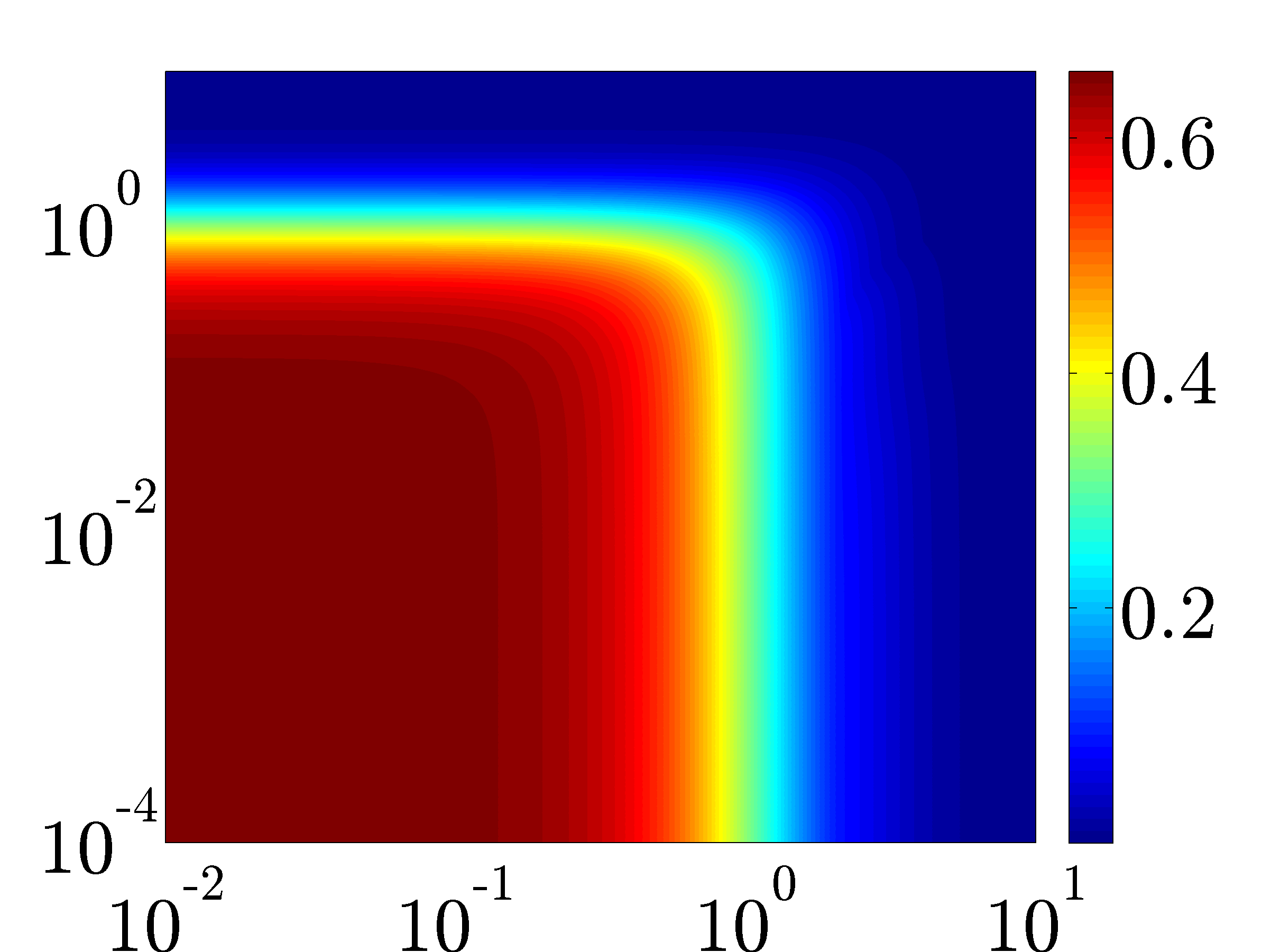}
				}
				\\
				{\large $k_z$}
				\\
				\subfigure[]{
				\label{fig.hinf-ud-We50-L10}
				}
			\end{tabular}
			&
			\hspace{-0.6cm}
			\begin{tabular}{c}
				$G(\bkappa;0.5,50,50)$
				\\
				{
				\includegraphics[width=0.33\columnwidth]
				{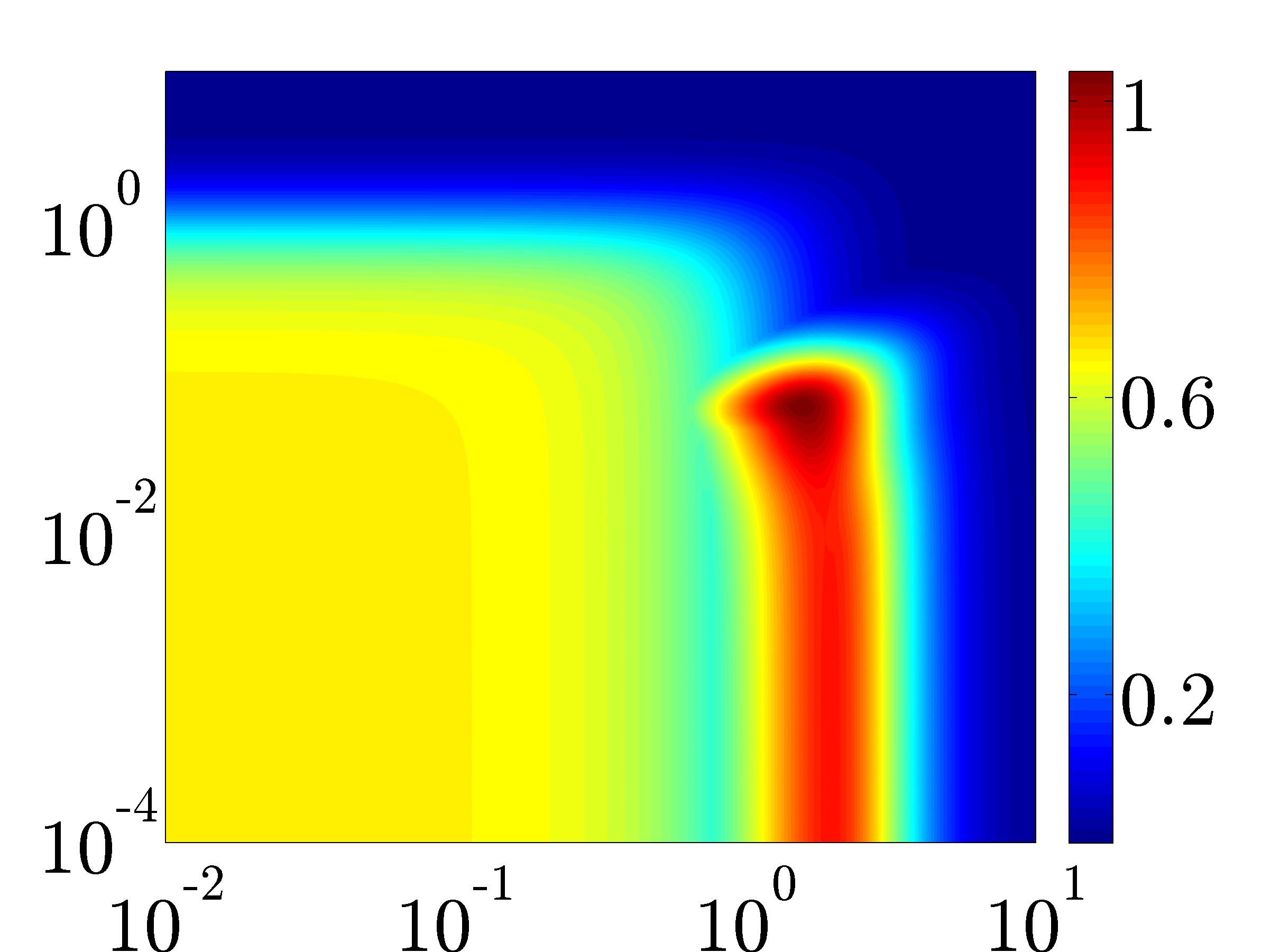}
				}
				\\
				{\large $k_z$}
				\\
				\subfigure[]{
				\label{fig.hinf-ud-We50-L50}
				}
			\end{tabular}
			&
			\hspace{-1.3cm}
			\begin{tabular}{c}
				$G(\bkappa;0.5,50,100)$
				\\
				{
				\includegraphics[width=0.33\columnwidth]
				{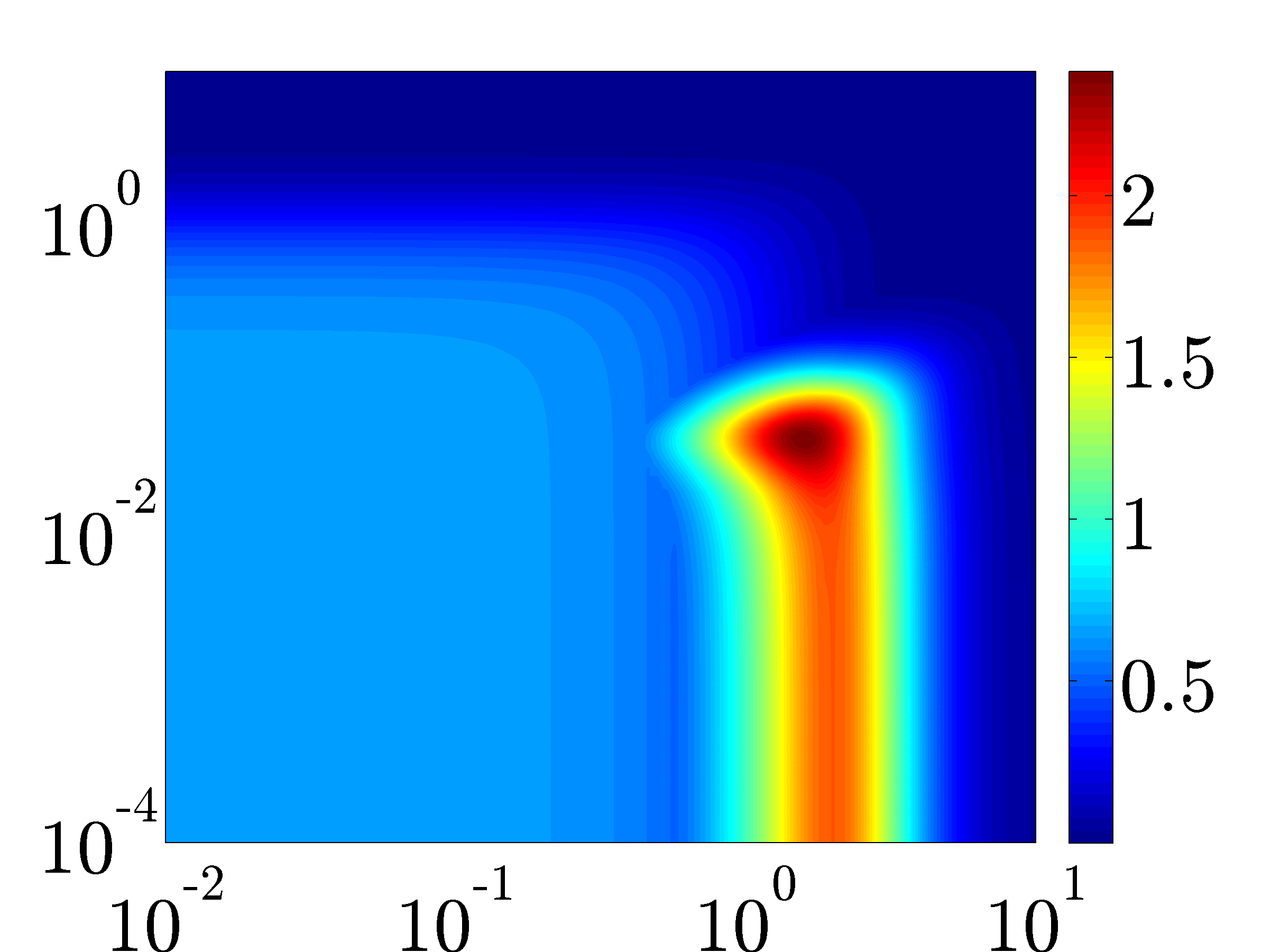}
				}
				\\
				{\large $k_z$}
				\\
				\subfigure[]{
				\label{fig.hinf-ud-We50-L100}
				}
			\end{tabular}
			\\
			\begin{sideways}
				\hspace{0.2cm}
				{\large $k_x$}
			\end{sideways}
			\hspace*{-0.4cm}
			\begin{tabular}{c}
				$G(\bkappa;0.5,100,10)$
				\\
				{
				\includegraphics[width=0.33\columnwidth]
				{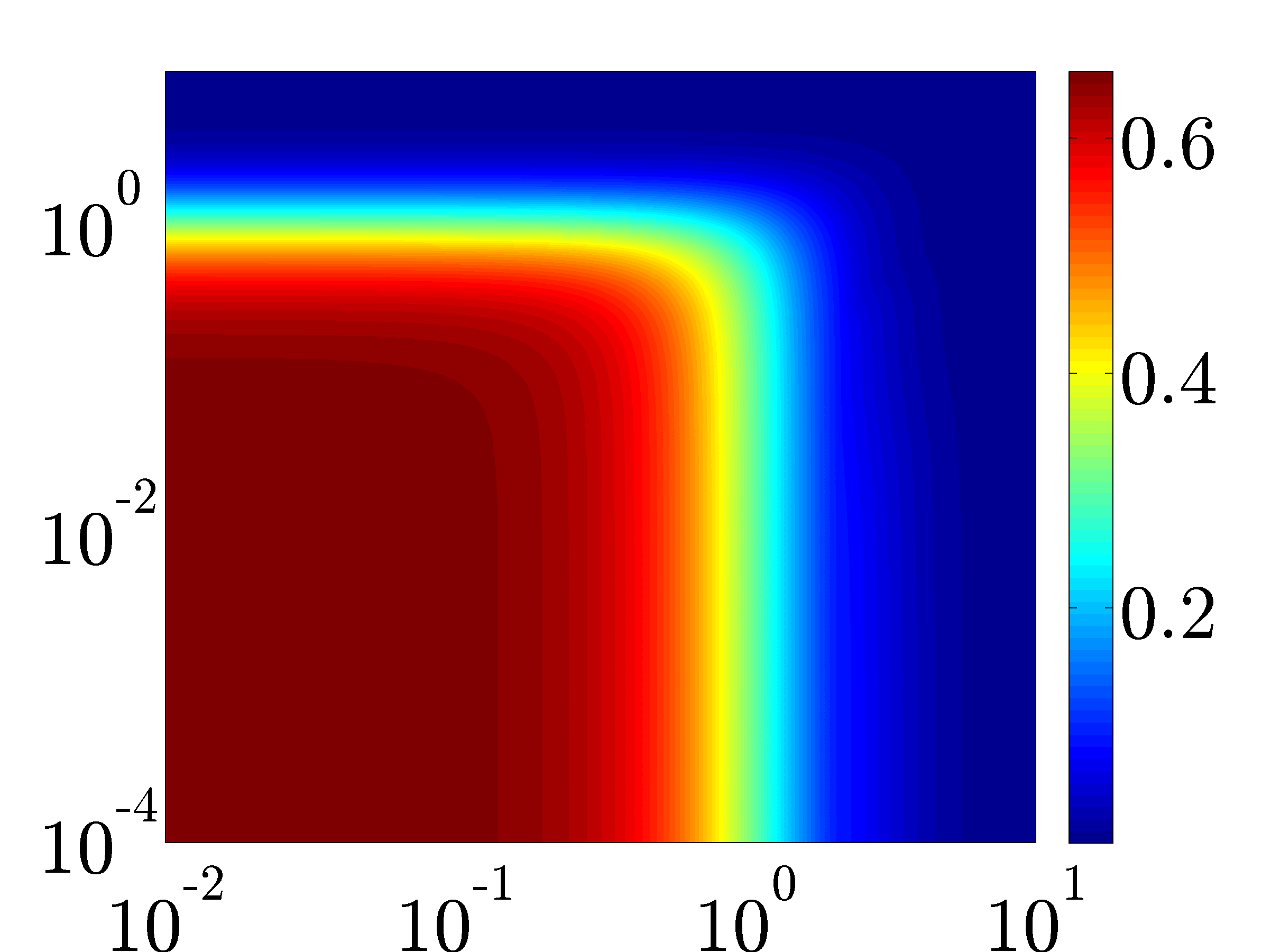}
				}
				\\
				{\large $k_z$}
				\\
				\subfigure[]{
				\label{fig.hinf-ud-We100-L10}
				}
			\end{tabular}
			&
			\hspace{-0.6cm}
			\begin{tabular}{c}
				$G(\bkappa;0.5,100,50)$
				\\
				{
				\includegraphics[width=0.33\columnwidth]
				{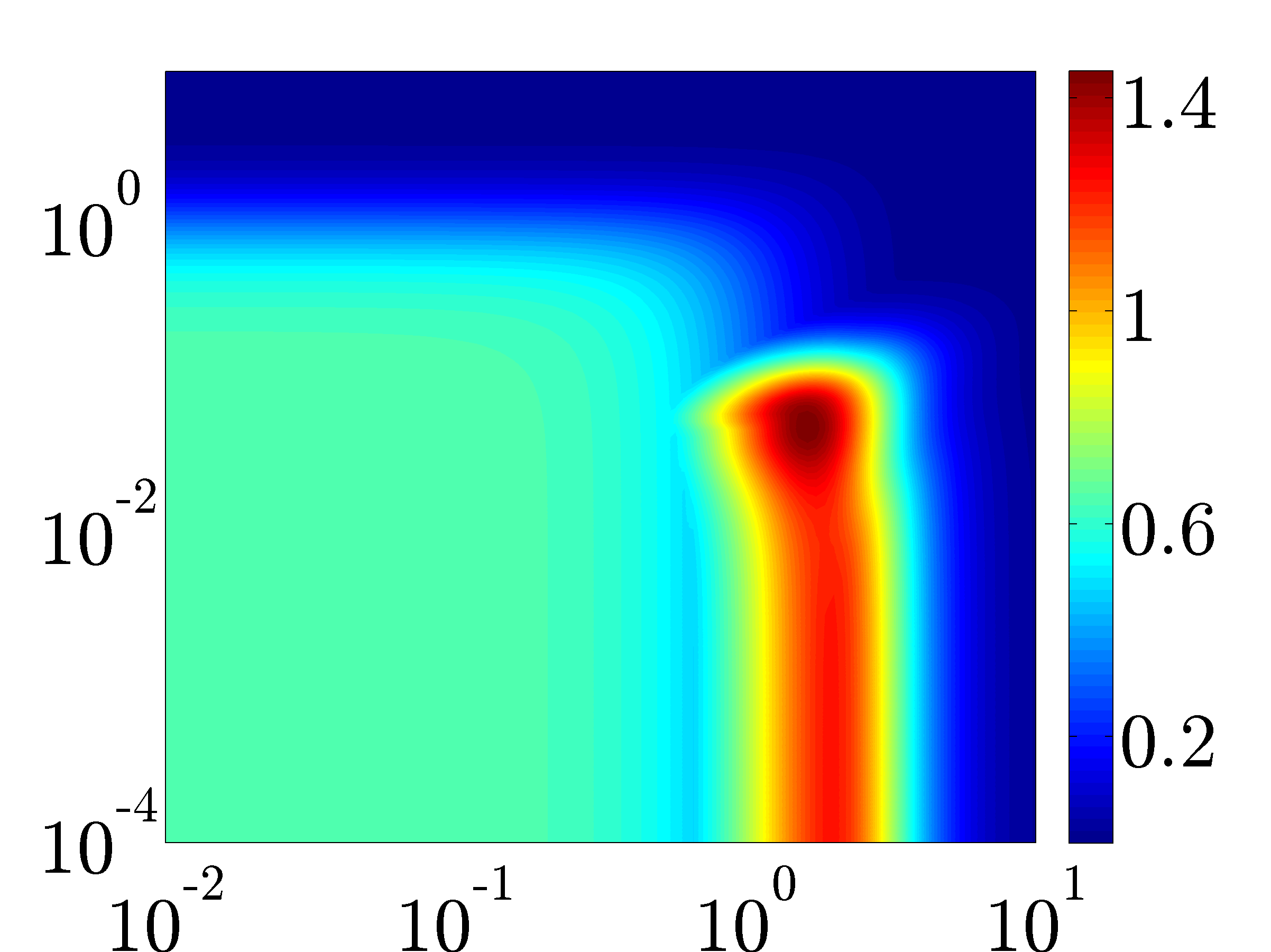}
				}
				\\
				{\large $k_z$}
				\\
				\subfigure[]{
				\label{fig.hinf-ud-We100-L50}
				}
			\end{tabular}
			&
			\hspace{-1.3cm}
			\begin{tabular}{c}
				$G(\bkappa;0.5,100,100)$
				\\
				{
				\includegraphics[width=0.33\columnwidth]
				{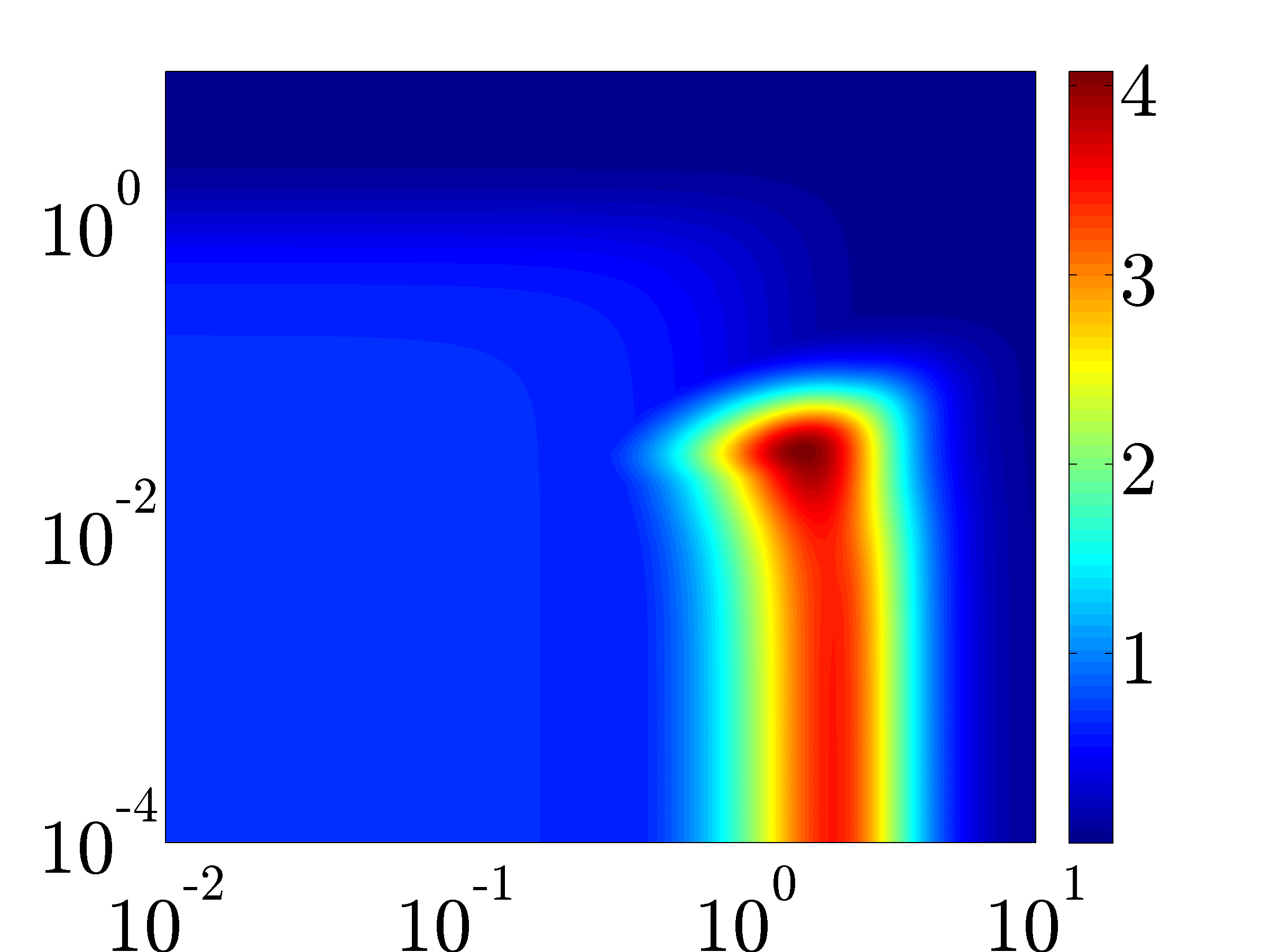}
				}
				\\
				{\large $k_z$}
				\\
				\subfigure[]{
				\label{fig.hinf-ud-We100-L100}
				}
			\end{tabular}
		\end{tabular}
	\end{center}
      	\caption{Worst-case amplification from $\bd$ to $\bv$ in Couette flow with $\beta = 0.5$, $\We = \{10, 50, 100\}$, and $L = \{10, 50, 100\}$: first row, $\We = 10$; second row, $\We = 50$; and third row, $\We = 100$.}
    	\label{fig.hinf-ud-kxkz}
 	\end{figure}
	
    \vspace*{-2ex}
\section{Frequency responses of 3D velocity fluctuations}
\label{sec.3d}

	In this section, we study the dynamics of three-dimensional velocity fluctuations in an inertialess shear-driven channel flow. In particular, we examine the worst-case amplification of deterministic disturbances and identify the corresponding wavenumbers that contain most energy. Our analysis shows that velocity fluctuations with large streamwise and ${\cal O}(1)$ spanwise length scales display the highest  sensitivity to disturbances (and consequently the lowest robustness to modeling imperfections). We further utilize the component-wise frequency responses~\citep{jovbamJFM05} to identify forcing components that have the strongest influence on the velocity fluctuations. In strongly elastic flows, we demonstrate that the wall-normal and spanwise forces have the highest impact, and that the streamwise velocity is most amplified by the system's dynamics.
	
	\begin{figure}
   	\begin{center}
       		\begin{tabular}{m{5cm}m{5cm}m{5cm}}
			\begin{sideways}
				\hspace{0.2cm}
				{\large $k_x$}
			\end{sideways}
			\hspace*{-0.4cm}
			\begin{tabular}{c}
				$G_{u1}(\bkappa;0.5,50,50)$
				\\
				{
				\includegraphics[width=0.33\columnwidth]
				{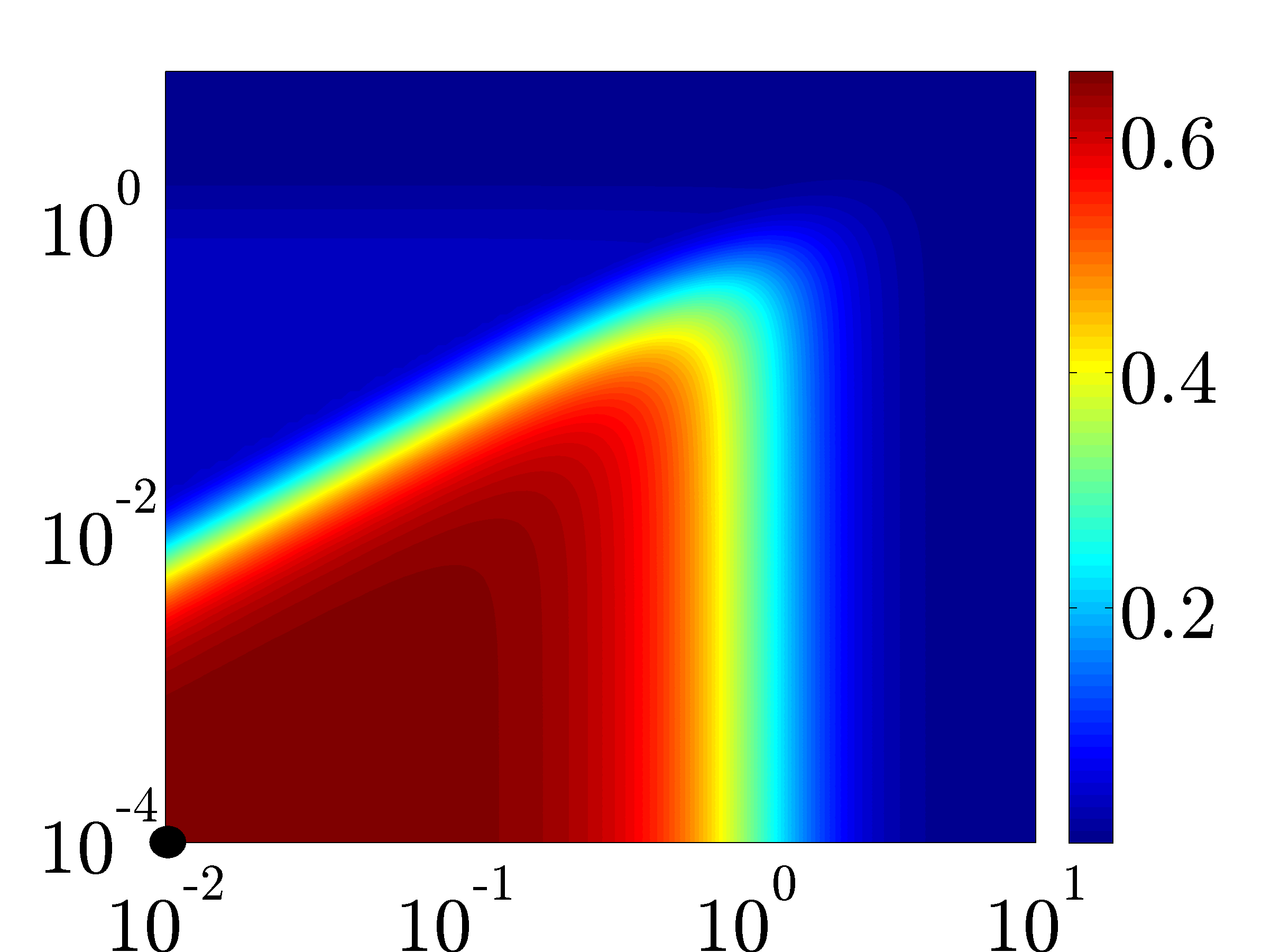}
				}
				\\
				{\large $k_z$}
				\\
				\subfigure[]{
				\label{fig.hinf-ud1}
				}
			\end{tabular}
			&
			\hspace{-0.6cm}
			\begin{tabular}{c}
				$G_{u2}(\bkappa;0.5,50,50)$
				\\
				{
				\includegraphics[width=0.33\columnwidth]
				{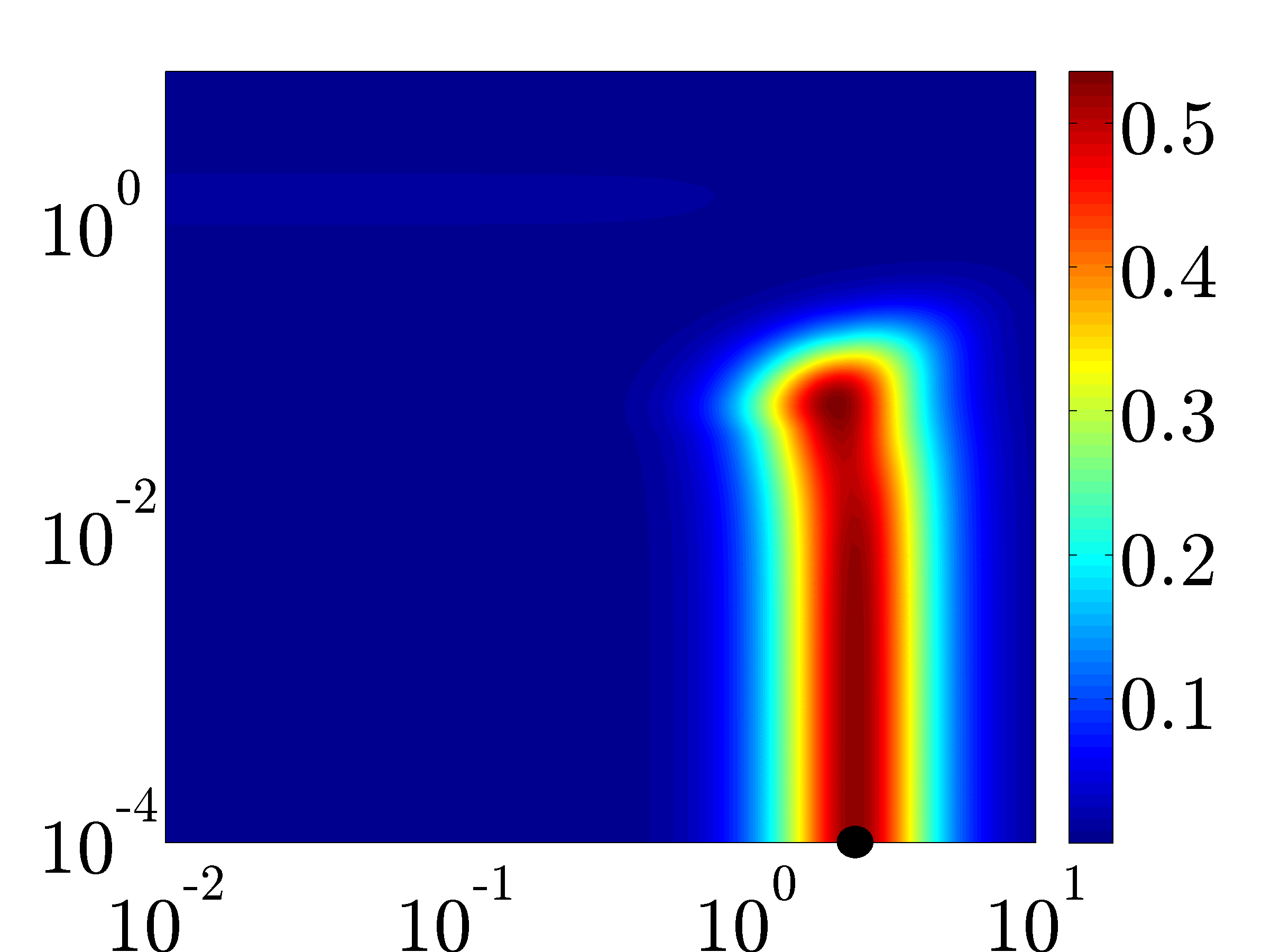}
				}
				\\
				{\large $k_z$}
				\\
				\subfigure[]{
				\label{fig.hinf-ud2}
				}
			\end{tabular}
			&
			\hspace{-1.3cm}
			\begin{tabular}{c}
				$G_{u3}(\bkappa;0.5,50,50)$
				\\
				{
				\includegraphics[width=0.33\columnwidth]
				{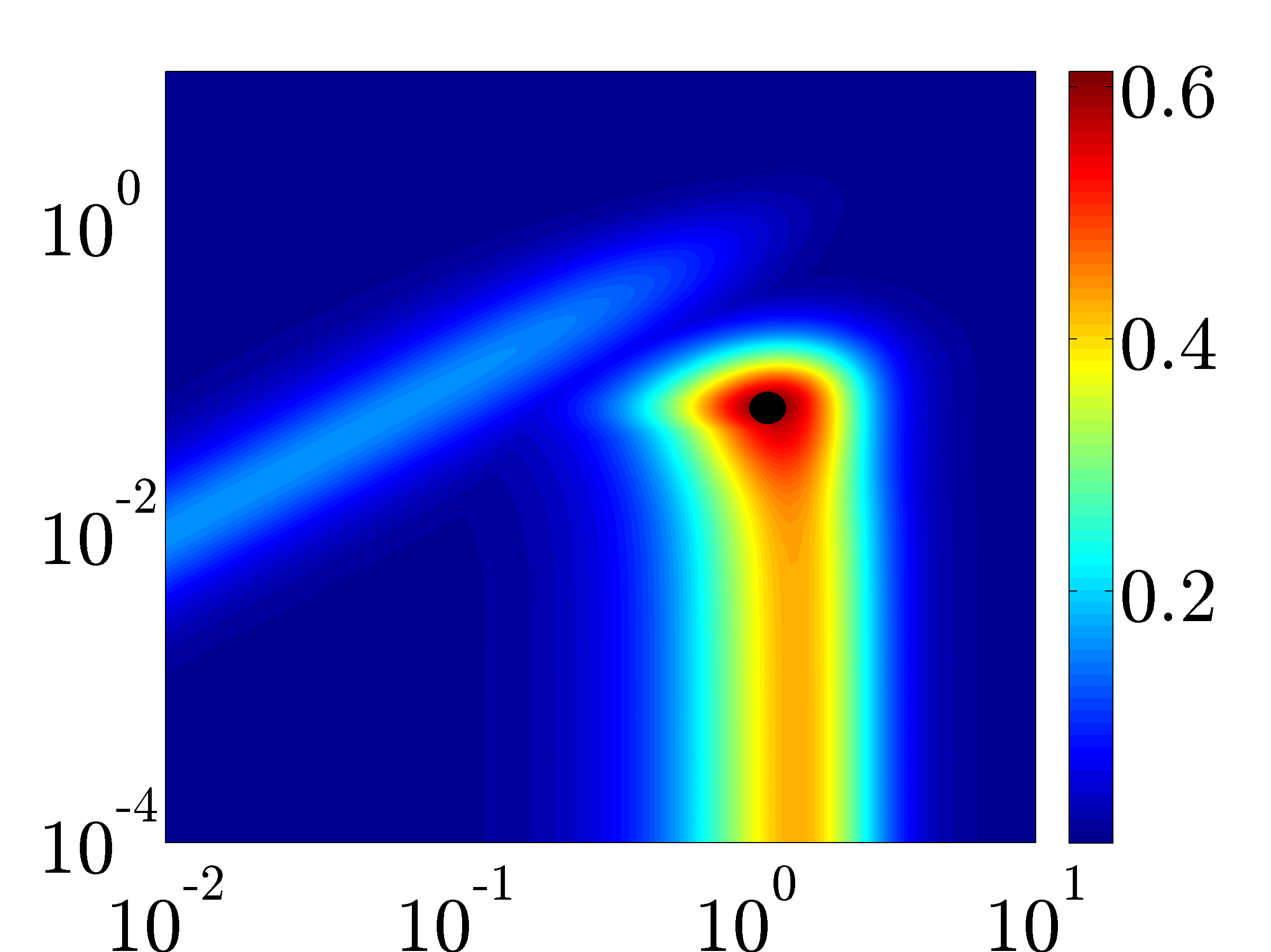}
				}
				\\
				{\large $k_z$}
				\\
				\subfigure[]{
				\label{fig.hinf-ud3}
				}
			\end{tabular}
			\\
			\begin{sideways}
				\hspace{0.2cm}
				{\large $k_x$}
			\end{sideways}
			\hspace*{-0.4cm}
			\begin{tabular}{c}
				$G_{v1}(\bkappa;0.5,50,50)$
				\\
				{
				\includegraphics[width=0.33\columnwidth]
				{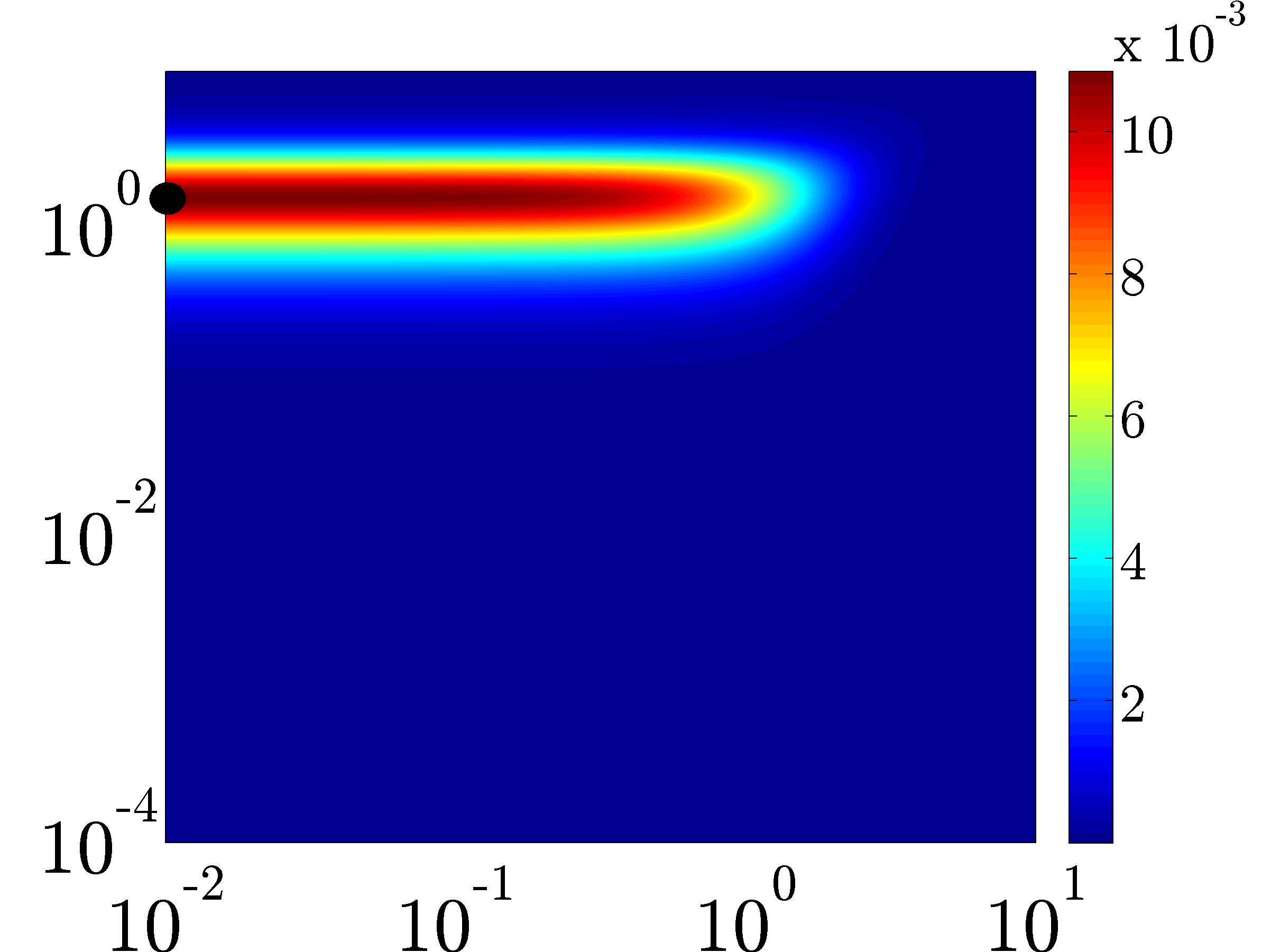}
				}
				\\
				{\large $k_z$}
				\\
				\subfigure[]{
				\label{fig.hinf-vd1}
				}
			\end{tabular}
			&
			\hspace{-0.6cm}
			\begin{tabular}{c}
				$G_{v2}(\bkappa;0.5,50,50)$
				\\
				{
				\includegraphics[width=0.33\columnwidth]
				{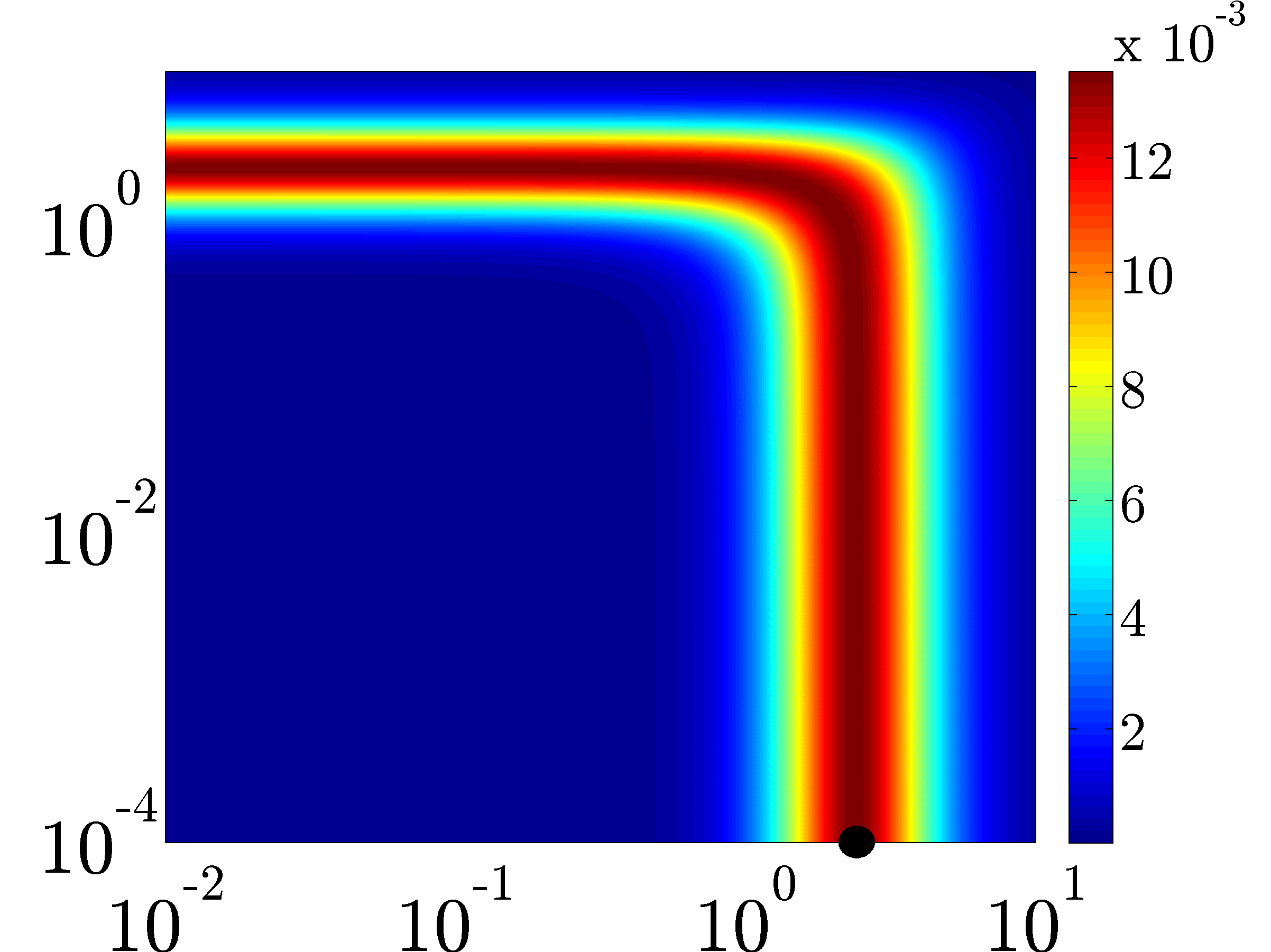}
				}
				\\
				{\large $k_z$}
				\\
				\subfigure[]{
				\label{fig.hinf-vd2}
				}
			\end{tabular}
			&
			\hspace{-1.3cm}
			\begin{tabular}{c}
				$G_{v3}(\bkappa;0.5,50,50)$
				\\
				{
				\includegraphics[width=0.33\columnwidth]
				{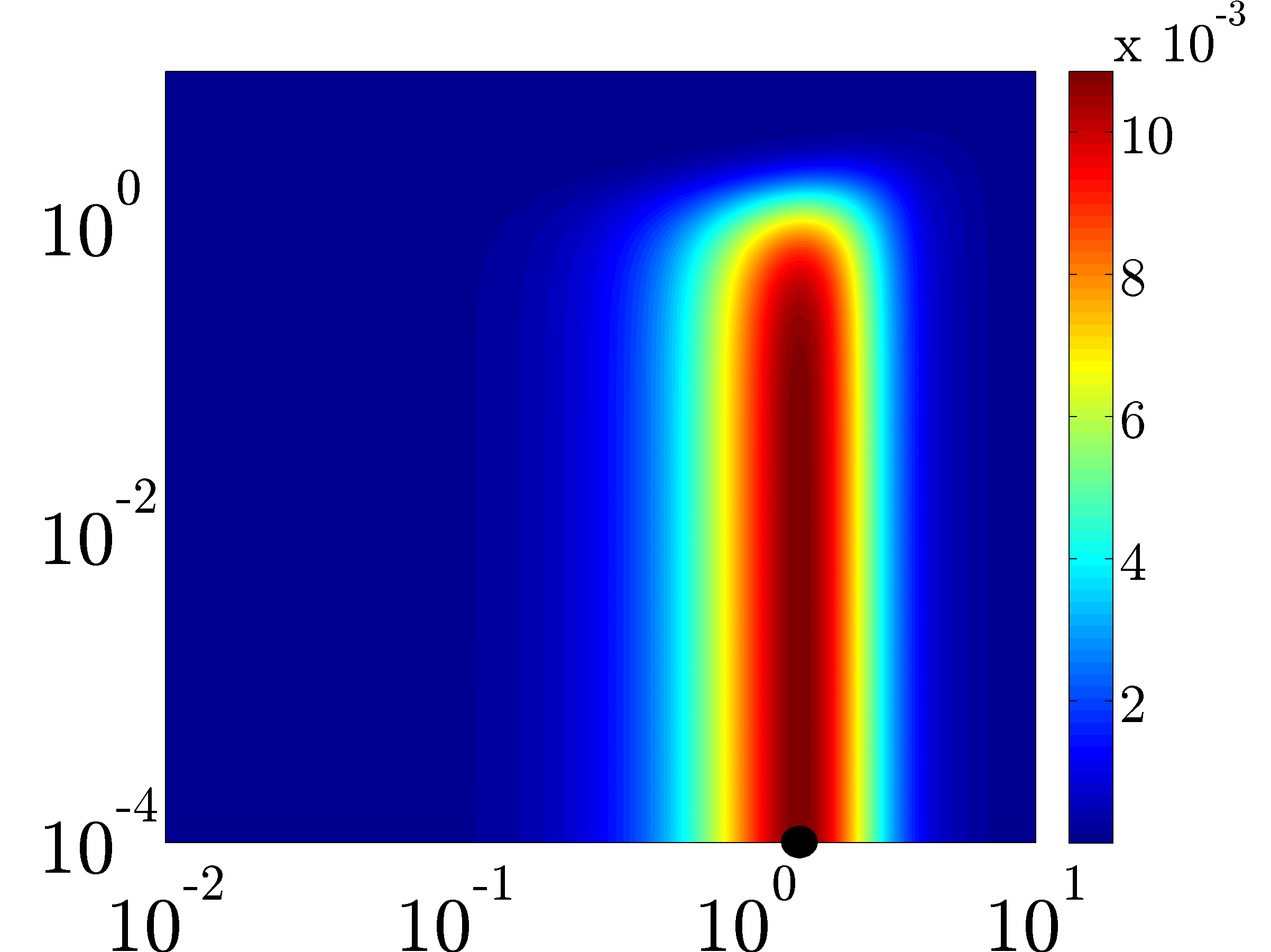}
				}
				\\
				{\large $k_z$}
				\\
				\subfigure[]{
				\label{fig.hinf-vd3}
				}
			\end{tabular}
			\\
			\begin{sideways}
				\hspace{0.2cm}
				{\large $k_x$}
			\end{sideways}
			\hspace*{-0.4cm}
			\begin{tabular}{c}
				$G_{w1}(\bkappa;0.5,50,50)$
				\\
				{
				\includegraphics[width=0.33\columnwidth]
				{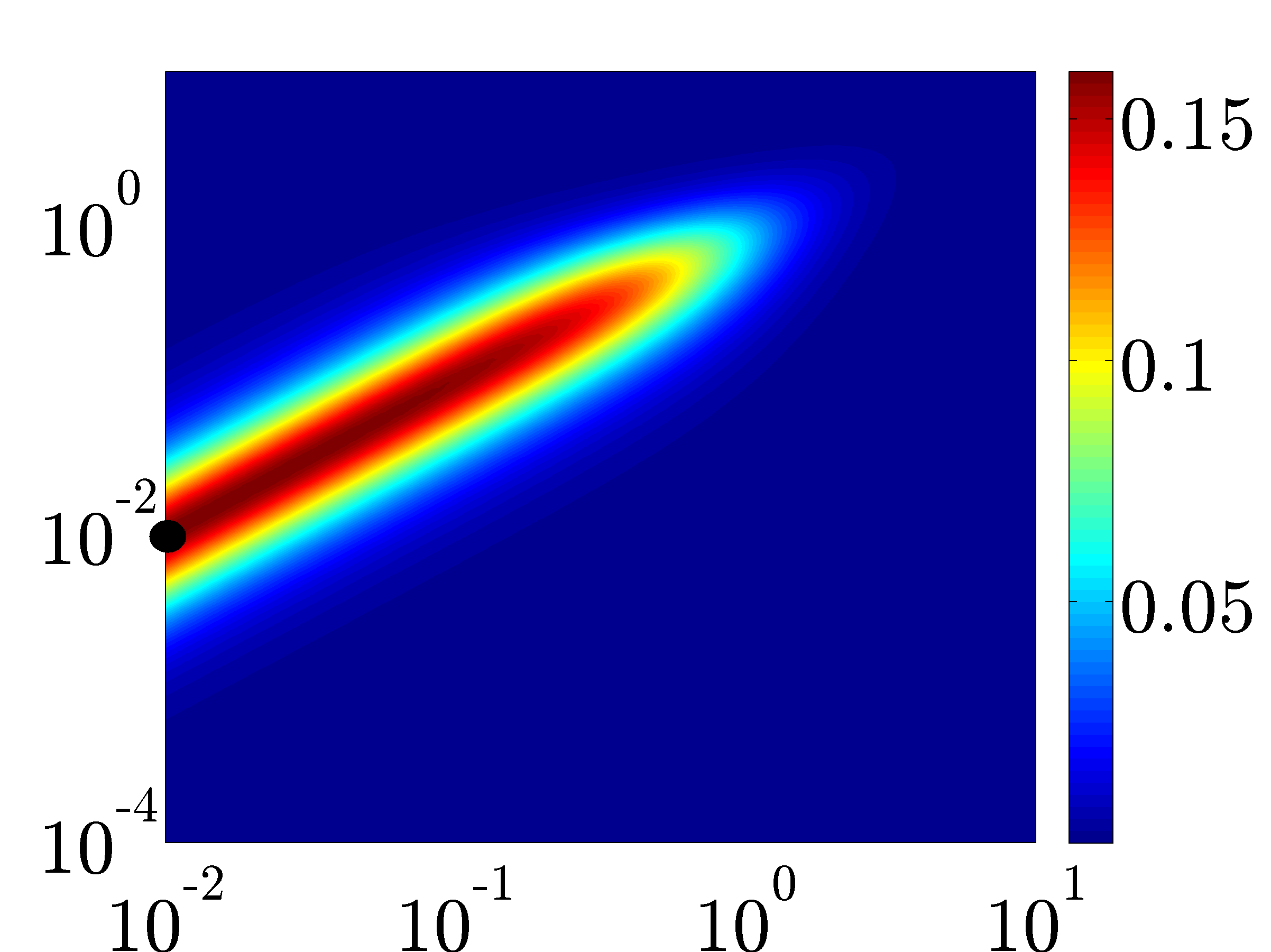}
				}
				\\
				{\large $k_z$}
				\\
				\subfigure[]{
				\label{fig.hinf-wd1}
				}
			\end{tabular}
			&
			\hspace{-0.6cm}
			\begin{tabular}{c}
				$G_{w2}(\bkappa;0.5,50,50)$
				\\
				{
				\includegraphics[width=0.33\columnwidth]
				{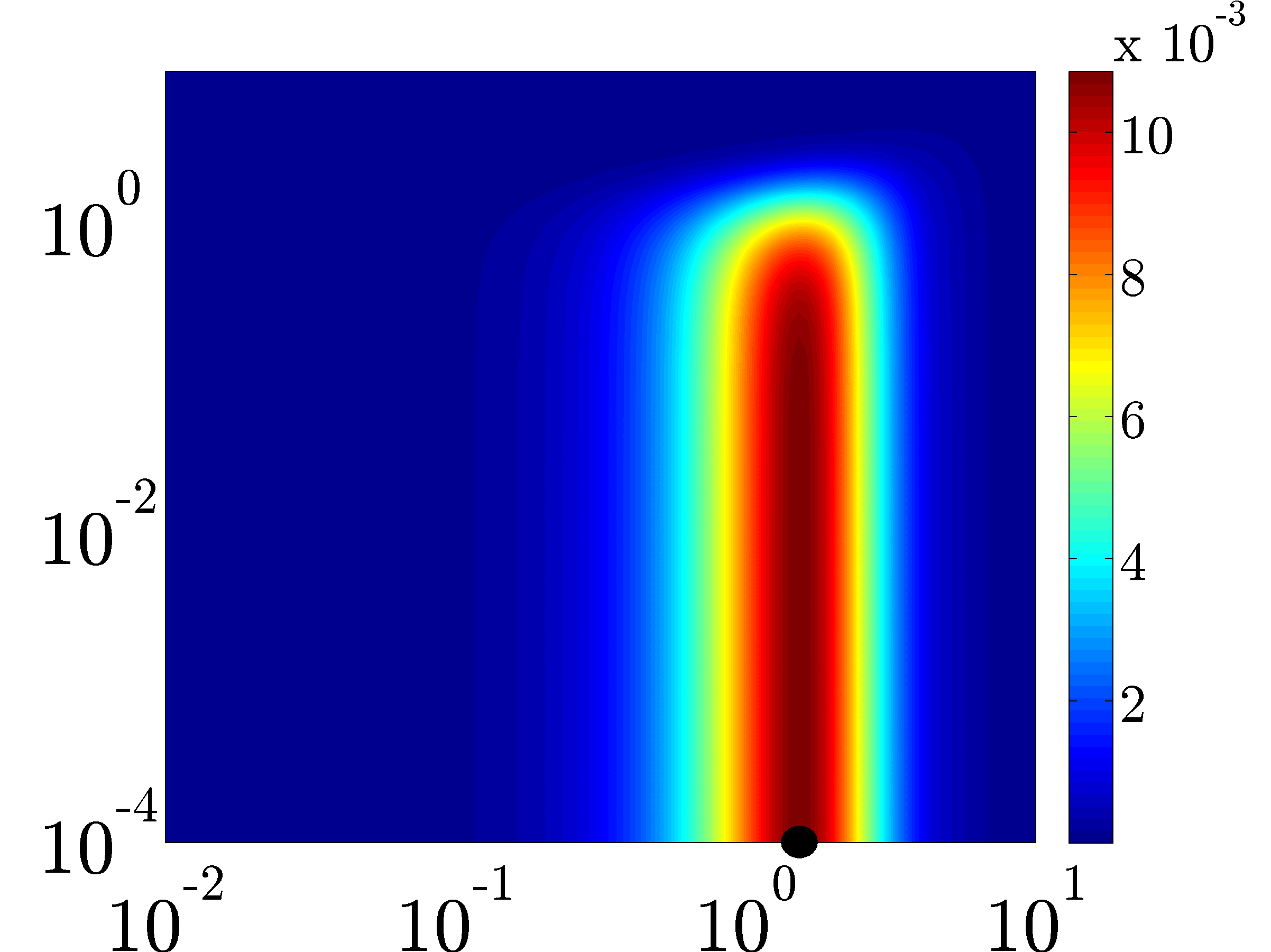}
				}
				\\
				{\large $k_z$}
				\\
				\subfigure[]{
				\label{fig.hinf-wd2}
				}
			\end{tabular}
			&
			\hspace{-1.3cm}
			\begin{tabular}{c}
				$G_{w3}(\bkappa;0.5,50,50)$
				\\
				{
				\includegraphics[width=0.33\columnwidth]
				{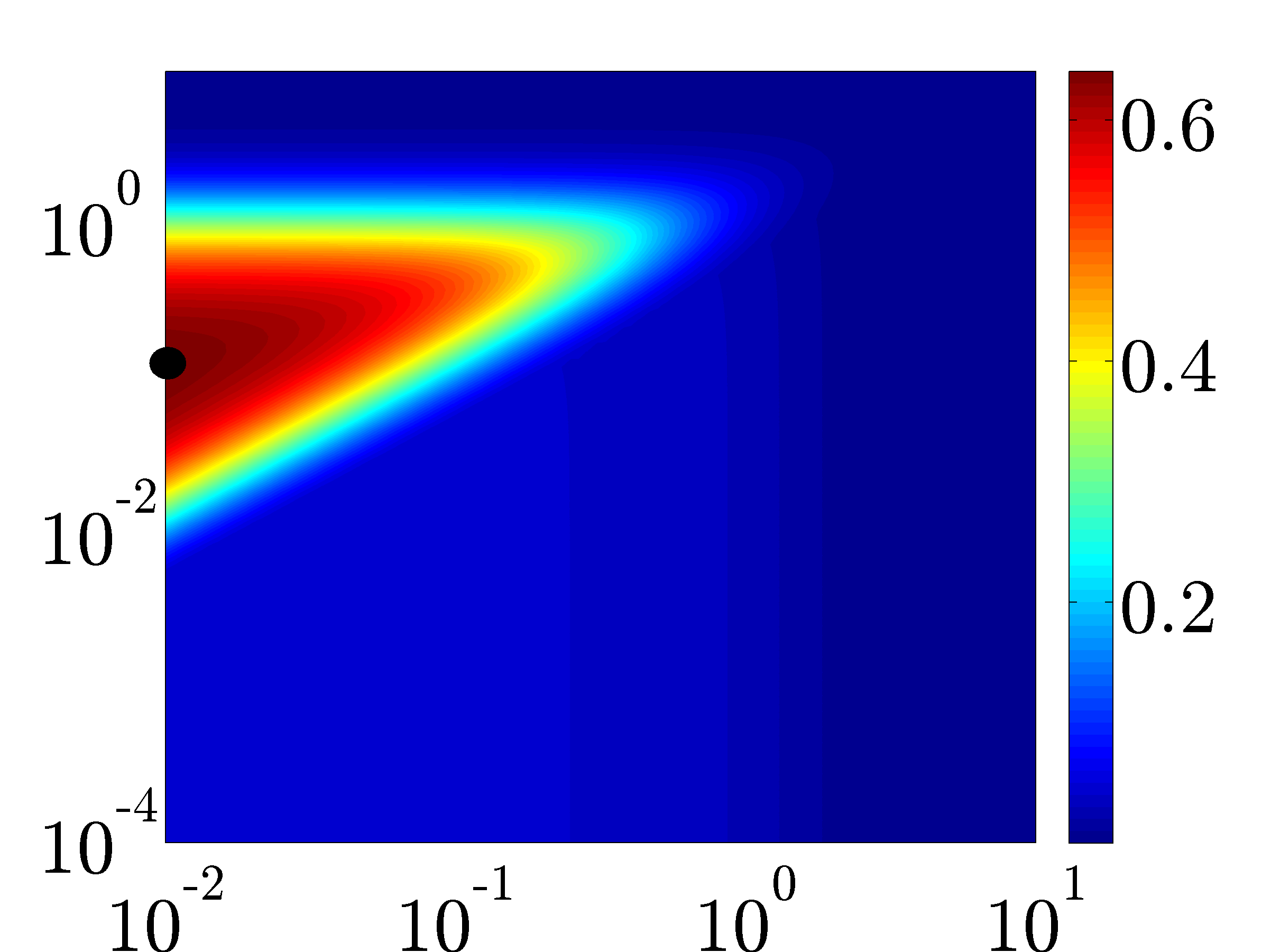}
				}
				\\
				{\large $k_z$}
				\\
				\subfigure[]{
				\label{fig.hinf-wd3}
				}
			\end{tabular}
		\end{tabular}
	\end{center}
      	\caption{Component-wise worst-case amplification from $d_j$ to $s$ in Couette flow with $s = \{ u, v, w\}$, $j = \{ 1, 2, 3\}$, $\beta = 0.5$, $\We = 50$, and $L = 50$. The symbol $(\bullet)$ identifies the largest value of the corresponding plot.}
    	\label{fig.hinf-uvw-123}
 	\end{figure}
	
	In the remainder of this section, we set the viscosity ratio to $\beta = 0.5$ and study the effect of the Weissenberg number, $\We$, and the maximum dumbbell extensibility, $L$, on the kinetic energy density. Figure~\ref{fig.hinf-ud-kxkz} shows the worst-case amplification of velocity fluctuations triggered by all three body forces in flows with $\We = \{ 10, 50, 100\}$ and $L = \{ 10, 50, 100\}$. For $L = 10$, the spatial frequency responses display low-pass filter features commonly seen in flows governed by viscous dissipation, with the peak amplification taking place at low wavenumbers. Furthermore, this spatial distribution remains almost unchanged as $\We$ increases from 10 to 100 (cf.\ figures~\ref{fig.hinf-ud-We10-L10},~\ref{fig.hinf-ud-We50-L10}, and~\ref{fig.hinf-ud-We100-L10}). However, as $L$ increases to $50$, the velocity fluctuations become more amplified with Weissenberg number and a dominant peak starts to appear in an isolated region around $k_x \approx \mathcal{O}(10^{-2})$ and $k_z \approx \mathcal{O}(1)$; see figures~\ref{fig.hinf-ud-We50-L50} and~\ref{fig.hinf-ud-We100-L50}. As $L$ increases to $100$, amplification with $\We$ increases even further. This indicates that in flows with large enough $L$ and $\We$, streamwise-elongated and spanwise-periodic flow fluctuations are the most amplified by deterministic body forces. Thus, in strongly elastic channel flows without inertia, streamwise-constant and nearly streamwise-constant fluctuations with a preferential spanwise length scale are most sensitive to external disturbances.
		
	We next study the component-wise frequency responses that quantify amplification from different forcing to different velocity components. This analysis facilitates identification of forcing components that are most effective in amplifying velocity fluctuations. In Couette flow with $\We = 50$ and $L = 50$, figure~\ref{fig.hinf-uvw-123} shows the worst-case amplification of the $9$ frequency response components in~(\ref{eq.uHd}). We see that the streamwise and spanwise velocity fluctuations are the most amplified. Furthermore, the maximum amplification of the streamwise velocity fluctuations triggered by $d_2$ and $d_3$ occurs around $k_x \approx \mathcal{O}(10^{-2})$ and $k_z \approx \mathcal{O}(1)$, respectively. This illustrates that streamwise velocity fluctuations are responsible for the most amplified region ($k_x \approx 10^{-2}$, $k_z \approx 2$) in figure~\ref{fig.hinf-ud-We50-L50}. In contrast, the square region around $k_x \approx 0$ and $k_z \approx 0$ in figure~\ref{fig.hinf-ud-We50-L50} arises from the responses of streamwise and spanwise velocity fluctuations to $d_1$ and $d_3$, respectively. We note that the wall-normal velocity experiences negligible amplification compared to that of the other two velocity components and hence, does not contribute to the large energy amplification in inertialess flows.
	
	The above results clearly illustrate the dominance of streamwise-constant and nearly streamwise-constant velocity fluctuations in strongly elastic Couette flow without inertia. The streamwise velocity is most amplified by disturbances, and this large response is caused by the wall-normal and spanwise body forces. Even though this section provides useful insight into the dynamics of inertialess channel flows of viscoelastic fluids, the scaling of energy amplification with $\We$ and $L$ cannot be deducted from our computations. For streamwise-constant velocity fluctuations, this issue is addressed in \S~\ref{sec.kx0} by developing explicit analytical expressions that quantify the dependence of the worst-case amplification on $\We$, $L$, $\beta$, and $k_z$.
	
    \vspace*{-2ex}
\section{Dynamics of streamwise-constant velocity fluctuations}
    \label{sec.kx0}

	Motivated by the observation that in strongly elastic flows the most amplified velocity fluctuations have large streamwise length-scales, we next examine the linearized model for fluctuations without streamwise variations, i.e., at $k_x = 0$. We use this model to establish an explicit scaling of the components of the frequency response operator with $\We$ and $L$, and to explain the observations made in~\S~\ref{sec.3d}. Since the largest amplification takes place at low temporal frequencies, we use analytical developments to show that in flows without temporal variations the worst-case amplification from $d_{2}$ and $d_{3}$ to $u$ scales linearly with the first normal stress difference $\bar{N}_{1}$ of the nominal flow. Consequently, this worst-case amplification scales quadratically with $\We$ as $L \rightarrow \infty$ and quadratically with $L$ as $\We \rightarrow \infty$. Therefore, even in flows with infinitely large polymer relaxation times
finite extensibility of polymer molecules limits the largest achievable amplification. Furthermore, the worst-case amplification from all other forcing to all other velocity components is both $\We$- and $L$-independent. We also present the spatial structures of the forcing and velocity fluctuation components that contribute to the above mentioned unfavorable scaling with $\bar{N}_1$, and demonstrate that the key physical mechanism involves interactions of polymer stress fluctuations in the ($y,z$)-plane with base shear.

	\subsection{Dependence of worst-case amplification on $\We$ and $L$}
	\label{sec.streamwise-constant}
		
		For fluctuations without streamwise variations, i.e.\ at $k_x = 0$, equations~(\ref{eq.psi}) and~(\ref{eq.veta}) simplify to
		\begin{subequations}
			\label{eq.veta-psi-2d3c}
		  	\begin{align}
				\label{eq.psiA}
		  		\dot{\bphi}_1
		  		& \; = \;
		  		-
		  		\bF_{11} \, \bphi_1
		  		\, + \,
		  		\bF_{1 v} \, v,
		  		\\[0.1cm]
				\label{eq.psiB}
		  		\dot{\bphi}_2
		  		& \; = \;
				-
		  		\bF_{22} \, \bphi_2
				\, + \,
		  		\bF_{21} \, \bphi_1
		  		\, + \,
		  		\bF_{2 v} \, v
		  		\, + \,
                \bF_{2 \eta} \, \eta,
				\\[0.1cm]
				\label{eq.v-2d3c}
				v
				& \; = \;
				\cfrac{1 \, - \, \beta}{\beta}
				\;
				\bC_{v 1} \, \bphi_1
				\, + \,
				\cfrac{1}{\beta}
				\;
				\bD_{v 2} \, d_2
				\, + \,
				\cfrac{1}{\beta}
				\;
				\bD_{v 3} \, d_3,
				\\[0.1cm]
				\label{eq.eta-2d3c}
				\eta
				& \; = \;
				\cfrac{1 \, - \, \beta}{\beta}
				\;
				\bC_{\eta 2} \, \bphi_2
				\, + \,
				\cfrac{1}{\beta}
				\;
				\bD_{\eta 1} \, d_1,
				\\[0.1cm]
				\label{eq.uvw-2d3c}
				\left[
				\begin{array}{c}
					u \\[0.1em]
					v \\[0.1em]
					w
				\end{array}
				\right]
				& \; = \;
				\left[
				\begin{array}{cc}
					0 & \bC_{u \eta} \\[0.1cm]
					\bI & 0 \\[0.1cm]
					\bC_{w v} & 0 \\[0.1cm]
				\end{array}
				\right]
				\left[
				\begin{array}{c}
					v \\[0.1cm]
					\eta
				\end{array}
				\right],
		  	\end{align}
		\end{subequations}
				where
    $
	\bphi_1
	=
	\left[
	\begin{array}{ccc}
	r_{22} & r_{23} & r_{33}
	\end{array}
	\right]^{T},
    $
    $
	\bphi_2
	=
	\left[
	\begin{array}{ccc}
	r_{13} & r_{12} & r_{11}
	\end{array}
	\right]^{T},
	$
$v$ is the wall-normal velocity, and $\eta$ is the wall-normal vorticity. On the other hand, the operators in~(\ref{eq.veta-psi-2d3c}) are given by
		\begin{equation*}
		\begin{array}{rcl}
			\bF_{11}
			& \!\! = \!\! &
			\left[
			\begin{array}{ccc}
				\bar{f} & 0 & 0 \\[0.2em]
				0 & \bar{f} & 0 \\[0.2em]
				0 & 0 & \bar{f}
			\end{array}
			\right],
			\;\;\;\;
			\bF_{22}
			\; = \;
			\left[
			\begin{array}{ccc}
				\bar{f} & 0 & 0 \\[0.3em]
				0 & \bar{f} & \We \, \bar{f} / \bar{L}^2 \\[0.3em]
				0 & -2 \, \We & \bar{f} + 2 \We^2 / \bar{L}^2
			\end{array}
			\right],
			\\[0.8cm]
			\bF_{1v}
			& \!\! = \!\! &
			\We
			\left[
			\begin{array}{c}
				2 \, \py \\[0.1em]
				\left( \mri / k_z \right) \left( \pyy \, + \, k_z^2 \right) \\[0.1em]
				-2 \, \py
			\end{array}
			\right],
			\;\;
			\bF_{21}
			\; = \;
			\We
			\left[
			\begin{array}{ccc}
				0 & 1 & 0
                \\[0.1cm]
				1 \, - \, \bar{f}/\bar{L}^2 & 0 & -\bar{f}/\bar{L}^2
                \\[0.1cm]
				-2 \, \We/\bar{L}^2 & 0 & -2 \, \We/\bar{L}^2
			\end{array}
			\right],
			\\[0.8cm]
			\bF_{2 v}
			& \!\! = \!\! &
			\cfrac{\We^2}{\bar{f}}
			\,
			\left[
			\begin{array}{c}
				\left( \mri / k_z \right) \pyy \\[0.2em]
				\py \\[0.2em]
				0
			\end{array}
			\right],
			\;\;
			\bF_{2 \eta}
			\; = \;
			\left[
			\begin{array}{c}
				1 \\[0.2em]
				-\We \left( \mri / k_z \right) \py \\[0.2em]
				-2 \left( \We^2 / \bar{f} \right) \left( \mri / k_z \right) \py	
			\end{array}
			\right],
			\\[0.7cm]	
			\bC_{v 1}
			& \!\! = \!\! &
			\left( \bar{f}/\We \right)
			\Delta^{-2}
			\left[
			\begin{array}{ccc}
				k_z^2 \, \py & \mri \, k_z \left( \pyy \, + \, k_z^2 \right) & -k_z^2 \, \py
			\end{array}
			\right],
			\\[0.2cm]
			\bC_{\eta 2}
			& \!\! = \!\! &
			\left( \bar{f}/\We \right)
			\Delta^{-1}
			\left[
			\begin{array}{ccc}
				k_z^2 & -\mri \, k_z \, \py & - \left( \We / \bar{L}^2 \right) \mri \, k_z \, \py
			\end{array}
			\right],
			\\[0.2cm]
			\bD_{v 2}
			& \!\! = \!\! &
			k_z^2 \Delta^{-2},
			\;\;
			\bD_{v 3} \; = \; \mri \, k_z \, \Delta^{-2} \, \py,
			\;\;
			\bD_{\eta 1} \; = \; -\mri \, k_z \, \Delta^{-1},
			\;\;
			\Delta \; = \; \pyy \, - \, k_z^2,
			\\[0.2cm]
			\bC_{u \eta}
			& \!\! = \!\! &
			- \mri / k_z,
			\;\;\;
			\bC_{w v}
			\; = \;
			\left( \mri / k_z \right) \py,
			\;\;
			\Delta^2 \; = \; \pyyyy \, - \, 2 \, k_z^2 \, \pyy \, + \, k_z^4.
		\end{array}
		\end{equation*}				
An evolution representation of the streamwise-constant model~(\ref{eq.veta-psi-2d3c}) can be obtained by eliminating the components of the conformation tensor from the equations. This is achieved by substituting the temporal Fourier transforms of~(\ref{eq.psiA}) and~(\ref{eq.psiB}) into~(\ref{eq.v-2d3c}) and~(\ref{eq.eta-2d3c}) and taking the inverse temporal Fourier transform of the resulting equations. We will show that this representation leads to convenient analytical expressions for the frequency response operator.

For streamwise-constant fluctuations the frequency response operator $\bH$ in~(\ref{eq.uHd}) simplifies to
		\begin{equation}
		\label{eq.uHd-2d3c}	
			\left[
			\begin{array}{c}
				u \\[0.1cm]
				v \\[0.1cm]
				w
			\end{array}
			\right]
			\; = \;
			\left[
			\begin{array}{ccc}
				\bH_{u 1} & \bH_{u 2} & \bH_{u 3}
				\\[0.1cm]
				0 & \bH_{v 2} & \bH_{v 3}
				\\[0.1cm]
				0 & \bH_{w 2} & \bH_{w 3}
			\end{array}
			\right]
			\!\!
			\left[
			\begin{array}{c}
				d_1 \\[0.1cm]
				d_2 \\[0.1cm]
				d_3
			\end{array}
			\right],
		\end{equation}
where the operators $\bH_{s j}$ are given by
        \begin{equation}
		\begin{array}{lrl}
			{\bH}_{v 2}(k_z, \omega; \beta, \We, L)
			& \!\! = \!\! &
			\cfrac{\mri \omega + \bar{f}}{\mri \omega \beta + \bar{f}} \,\, \bD_{v 2},
			\;\;\;
			{\bH}_{w 2}(k_z, \omega; \beta, \We, L)
			\; = \;
			\cfrac{\mri \omega + \bar{f}}{\mri \omega \beta + \bar{f}}
            \,\,
            \bC_{w v} \, \bD_{v 2},
			\\[0.4cm]
			{\bH}_{v 3}(k_z, \omega; \beta, \We, L)
			& \!\! = \!\! &
			\cfrac{\mri \omega + \bar{f}}{\mri \omega \beta + \bar{f}} \,\, \bD_{v 3},
			\;\;\;
			{\bH}_{w 3}(k_z, \omega; \beta, \We, L)
			\; = \;
			\cfrac{\mri \omega + \bar{f}}{\mri \omega \beta + \bar{f}} \,\, \bar{\bC}_{w v} \, \bD_{v 3},
			\\[0.4cm]
			{\bH}_{u 1}(k_z, \omega; \beta, \We, L)
			& \!\! = \!\! &
			\cfrac{\mri \omega + \bar{f}}{\mri \omega \beta + \bar{f}} \,\, \bE_{uu}^{\text{--}1} \bC_{u \eta} \, \bD_{\eta 1},
			\\[0.5cm]
			{\bH}_{u 2}(k_z, \omega; \beta, \We, L)
			& \!\! = \!\! &
			\bE_{uu}^{\text{--}1} \, \bE_{uv} \, \bD_{v 2},
			\;\;\;
			{\bH}_{u 3}(k_z, \omega; \beta)
			\; = \;
			\bE_{uu}^{\text{--}1} \, \bE_{uv} \, \bD_{v 3},
		\end{array}
	\label{eq.H-kx0-component}
	\end{equation}
	with
	 \begin{subequations}
		 \begin{align}
		 	\label{eq.Huu-kx0}
			\bE_{uu}
			\;\; = \;\; &
			\bI \, - \, \cfrac{2 \, \We^2 \left( \beta - 1\right)}{\bar{L}^2} \, \cfrac{2 \mri \omega \bar{f}  - \omega^2}{\left( \mri \omega \beta \, + \, \bar{f} \right) \left( \zeta_0 \, - \, \omega^2 \, + \, \mri \omega \zeta_1 \right)} \; \Delta^{-1} \, \pyy,
			\\[0.3cm]
			\label{eq.Huu-kx0}
			\bE_{uv}
			\;\; = \;\; &
			\cfrac{\We \bar{f} \left( \beta - 1 \right)}{\left( \mri \omega \beta + \bar{f} \right)^2}
			\, + \,
			\cfrac{2 \, \We^3 \left( \beta - 1\right)}{\bar{L}^2} \, \cfrac{3 \mri \omega \bar{f}  - \omega^2}{\left( \mri \omega \beta \, + \, \bar{f} \right)^2 \left( \zeta_0 \, - \, \omega^2 \, + \, \mri \omega \zeta_1 \right)} \; \Delta^{-1} \, \pyy,
			\\[0.2cm]
			\zeta_0
			\;\; = \;\; &
			\bar{f}^2 \, + \, \cfrac{4 \We^2 \bar{f}}{\bar{L}^2},
			\;\;\;\;
			\zeta_1
			\; = \;
			2 \bar{f} \, + \, \cfrac{2 \We}{\bar{L}^2}.
		\end{align}
	\label{eq.Huu-Huv-kx0}
	\end{subequations}
	
In~\eqref{eq.H-kx0-component}, we have successfully separated the temporal and spatial parts of the frequency response operators from $d_2$ and $d_3$ to $v$ and $w$. The absence of inertia induces simple temporal dependence of $\bH_{s j}$ with $\{ s = v, w; \, j = 2, 3\}$ and facilitates analytical determination of the temporal frequency $\omega$ at which the largest worst-case amplification takes place. These four frequency response operators exhibit high-pass temporal characteristics, with the peak amplification taking place at infinite frequency, $\omega = \infty$. The addition of a small amount of inertia would introduce roll-off at high temporal frequencies, thereby shifting the peak amplification to finite temporal frequency~\citep{jovkumJNNFM11}.
	
		These observations allow us to obtain explicit expressions for the worst-case amplification from $d_2$ and $d_3$ to $v$ and $w$. For example, the worst-case amplification from $d_2$ to $v$ is given by
		\begin{equation*}
			\begin{array}{rcl}
				G_{v 2}(k_z; \beta)
				& \!\! = \!\! &
				{\ds \sup_{\omega} \, \sigma_{\max}^{2}}
				\left( \bH_{v 2}(k_z,\omega; \We,\beta,L) \right)
				\\[0.3cm]
				& \!\! = \!\! &
				\left( 1/ \beta^{2} \right)
				\sigma_{\max}^{2}
				\left( \bD_{v 2} \right)
				\\[0.15cm]
				& \!\! = \!\! &
				\left( 1/ \beta^{2} \right) g_{v 2} (k_z),
			\end{array}
		\end{equation*}
		where the function $g_{v 2}$, which is independent of $\We$, $\beta$, and $L$, captures the spanwise frequency response (from $d_2$ to $v$). A similar procedure yields the following expressions for the worst-case amplification from $d_2$ and $d_3$ to $v$ and $w$
		\begin{equation}
        		\label{eq.hinf-g-vw-d23}
			\left[
			\begin{array}{cc}
				G_{v 2} (k_z; \beta)
				&
				G_{v 3} (k_z; \beta)
				\\[0.1cm]
				G_{w 2} (k_z; \beta)
				&
				G_{w 3} (k_z; \beta)
			\end{array}
			\right]
			~ = ~
			\left[
			\begin{array}{cc}
				g_{v 2} (k_z) / \beta^2
				&
				g_{v 3} (k_z) / \beta^2
				\\[0.1cm]
				g_{w 2} (k_z) / \beta^2
				&
				g_{w 3} (k_z) / \beta^2
			\end{array}
			\right],
		\end{equation}
		where the functions $g$ represent the $\beta$-, $\We$-, and $L$-independent spanwise frequency responses. We note that, in Couette flow without inertia, the worst-case amplification of the four components in~\eqref{eq.hinf-g-vw-d23} is equivalent for Oldroyd-B and FENE-CR fluids.

\begin{figure}
	   	\begin{center}
	       		\begin{tabular}{ccc}
				\hspace{-0.3cm}
				\begin{tabular}{c}
					$g_{v 2}(k_z)$
					\\
					{
					\includegraphics[width=0.33\columnwidth]
					{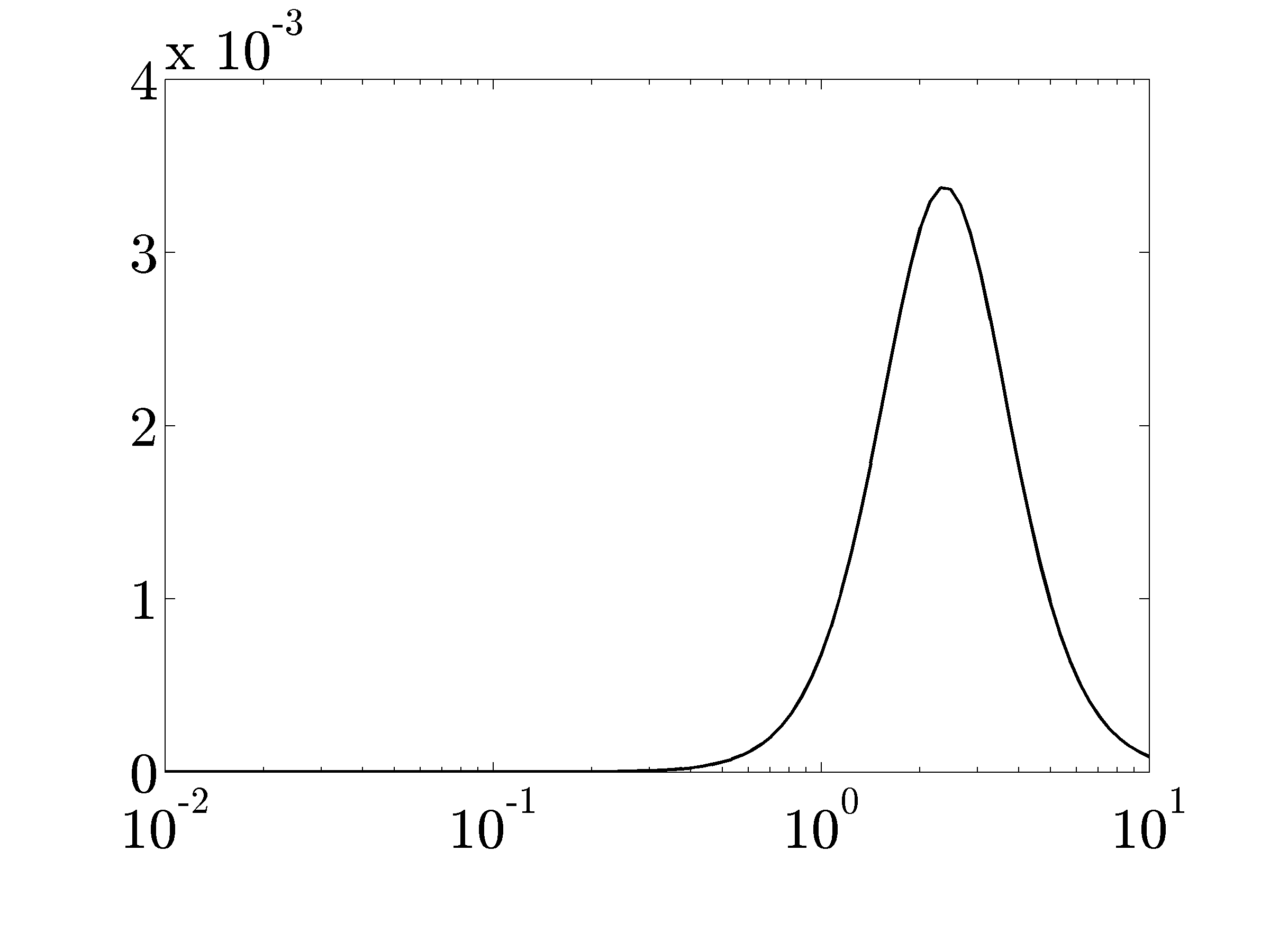}
					}
					\\[-0.2cm]
					{\large $k_z$}
					\\
					\subfigure[]
					{
					\label{fig.hinf-vd2-kx0}
					}
				\end{tabular}
				&
				\hspace{-0.6cm}
				\begin{tabular}{c}
					$g_{v 3}(k_z)$, $g_{w 2}(k_z)$
					\\
					{
					\includegraphics[width=0.33\columnwidth]
					{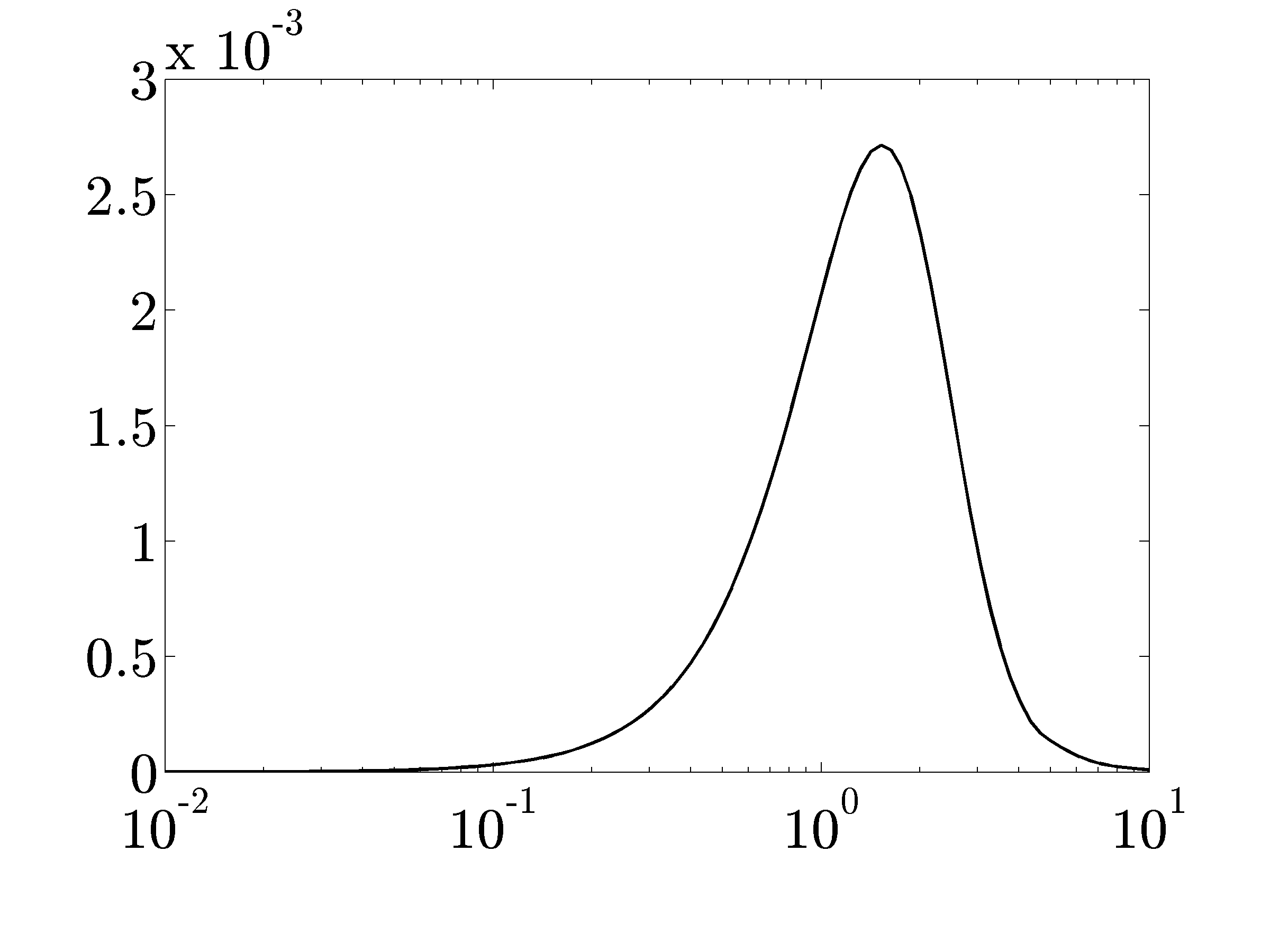}
					}
					\\[-0.2cm]
					{\large $k_z$}
					\\
					\subfigure[]
					{
					\label{fig.hinf-vd3-kx0}
					}
				\end{tabular}
				&
				\hspace{-0.6cm}
				\begin{tabular}{c}
					$g_{w 3}(k_z)$
					\\
					{
					\includegraphics[width=0.33\columnwidth]
					{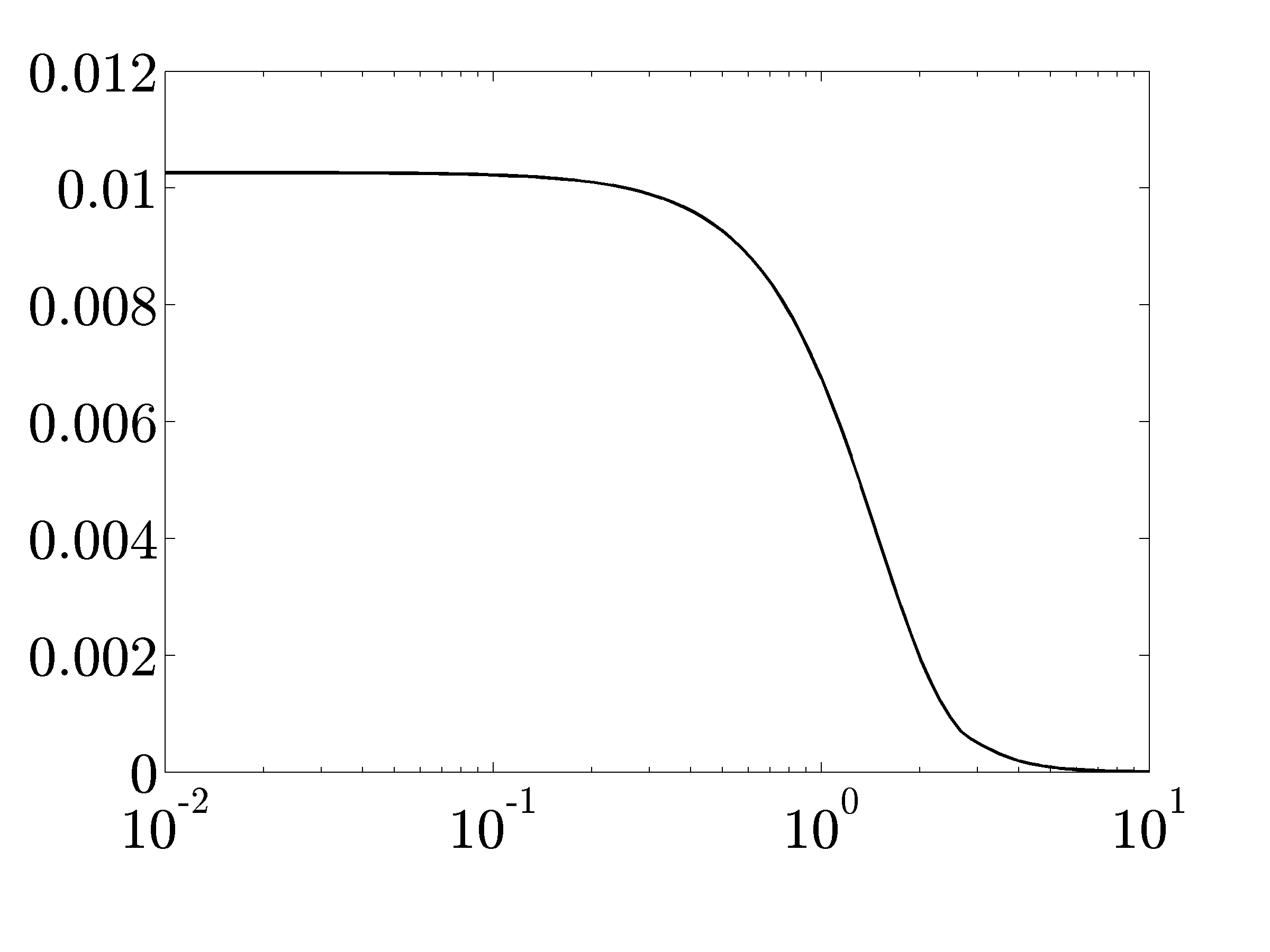}
					}
					\\[-0.2cm]
					{\large $k_z$}
					\\
					\subfigure[]
					{
					\label{fig.hinf-wd3-kx0}
					}
				\end{tabular}
			\end{tabular}
		\end{center}
		\caption{Functions characterizing worst-case amplification from $d_2$ and $d_3$ to $v$ and $w$ at $k_x = 0$; $g_{s j}(k_z)$ with $\{ s = v, w$; $j = 2, 3\}$.}
	    	\label{fig.hinf-vw23-kx0}
	 	\end{figure}

		The functions $g_{s j}(k_z)$ with $\{ s = v, w$; $j = 2, 3\}$, are shown in figure~\ref{fig.hinf-vw23-kx0}. We see that $g_{v 2}$ and $g_{v 3}$ exhibit similar trends with peaks at $k_z \approx {\cal O}(1)$; we also note that $g_{w 2} = g_{v 3}$. In contrast, $g_{w 3}$ has a low-pass shape with maximum occurring at $k_z = 0$. The peak value of this function is about four times larger than the peak values of $g_{v2}$ and $g_{v 3}$.
		
\begin{figure}
	   	\begin{center}
	       		\begin{tabular}{ccc}
				\hspace{-0.3cm}
				\begin{tabular}{c}
					$\sigma^{2}_{\max} \left( \bH_{u 1}\right)$
					\\
					{
					\includegraphics[width=0.33\columnwidth]
					{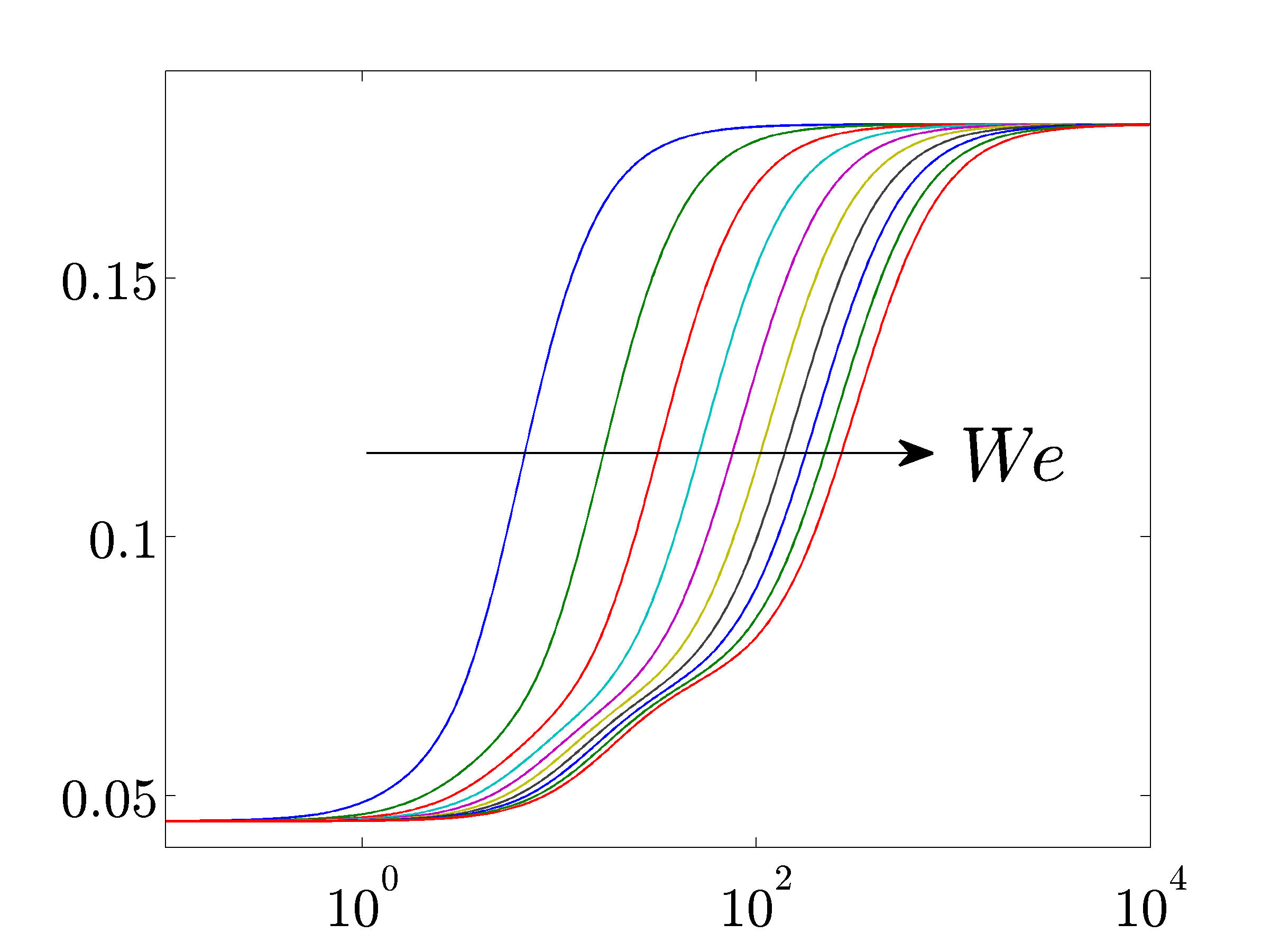}
					}
					\\
					{\large $\omega$}
					\\
					\subfigure[]
					{
					\label{fig.Smax-ud1-kx0-L10}
					}
				\end{tabular}
				&
				\hspace{-0.7cm}
				\begin{tabular}{c}
					$\sigma^{2}_{\max} \left( \bH_{u 2}\right)$
					\\
					{
					\includegraphics[width=0.33\columnwidth]
					{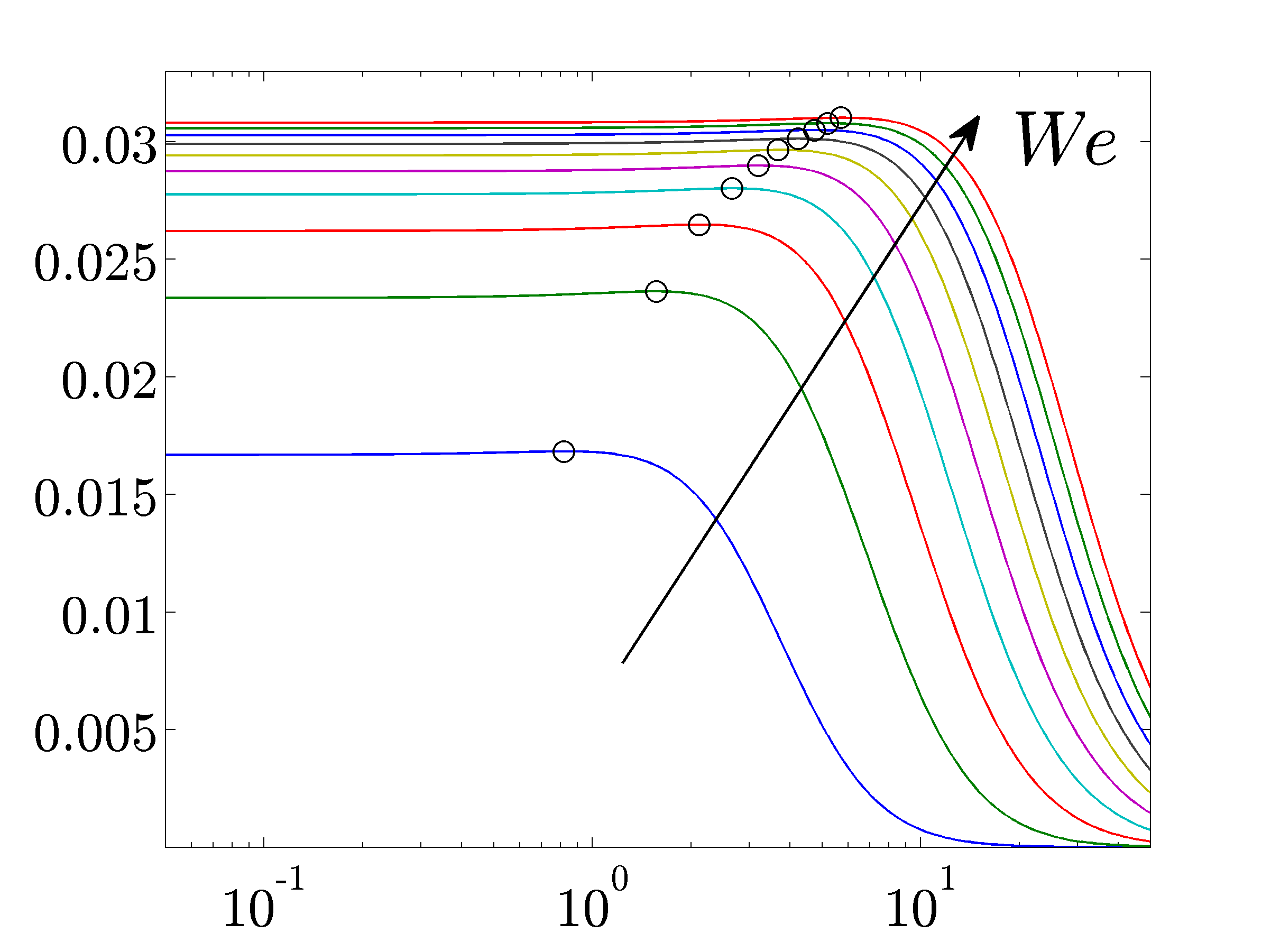}
					}
					\\
					{\large $\omega$}
					\\
					\subfigure[]
					{
					\label{fig.Smax-ud2-kx0-L10}
					}
				\end{tabular}
				&
				\hspace{-0.7cm}
				\begin{tabular}{c}
					$\sigma^{2}_{\max} \left( \bH_{u 3}\right)$
					\\
					{
					\includegraphics[width=0.33\columnwidth]
					{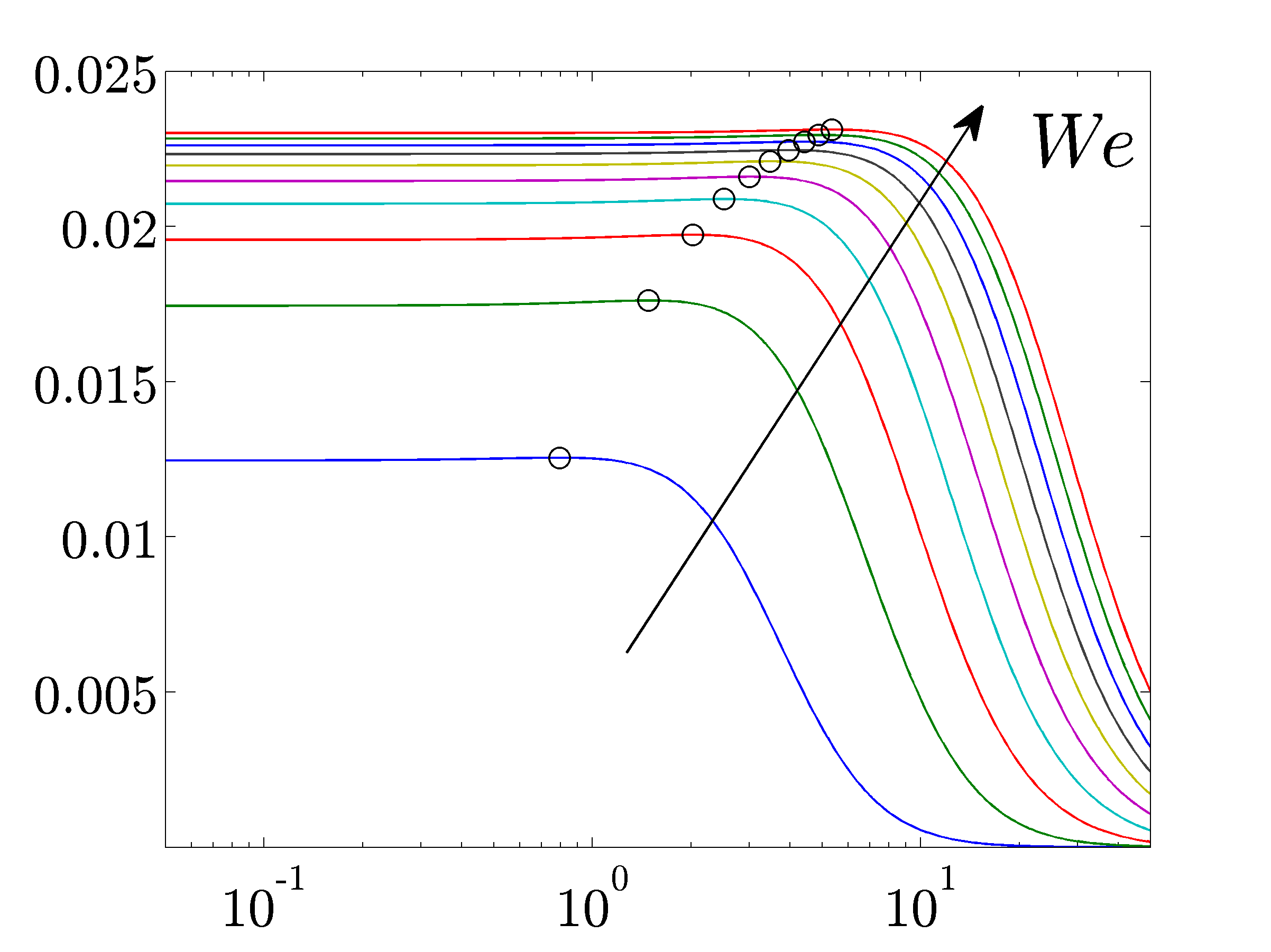}
					}
					\\
					{\large $\omega$}
					\\
					\subfigure[]
					{
					\label{fig.Smax-ud3-kx0-L10}
					}
				\end{tabular}
			\end{tabular}
		\end{center}
		\caption{Maximum singular values of the frequency response operators from $d_j$ to $u$ as a function of $\omega$ in streamwise-constant Couette flow with $j = \{ 1, 2, 3 \}$, $k_z = 1.5$, $\beta = 0.5$, $L = 10$, and $\We = [ 10, 100 ]$. The symbol ($\circ$) identifies the peak values of the corresponding curves.}
	    	\label{fig.Smax-u123-kx0}
	 	\end{figure}

		We next examine responses of streamwise velocity to different forcing components. We first analyze the temporal characteristics of the frequency response operators $\bH_{uj}$ with $j = \{1, 2, 3\}$. Since the largest amplification of $u$ arising from $\bd$ takes place at $k_z \approx {\cal O}(1)$, in figure~\ref{fig.Smax-u123-kx0} we show the temporal frequency dependence of $\sigma_{\max} \left( \bH_{uj} \right)$ for $k_z = 1.5$, $\beta = 0.5$ and $L = 10$. Figure~\ref{fig.Smax-ud1-kx0-L10} shows high-pass temporal features of $\sigma_{\max} \left( \bH_{u1} \right)$ with its maximum value taking place at $\omega = \infty$; furthermore, this peak value is independent of the Weissenberg number and the maximum extensibility of the polymer molecules $L$. In contrast, $\sigma_{\max} \left( \bH_{u2} \right)$ and $\sigma_{\max} \left( \bH_{u3} \right)$ have low-pass characteristics and attain their largest values (which depend on both $\We$ and $L$) at low temporal frequencies.
		
\begin{figure}
	   	\begin{center}
	       		\begin{tabular}{ccc}
				\hspace{-0.3cm}
				\begin{tabular}{c}
					$G_{u 1}(k_z; 0.5, \We, 10)$
					\\
					{
					\includegraphics[width=0.33\columnwidth]
					{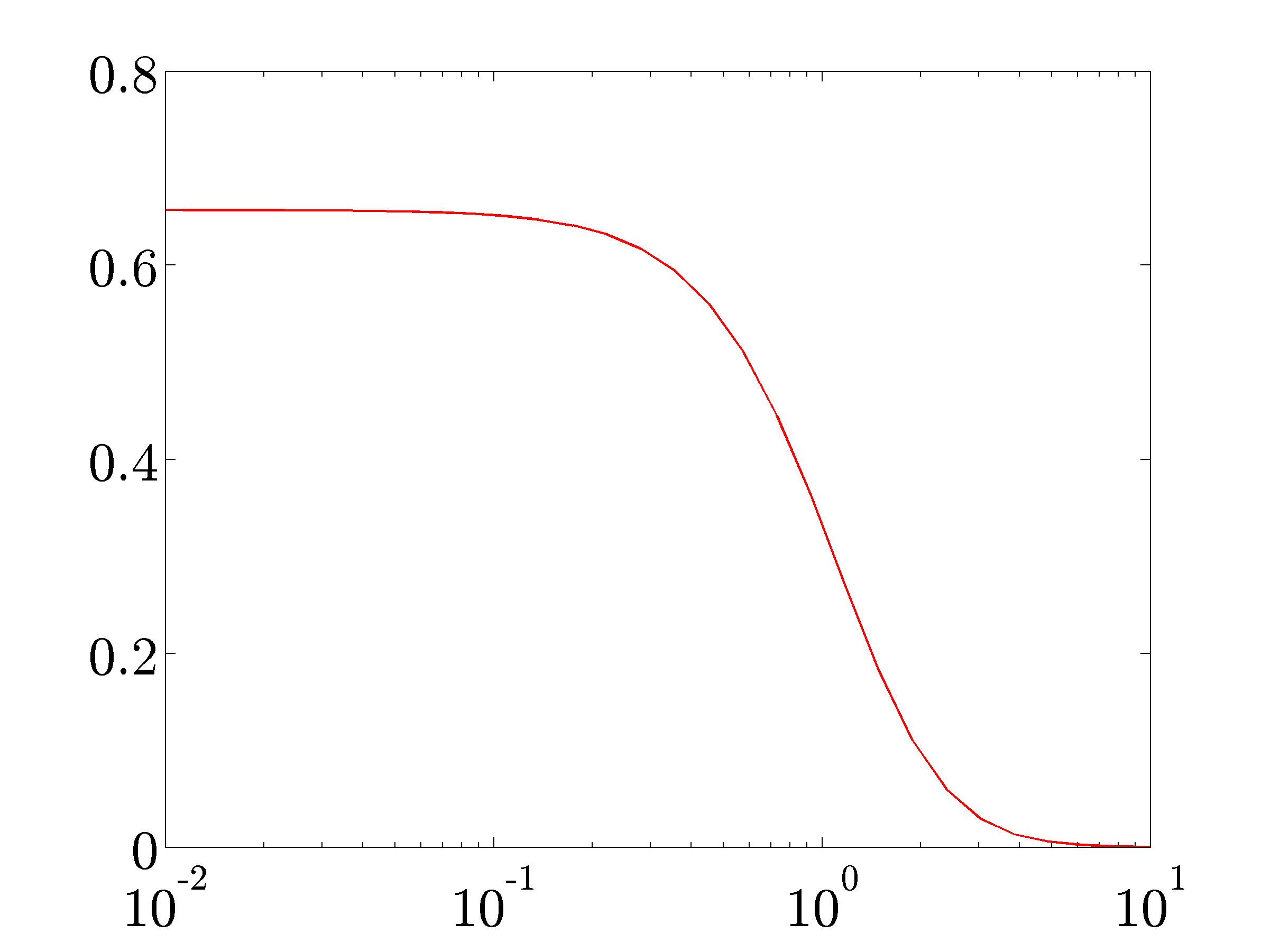}
					}
					\\
					{\large $k_z$}
					\\
					\subfigure[]
					{
					\label{fig.hinf-ud1-kx0-L10}
					}
				\end{tabular}
				&
				\hspace{-0.7cm}
				\begin{tabular}{c}
					$G_{u 2}(k_z; 0.5, \We, 10)$
					\\
					{
					\includegraphics[width=0.33\columnwidth]
					{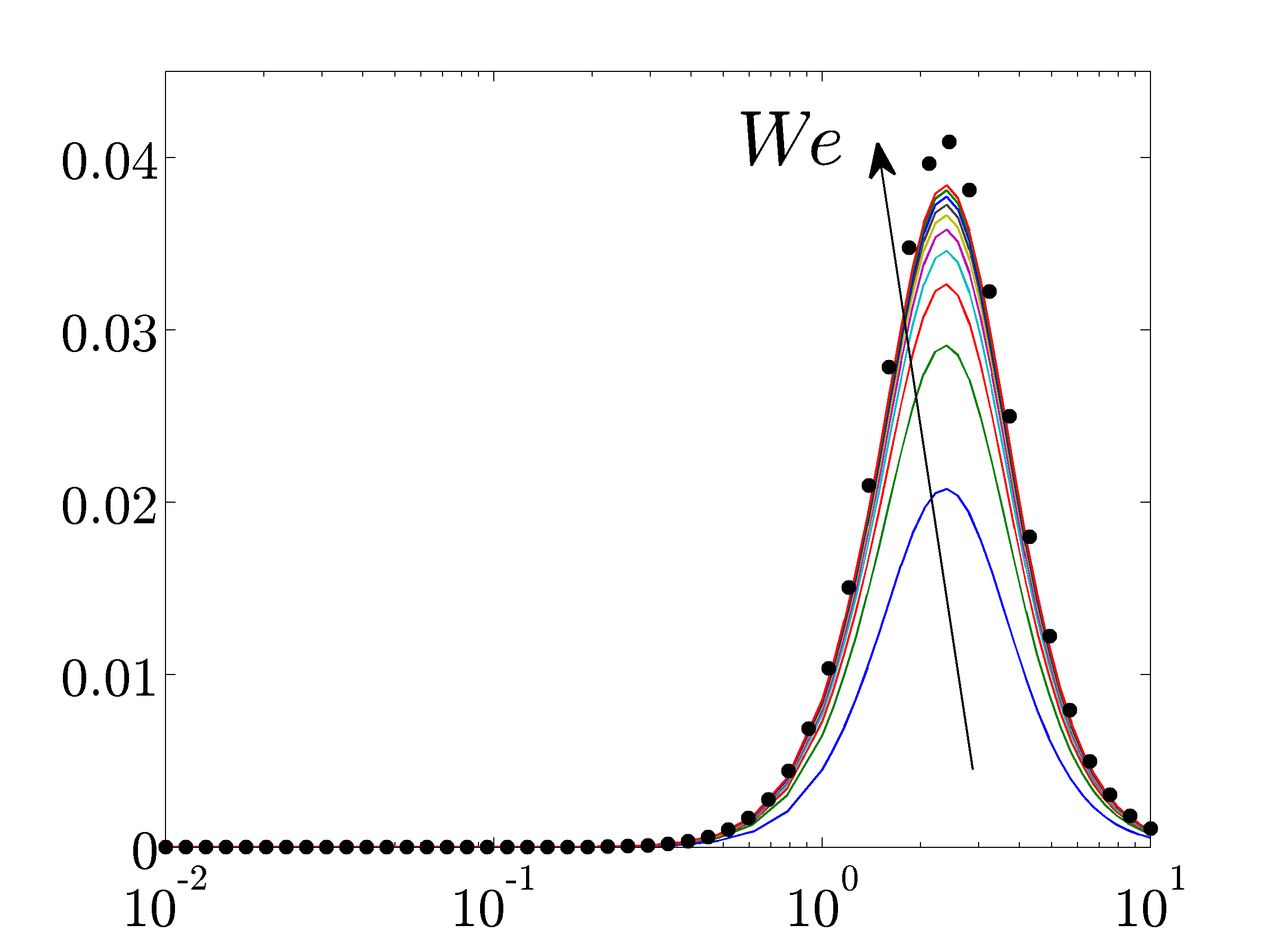}
					}
					\\
					{\large $k_z$}
					\\
					\subfigure[]
					{
					\label{fig.hinf-ud2-kx0-L10}
					}
				\end{tabular}
				&
				\hspace{-0.7cm}
				\begin{tabular}{c}
					$G_{u 3}(k_z; 0.5, \We, 10)$
					\\
					{
					\includegraphics[width=0.33\columnwidth]
					{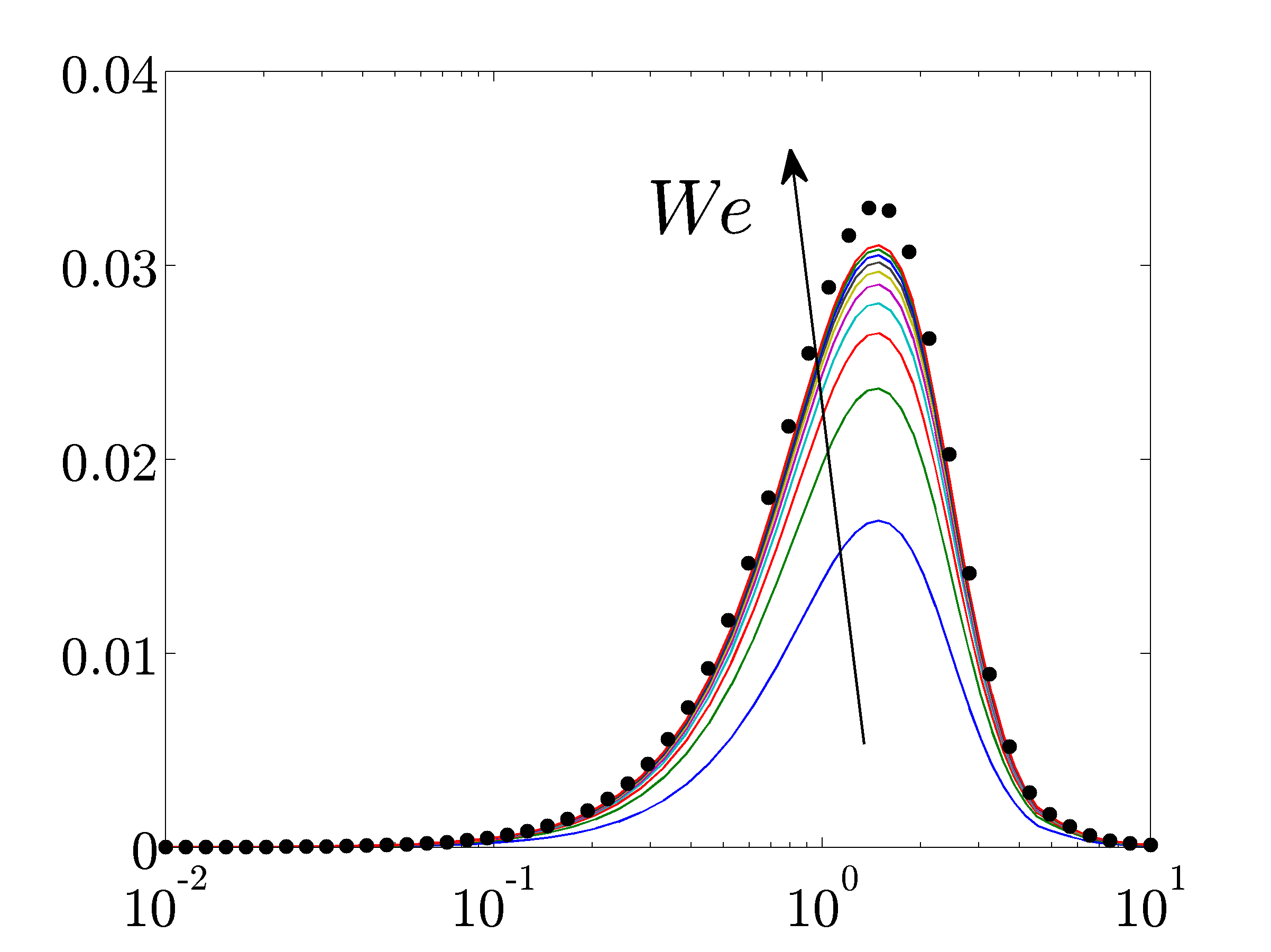}
					}
					\\
					{\large $k_z$}
					\\
					\subfigure[]
					{
					\label{fig.hinf-ud3-kx0-L10}
					}
				\end{tabular}
				\\
				\hspace{-0.3cm}
				\begin{tabular}{c}
					$G_{u 1}(k_z; 0.5, \We, 100)$
					\\
					{
					\includegraphics[width=0.33\columnwidth]
					{Gu1_fenecr_kx0_kz_We}
					}
					\\
					{\large $k_z$}
					\\
					\subfigure[]
					{
					\label{fig.hinf-ud1-kx0-L100}
					}
				\end{tabular}
				&
				\hspace{-0.7cm}
				\begin{tabular}{c}
					$G_{u 2}(k_z; 0.5, \We, 100)$
					\\
					{
					\includegraphics[width=0.33\columnwidth]
					{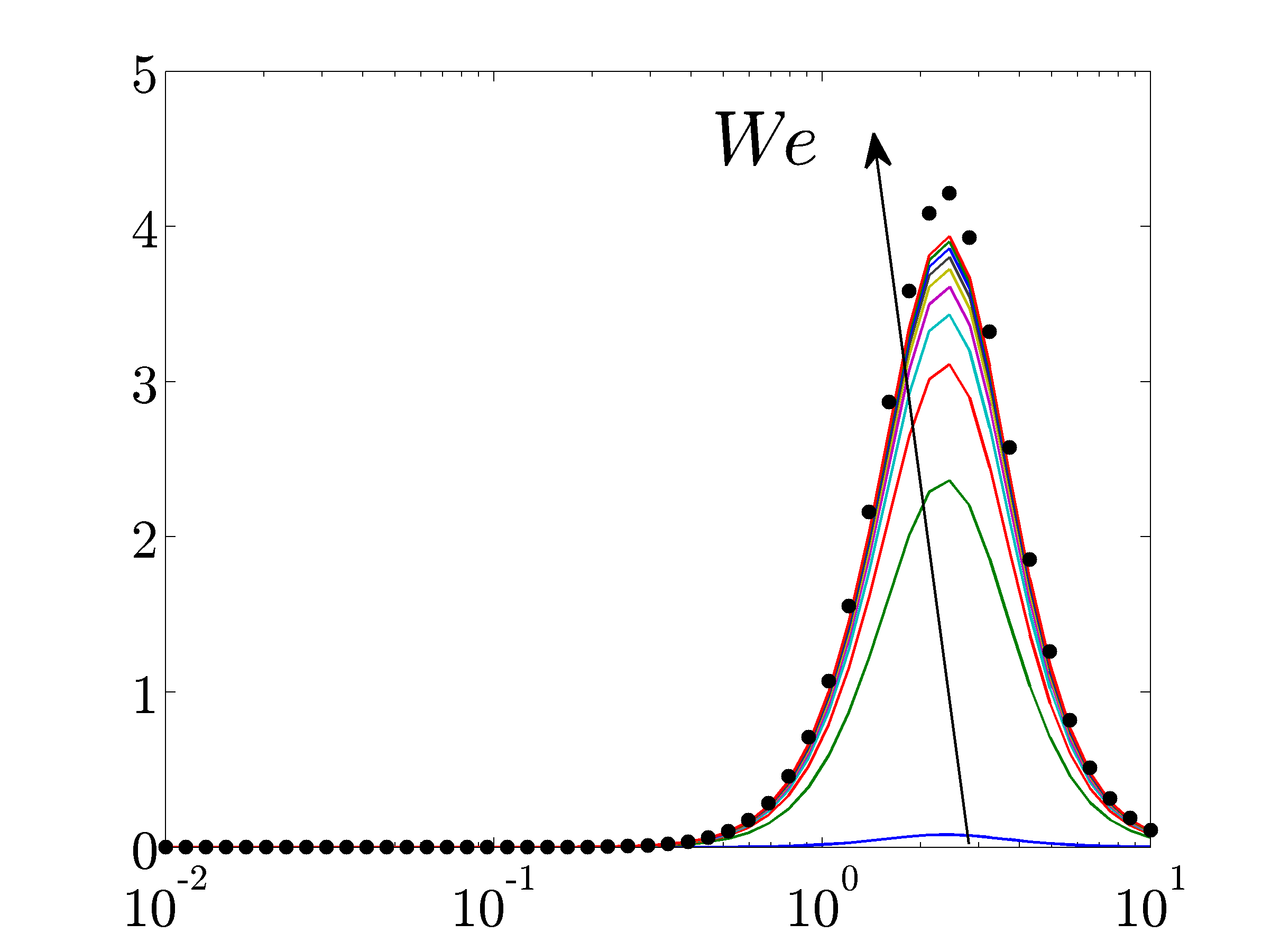}
					}
					\\
					{\large $k_z$}
					\\
					\subfigure[]
					{
					\label{fig.hinf-ud2-kx0-L100}
					}
				\end{tabular}
				&
				\hspace{-0.7cm}
				\begin{tabular}{c}
					$G_{u 3}(k_z; 0.5, \We, 100)$
					\\
					{
					\includegraphics[width=0.33\columnwidth]
					{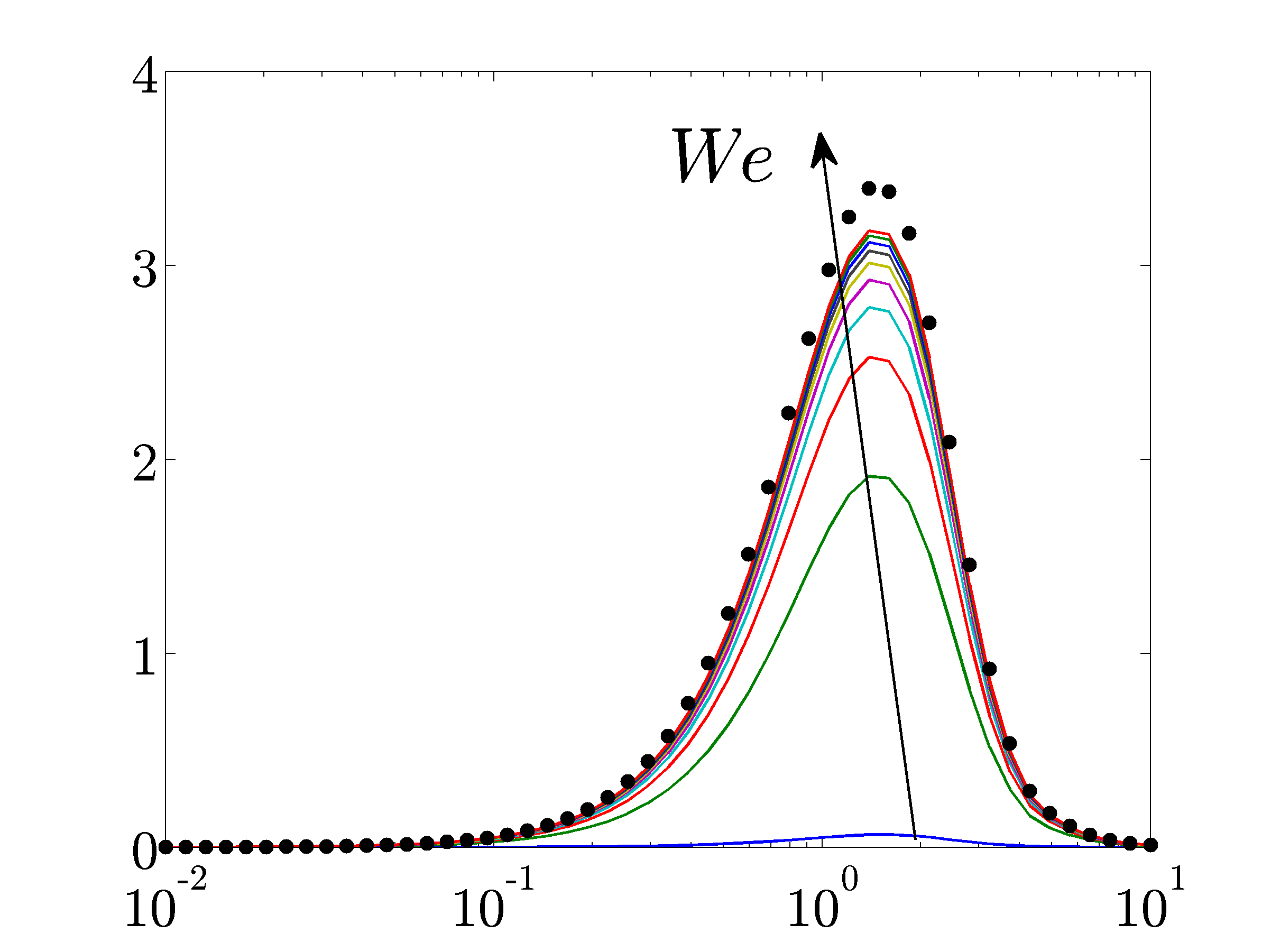}
					}
					\\
					{\large $k_z$}
					\\
					\subfigure[]
					{
					\label{fig.hinf-ud3-kx0-L100}
					}
				\end{tabular}
			\end{tabular}
		\end{center}
		\caption{Worst-case amplification from $d_j$ to $u$ in streamwise-constant Couette flow: first row, $L = 10$ and $\We = [ 10, 100 ]$; and second row, $L = 100$ and $\We = [ 10, 1000 ]$. The symbol $(\bullet)$ shows the worst-case amplification from $d_j$ to $u$ in the limit of infinitely large $\We$.}
	    	\label{fig.hinf-u123-kx0}
	 	\end{figure}

		Figure~\ref{fig.hinf-u123-kx0} shows the $k_z$-dependence of the functions $G_{u j}$ that quantify the worst-case amplification from different forcing components to the streamwise velocity for $L = \{10, 100\}$ and for multiple values of the Weissenberg number. Compared to the wall-normal and spanwise velocity fluctuations, streamwise velocity is more amplified by disturbances; cf.\ figures~\ref{fig.hinf-vw23-kx0} and~\ref{fig.hinf-u123-kx0}. As evident from figures~\ref{fig.hinf-ud1-kx0-L10} and~\ref{fig.hinf-ud1-kx0-L100}, $G_{u 1}$ has high values for low spanwise wavenumbers and is independent of both $\We$ and $L$. On the other hand, $G_{u2}$ and $G_{u3}$ achieve their peaks at $k_z \approx \cO(1)$. For a fixed value of $L$ these two functions increase with the Weissenberg number, and the largest amplification takes place in the limit of infinitely large $\We$. In contrast to the Oldroyd-B fluids, finite extensibility of the nonlinear springs in the FENE-CR model induces finite values of $G_{u2}$ and $G_{u3}$ even at arbitrarily large Weissenberg numbers. Furthermore, for a fixed value of $\We$, worst-case amplification increases with $L$ and the largest amplification is obtained in the Oldroyd-B limit (as $L \rightarrow \infty$). Analytical explanation for these observations is provided below.
		
		We next present explicit expressions for the worst-case amplification from different forcing components to the streamwise velocity. We summarize our major findings here and relegate derivations to Appendix~\ref{sec.app-derivation}. We first consider the worst-case response of the streamwise velocity in the presence of the streamwise body forcing. For all values of $\We$ and $L$, our computations indicate that the worst-case amplification from $d_1$ to $u$ takes place at $\omega = \infty$. Consequently, in the limit of infinitely large $\omega$ we have
		\begin{equation}
		\begin{array}{rcl}
			G_{u1} ( k_z; \beta )
			& \!\! = \!\! &
			{\ds \lim_{\omega \, \rightarrow \, \infty}
			\sigma^2_{\max} \left( \bH_{u1}(k_z,\omega; \beta, \We, L) \right)}
			\\[0.3cm]
			& \!\! = \!\! &
			\left( 1/\beta^2 \right) \sigma^2_{\max} \left( \bC_{u \eta} \, \bD_{\eta 1} \right)
			\; = \;
			\left( 1/\beta^2 \right) g_{u1} (k_z),
		\end{array}
		\end{equation}
		where $g_{u1}$ is the spanwise frequency response from $d_1$ to $u$. Note that $G_{u1}$ is independent of both $\We$ and $L$ which is in agreement with the observations made in figures~\ref{fig.hinf-ud1-kx0-L10} and~\ref{fig.hinf-ud1-kx0-L100}.
		
\begin{figure}
	   	\begin{center}
			\begin{tabular}{c}
				{\large $g_{u 2}(k_z)$, $g_{u 3}(k_z)$}
				\\
				{
				\includegraphics[width=0.5\columnwidth]
				{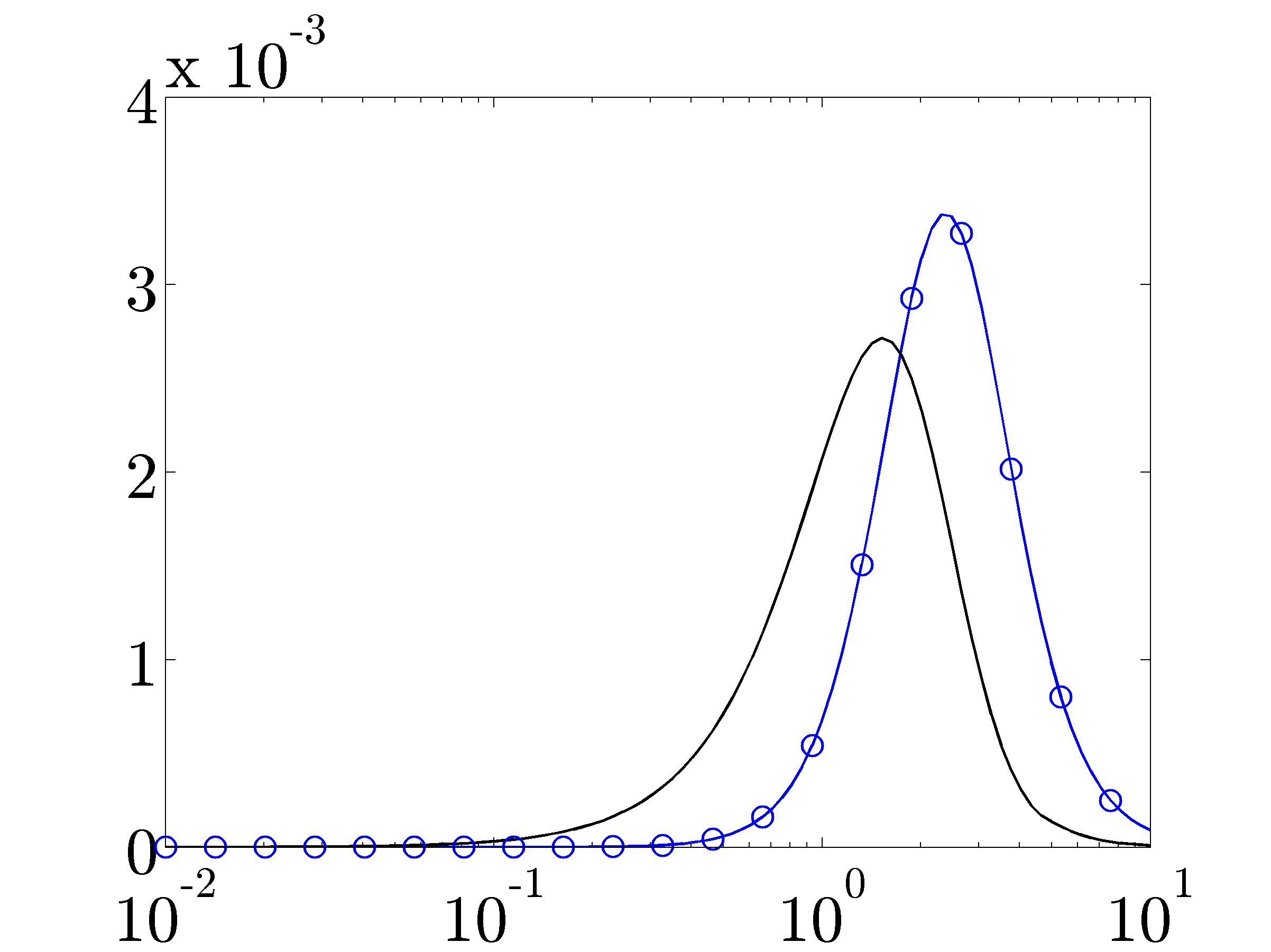}
				}
				\\
				{\Large $k_z$}
			\end{tabular}
		\end{center}
		\caption{Functions characterizing worst-case amplification from $d_2$ and $d_3$ to $u$ at $k_x = 0$ and $\omega = 0$: $g_{u 2}$ ($\circ$); and $g_{u 3}$ (solid).}
	    	\label{fig.hinf-u23-kx0}
	 	\end{figure}

		Derivation of the analytical expressions for $G_{u2}$ and $G_{u3}$ is more challenging because the worst-case amplification of $u$ arising from $d_2$ and $d_3$ depends on both $\We$ and $L$. However, since our computations demonstrate that the worst-case amplification from $d_2$ and $d_3$ to $u$ takes place at low temporal frequencies, the essential features can be captured by analyzing the corresponding frequency responses at $\omega = 0$. Thus, the worst-case amplification of $u$ caused by $d_2$ and $d_3$ can be reliably approximated by
		\begin{equation}
		\label{eq.Guj-omega0}
		\begin{array}{rcl}
			G_{u j} \left( k_z; \beta, \We, L \right)
			& \!\! \approx \!\! &
            \sigma^2_{\max} \left( \bH_{u j}(k_z,0; \beta, \We, L) \right)
            \\[0.2cm]
			& \!\! = \!\! &
			\left( \We/\bar{f} \right)^2
			\left( 1 - \beta \right)^2
			g_{u j}(k_z)
			\\[0.2cm]
			& \!\! = \!\! &
			{\ds
			\cfrac{1}{2}
            \;
            \bar{N}_{1}
			\left( 1 - \beta \right)^2
			g_{u j}(k_z),
			\;\;\;
			j \; = \; \{ 2, 3 \},
			}
		\end{array}
		\end{equation}
		where the functions $g_{u j}$ with $j = \{ 2, 3 \}$ quantify the spanwise frequency responses from $d_{2}$ and $d_{3}$ to $u$; see figure~\ref{fig.hinf-u23-kx0}. Equation~\eqref{eq.Guj-omega0} shows that the worst-case amplification of the streamwise velocity fluctuations scales linearly with the first normal stress difference $\bar{N}_{1}$ of the nominal Couette flow. Hence, even in the absence of inertia, velocity fluctuations can experience large amplification in flows with large normal stress difference.
		
		Furthermore, table~\ref{table.Gu23-limit} shows explicit expressions for $G_{u2}$ and $G_{u3}$ in the limit of infinitely large $\We$ (or infinitely large $L$). In these two cases, the first normal stress difference of the base Couette flow is given by
		\[
			{\ds \lim_{L \, \rightarrow \, \infty} \bar{N}_{1} }
			\; = \;
			2 \, \We^{2},
			\;\;\;
			{\ds \lim_{\We \, \rightarrow \, \infty} \bar{N}_{1} }
			\; = \;
			\bar{L}^2.
		\]
		For Oldroyd-B fluids polymer molecules are modeled by infinitely extensible linear springs and both $G_{u2}$ and $G_{u3}$ scale quadratically with the Weissenberg number. On the other hand, as $\We \rightarrow \infty$, $G_{u2}$ and $G_{u3}$ scale quadratically with the maximum extensibility of the polymer molecules $L$. We conclude that -- even for infinitely large polymer relaxation times -- energy amplification of velocity fluctuations in inertialess Couette flow of viscoelastic fluids is bounded by the maximum extensibility of nonlinear dumbbells.
		
		\begin{table}
		\centering
		\begin{tabular}{c | c | c}
			& $L \longrightarrow \infty$ & $\We \longrightarrow \infty$ \\ \hline
			\hspace{0.45cm} $G_{u2}\left( k_z; \beta, \cdot , \cdot \right)$ \hspace{0.45cm}
			&
			\hspace{0.45cm} $\We^2 \left( 1 - \beta \right)^2 g_{u2}( k_z )$ \hspace{0.45cm}
			&
			\hspace{0.45cm} $0.5 \, \bar{L}^2 \left( 1 - \beta \right)^2 g_{u2}( k_z )$ \hspace{0.45cm}
			\\[0.1cm] \hline
			\hspace{0.45cm} $G_{u3}\left( k_z; \beta, \cdot , \cdot \right)$ \hspace{0.45cm}
			&
			\hspace{0.45cm} $\We^2 \left( 1 - \beta \right)^2 g_{u3}( k_z )$ \hspace{0.45cm}
			&
			\hspace{0.45cm} $0.5 \, \bar{L}^2 \left( 1 - \beta \right)^2 g_{u3}( k_z )$ \hspace{0.45cm}
			\\[0.1cm]
		\end{tabular}
		 \caption{Worst-case amplification of streamwise velocity fluctuations arising from the wall-normal and spanwise forces in the limit of infinitely large maximum extensibility of polymer chains $L$ or infinitely large Weissenberg number $\We$.}
      		\label{table.Gu23-limit}
		\end{table}

		\subsection{Dominant flow structures}

		In this section, we present the spatial structures of the wall-normal and spanwise body forces that induce the largest amplification in streamwise velocity. We also discuss the ($y,z$)-dependence of the resulting streamwise velocity fluctuations. These structures are purely harmonic in the spanwise direction with period determined by the value of $k_z$ at which the functions $g_{u2}$ and $g_{u3}$ attain their maxima ($k_z \approx 2.5$ and $k_z \approx 1.5$, respectively). Furthermore, since the worst-case amplification occurs at $\omega = 0$, these structures are constant in time and their wall-normal profiles are determined by the principal singular functions of the frequency response operators that map $d_2$ and $d_3$ to $u$.

		\begin{figure}
	   	\begin{center}
	       		\begin{tabular}{m{6cm}m{6cm}}
				\begin{sideways}
					\hspace{0.6cm}
					{\large $y$}
				\end{sideways}
				\hspace*{-0.3cm}
				\begin{tabular}{c}
					{
					\includegraphics[width=0.45\columnwidth]
					{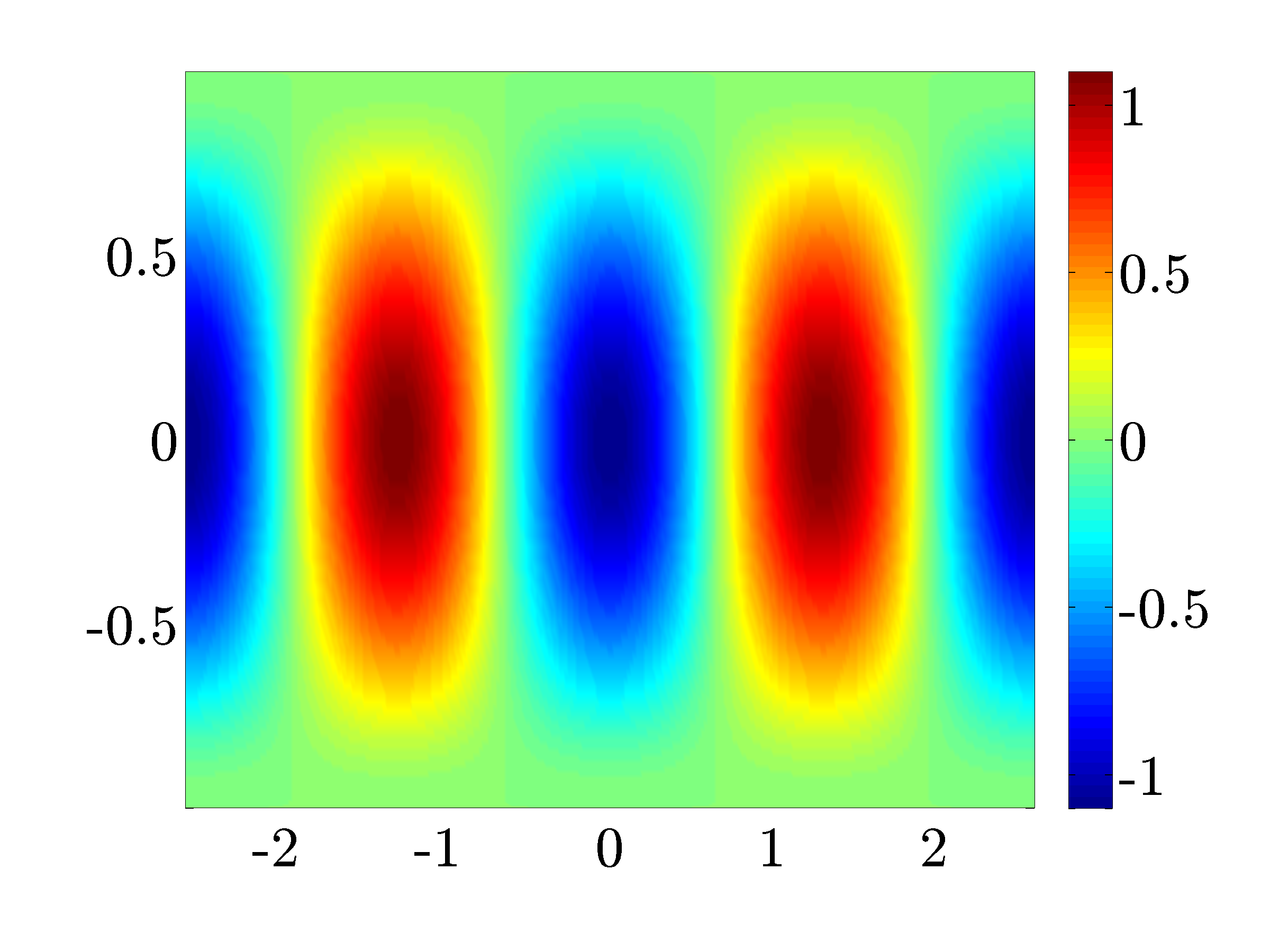}
					}
					\\[-0.2cm]
					{\large $z$}
					\\
					\subfigure[]{
					\label{fig.d2-structure-kx0-cout}
					}
				\end{tabular}
				&
				\begin{tabular}{c}
					{
					\includegraphics[width=0.45\columnwidth]
					{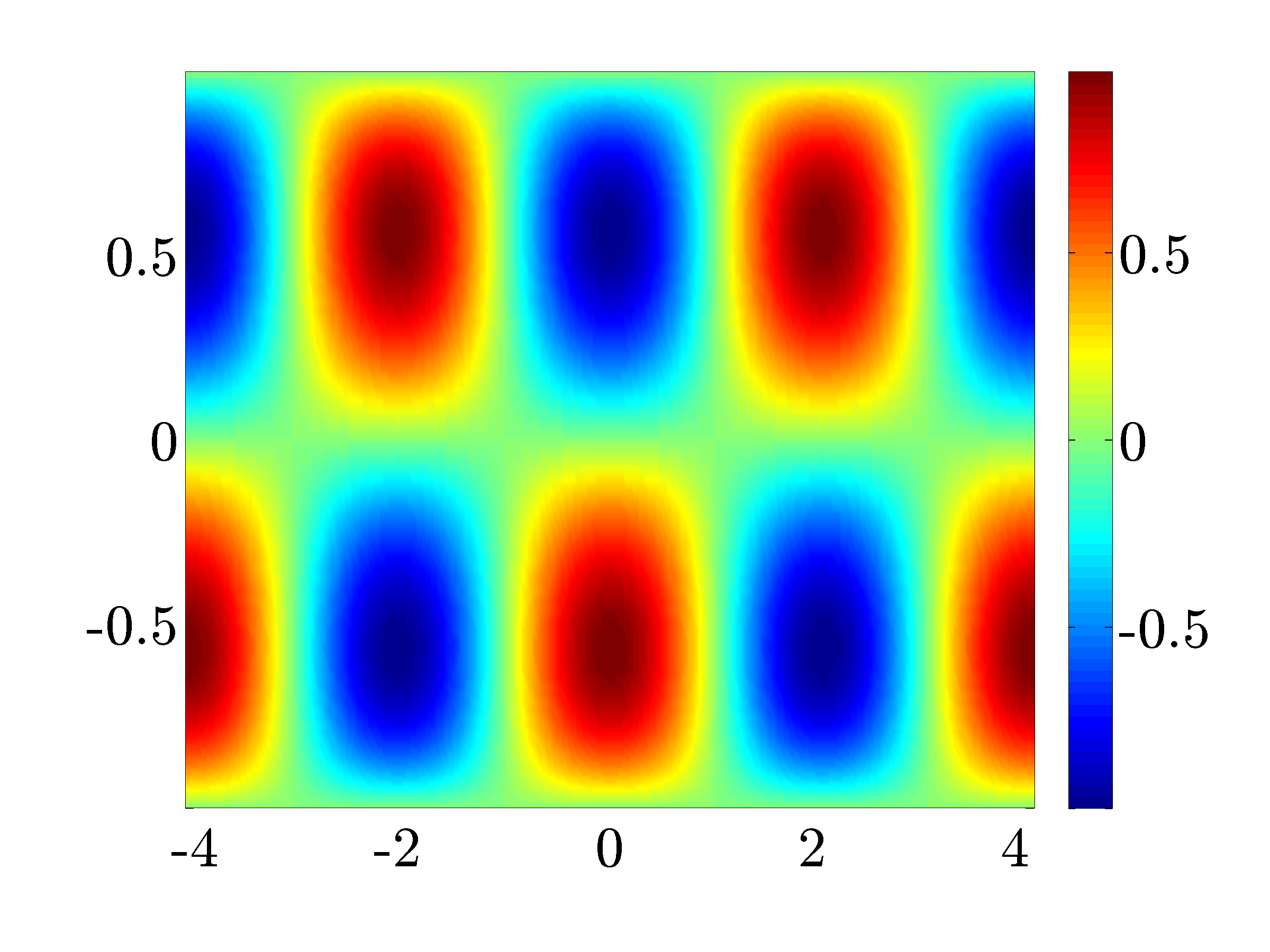}
					}
					\\[-0.2cm]
					{\large $z$}
					\\
					\subfigure[]{
					\label{fig.d3-structure-kx0-cout}
					}
				\end{tabular}
				\\[-0.15cm]
				\begin{sideways}
					\hspace{0.6cm}
					{\large $y$}
				\end{sideways}
				\hspace*{-0.3cm}
				\begin{tabular}{c}
					{
					\includegraphics[width=0.45\columnwidth]
					{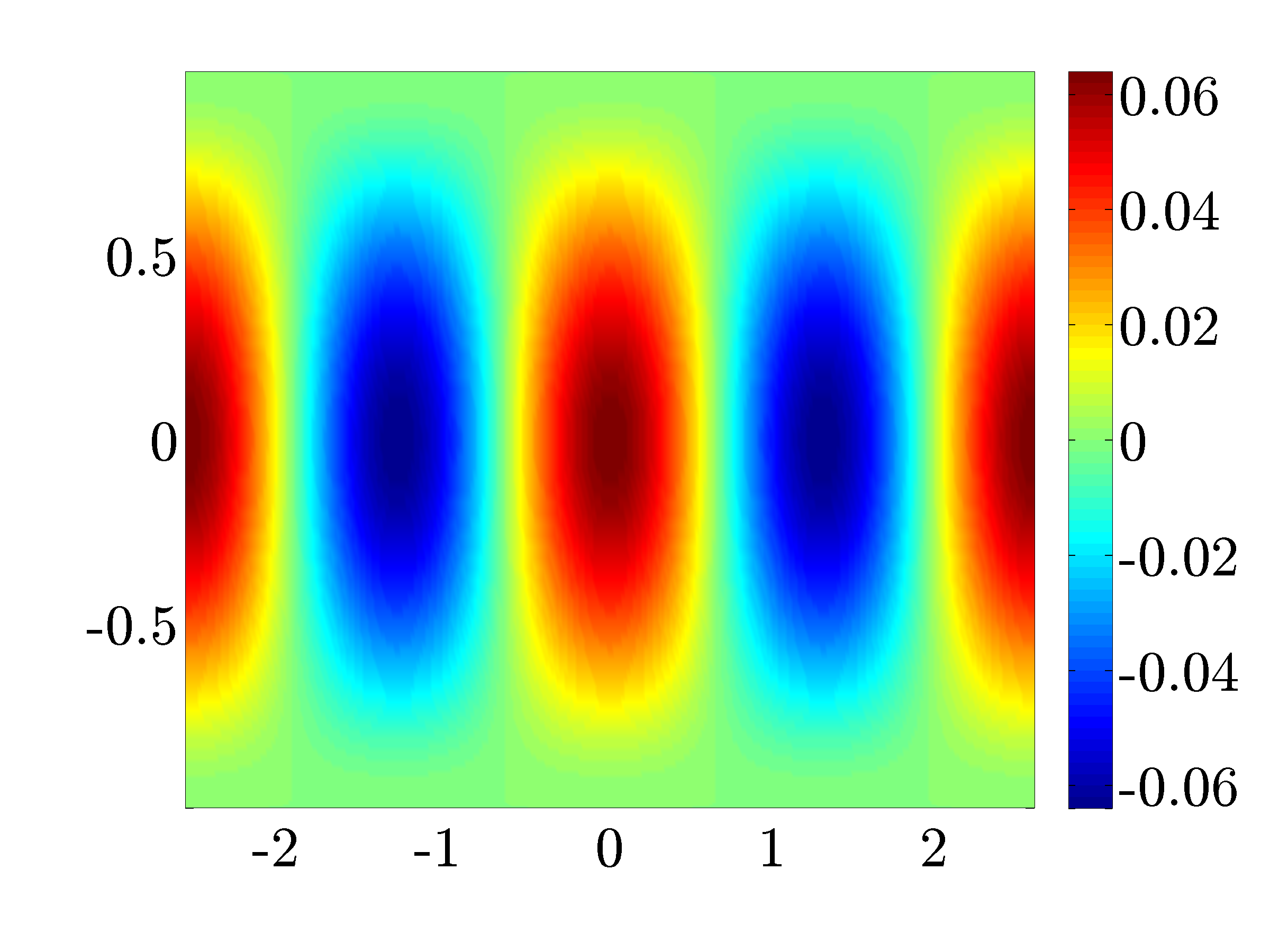}
					}
					\\[-0.2cm]
					{\large $z$}
					\\
					\subfigure[]{
					\label{fig.ud2-structure-kx0-cout}
					}
				\end{tabular}
				&
				\begin{tabular}{c}
					{
					\includegraphics[width=0.45\columnwidth]
					{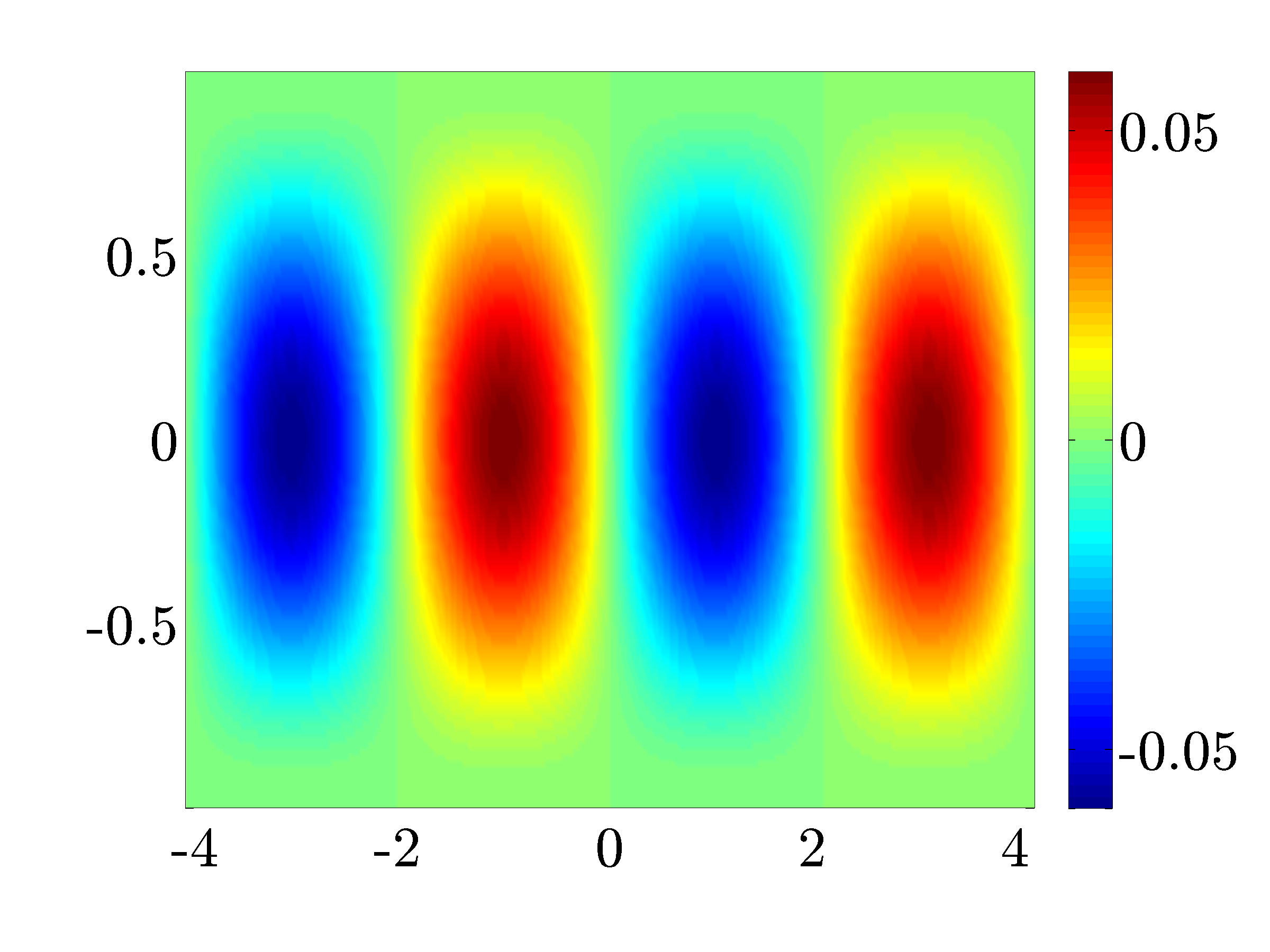}
					}
					\\[-0.2cm]
					{\large $z$}
					\\
					\subfigure[]{
					\label{fig.ud3-structure-kx0-cout}
					}
				\end{tabular}
			\end{tabular}
		\end{center}
		\caption{First row: body forcing fluctuations in $d_2$ and $d_3$ leading to the largest amplification of $u$. Second row: streamwise velocity fluctuations obtained by forcing the inertialess Couette flow with the body force fluctuations shown in (a) and (b), respectively.}
	    	\label{fig.d23-structure}
	 	\end{figure}
		
		Fluctuations in $d_2$ and $d_3$ that lead to the largest amplification of $u$ are shown in figures~\ref{fig.d2-structure-kx0-cout} and~\ref{fig.d3-structure-kx0-cout}. The wall-normal forcing is symmetric with respect to the channel centerline with the peak value located at the center of the channel; in contrast, the spanwise forcing is antisymmetric with respect to the channel centerline.

		Figures~\ref{fig.ud2-structure-kx0-cout} and~\ref{fig.ud3-structure-kx0-cout} illustrate the spatial structures of streamwise velocity induced by the body forcing fluctuations shown in figures~\ref{fig.d2-structure-kx0-cout} and~\ref{fig.d3-structure-kx0-cout}. Both body forces yield a symmetric response in $u$ with vortices occupying the entire channel width. We note that the dominant flow structures shown in figure~\ref{fig.d23-structure} do not exhibit significant deviation with $\beta$, $\We$, and $L$. Furthermore, we observe striking similarity between these flow structures and flow structures resulting from the analysis of stochastically forced Couette flow of Oldroyd-B fluids~\citep{jovkumJNNFM11}. Although these spatial structures may not match the full complexity of flow patterns produced in experiments and direct numerical simulations, our analysis identifies dynamical features that are likely to play an important role in shear-driven channel flows of viscoelastic fluids.
		
	\subsection{Physical mechanisms}
	
We next discuss the physical mechanisms responsible for strong influence of wall-normal and spanwise forces on streamwise velocity fluctuations in inertialess channel flows. As shown in Appendix~\ref{sec.app-derivation}, in the absence of streamwise forcing and streamwise variations in flow fluctuations, $u (y,z,t)$ evolves according to
	\begin{equation}
	\begin{array}{rcl}
		\Delta \dot{u}
		& \!\! = \!\! &
		- \dfrac{\bar{f}}{\beta} \,
		\Bigg\{
		\Delta u
		\, + \,
		\left( 1 - \beta \right)
		\,
		\big(
		\partial_{y}
		\left( U' r_{22} \right)
		\, + \,
		\partial_{z}
		\left( U' r_{23} \right)
		\big)
		\, + \,
		\cfrac{1 - \beta}{\bar{L}^2}
		\;
		\py
		\dot{r}_{11}
		\Bigg\}
		\\[0.6cm]
		& \!\! = \!\! &
		- \dfrac{\bar{f}}{\beta} \,
		\Bigg\{
		\Delta u
		\, + \,
        \sqrt{\dfrac{\bar{N}_1}{2}}
		\left( 1 - \beta \right)
		\,
		\big(
		\partial_{y}
		\left( U' \tau_{22} \right)
		\, + \,
		\partial_{z}
		\left( U' \tau_{23} \right)
		\big)
		~ +
		\\[0.6cm]
		&&
		\cfrac{1 - \beta}{\bar{L}^2}
		\left(
		\cfrac{2 \We^2}{\bar{f}} \; \pyy u
		\, + \,
		2 \We \, \py \left( U' r_{12} \right)
		\, - \,
		\left(
		\bar{f}
		\, + \,
		\dfrac{2 \We^2}{\bar{L}^2}
		\right)
		\partial_{y}
		r_{11}
		\right)
		\Bigg\}.
	\end{array}
	\label{eq.u-SOB-tau}
	\end{equation}
Therefore, even in the absence of inertia, the source term in the evolution equation for streamwise velocity is provided by the interactions between polymer stress fluctuations (in the wall-normal/spanwise plane) with the base shear $U'$. Furthermore, the finite extensibility of the polymer chains introduces additional source terms that are not present in the Oldroyd-B model; physically, these additional terms originate from: (i) the interaction between the streamwise shear component of the conformation tensor $r_{12}$ with the base shear; and (ii) the wall-normal gradient of the streamwise component of the conformation tensor $r_{11}$. Relative to the terms that are already present in the Oldroyd-B model, the influence of these additional terms (that arise from temporal changes in $r_{11}$) is much weaker.

We have demonstrated in~\S~\ref{sec.streamwise-constant} that the the essential features of the worst-case amplification from $d_{2}$ and $d_3$ to $u$ can be captured by analyzing the corresponding frequency responses at $\omega = 0$. At zero temporal frequency, equation~\eqref{eq.u-SOB-tau} simplifies to a static-in-time relation between the streamwise velocity and the fluctuating components of the polymer stress tensor,
	\begin{equation}
		\Delta u
		\; = \;
        - \, \sqrt{{\bar{N}_{1}}/{2}}
        \,
		\left( 1 - \beta \right)
		\big(
		\partial_{y}
		\left( U' \tau_{22} \right)
		\, + \,
		\partial_{z}
		\left( U' \tau_{23} \right)
		\big).
	\label{eq.u-SOB-tau-omega0}
	\end{equation}
	Furthermore, $\tau_{22}$ and $\tau_{23}$ are proportional to the spatial gradients of the ($y,z$)-plane streamfunction $\psi$ (i.e., $v = \partial_z \psi$, $w = -\partial_y \psi$),
	\begin{equation}
	\label{eq.r22-r23}
		\tau_{22}
		\; = \;
		2 \, \pyz \psi,
		\;\;\;
		\tau_{23}
		\; = \;
		-\left( \pyy - \pzz \right) \psi,
	\end{equation}
	and $\psi$ is induced by the action of the wall-normal and spanwise forces
	\begin{equation}
	\label{eq.psi-d2d3}
		\psi
		\; = \;
		\Delta^{-2}
		\left[
		\begin{array}{cc}
			-\pz & \py
		\end{array}
		\right]
		\left[
		\begin{array}{c}
			d_{2} \\[0.1cm]
			d_{3}
		\end{array}
		\right].
	\end{equation}
Finally, by substituting~\eqref{eq.r22-r23} into~(\ref{eq.u-SOB-tau-omega0}) we obtain the following expression
	\beq
	\begin{array}{rcl}
		\Delta u
		& \!\! = \!\! &
		- \,
        \ds{\sqrt{{\bar{N}_{1}}/{2}}}
		\,
		\left(1 - \beta \right)
		\pz
		\,
		\Delta
		\psi
		\\[0.25cm]
		& \!\! = \!\! &
		\ds{\sqrt{{\bar{N}_{1}}/{2}}}
		\,
		\left(1 - \beta \right)
		\pz
        \,
		\omega_x,
	\end{array}
	\label{eq.u-SOB-psi}
	\eeq
that relates fluctuations in the streamwise velocity $u$ and the streamwise vorticity $\omega_x$ in inertialess Couette flow of FENE-CR fluids without streamwise and temporal variations (i.e., at $k_x = 0$ and $\omega = 0$).
    	
This demonstrates that $\cO (1)$ fluctuations in streamwise vorticity induce $\cO (\sqrt{\bar{N}_{1}})$ fluctuations in streamwise velocity through a viscoelastic equivalent of the well-known lift-up mechanism. In contrast to Newtonian fluids, where vortex tilting induces large amplification, the lift-up mechanism in viscoelastic fluids originates from interactions between polymer stress fluctuations in the ($y,z$)-plane with background shear~\citep{jovkumJNNFM11}. In the absence of inertia, a static-in-time momentum equation relates the wall-normal and spanwise velocity fluctuations (and consequently the streamwise vorticity) to the polymer stress fluctuations $\tau_{22}$, $\tau_{23}$, and $\tau_{33}$. Interactions of these polymer stress fluctuations with background shear induce the energy transfer from the mean flow to fluctuations and redistribute momentum in the ($y,z$)-plane through a movement of the low speed fluid (away from the wall) and the high speed fluid (towards the wall). This momentum exchange is responsible for the generation of alternating regions of high and low streamwise velocity (relative to the mean flow), and it is facilitated by large normal stress difference $\bar{N}_1$, low viscosity ratios $\beta$, strong base shear $U'$, and strong spatial variations in streamwise vorticity fluctuations. As in streamwise-constant inertial flows of Newtonian fluids, this amplification disappears either in the absence of spanwise variations in flow fluctuations or in the absence of the background shear.
	
	\vspace*{-2ex}
\section{Dynamics of streamwise-constant polymer stress fluctuations}
\label{sec.polymer-stress-2d3c}
	
	Although we have so far confined our attention to the dynamics of velocity fluctuations, it is worth noting that polymer stress fluctuations can also experience significant amplification even in the absence of inertia. Since our computations (not shown here) demonstrate that largest responses in polymer stress fluctuations are induced by streamwise-constant deterministic forcing, we next examine the responses from body forcing to polymer stress fluctuations without streamwise-variations. We use analytical developments to show that the wall-normal and spanwise forces induce the largest amplification of the polymer stress fluctuations. The worst-case amplification obtained in the presence of these two forcing components takes place at $\omega = 0$ and it is proportional to: $\bar{N}_{1}$ for $\tau_{13}$ and $\tau_{12}$; and $\bar{N}_{1}^2 / \left(1 + \bar{N}_{1}/\bar{L}^2 \right)^2$ for $\tau_{11}$. Furthermore, the worst-case amplification from all forcing components to $\tau_{22}$, $\tau_{23}$, and $\tau_{33}$ is independent of $\beta$, $\We$, and $L$. We also illustrate that the worst-case amplification from $d_2$ and $d_3$ to $\tau_{11}$ scales
    \bi
    \item quartically with $\We$ as $L \rightarrow \infty$;

    \item quartically with $L$ as $\We \rightarrow \infty$.
    \ei
		
Following a sequence of straightforward algebraic manipulations, the frequency response operator $\bG$ that maps body forcing fluctuations $d_1$, $d_2$, and $d_3$ to polymer stress fluctuations can be expressed as
	\begin{equation}
		\left[
		\begin{array}{c}
			\tau_{22} \\[0.1cm]
			\tau_{23} \\[0.1cm]
			\tau_{33} \\[0.1cm]
			\tau_{13} \\[0.1cm]
			\tau_{12} \\[0.1cm]
			\tau_{11}
		\end{array}
		\right]
		\; = \;
		\left[
		\begin{array}{ccc}
			0 & \bG_{12} & \bG_{13} \\[0.1cm]
			0 & \bG_{22} & \bG_{23} \\[0.1cm]
			0 & \bG_{32} & \bG_{33} \\[0.1cm]
			\bG_{41} & \bG_{42} & \bG_{43} \\[0.1cm]
			\bG_{51} & \bG_{52} & \bG_{53} \\[0.1cm]
			\bG_{61} & \bG_{62} & \bG_{63}
		\end{array}
		\right]
		\,
		\left[
		\begin{array}{c}
			d_{1} \\[0.1cm]
			d_{2} \\[0.1cm]
			d_{3}
		\end{array}
		\right],
	\label{eq.H-psi}
	\end{equation}
	where the streamwise-constant frequency response operators $\bG_{\ell j}$ are given in Appendix~\ref{sec.app-Gop-2d3c}.
			
	We note that all components of the frequency response operator in~(\ref{eq.H-psi}) exhibit roll-off at high temporal frequencies, thereby indicating that the largest singular value of each component of $\bG$ peaks at finite temporal frequency. In particular, the worst-case amplification from $d_2$ and $d_3$ to $\tau_{22}$, $\tau_{23}$, and $\tau_{33}$ takes place at $\omega = 0$ and is determined by the following
$\beta$-, $\We$-, and $L$-independent functions
	\begin{equation}
	\label{eq.hinf-tau1-d23-2d3c}
		\begin{array}{rcl}
			\left[
			\begin{array}{rr}
				G_{1 2} (k_z)
				&
				G_{1 3} (k_z)
				\\[0.1cm]
				G_{2 2} (k_z)
				&
				G_{2 3} (k_z)
				\\[0.1cm]
				G_{3 2} (k_z)
				&
				G_{3 3} (k_z)
			\end{array}
			\right]
			& \!\! = \!\! &
                			\left[
			\begin{array}{rr}
				g_{1 2} (k_z)
				&
				g_{1 3} (k_z)
				\\[0.1cm]
				g_{2 2} (k_z)
				&
				g_{2 3} (k_z)
				\\[0.1cm]
				g_{3 2} (k_z)
				&
				g_{3 3} (k_z)
			\end{array}
			\right].
		\end{array}
	\end{equation}

\begin{figure}
   	\begin{center}
       		\begin{tabular}{cc}
			\hspace{-0.3cm}
			\begin{tabular}{c}
				{\large $g_{12}(k_z)$, $g_{13}(k_z)$, $g_{32}(k_z)$, $g_{33}(k_z)$}
				\\
				{
				\includegraphics[width=0.45\columnwidth]
				{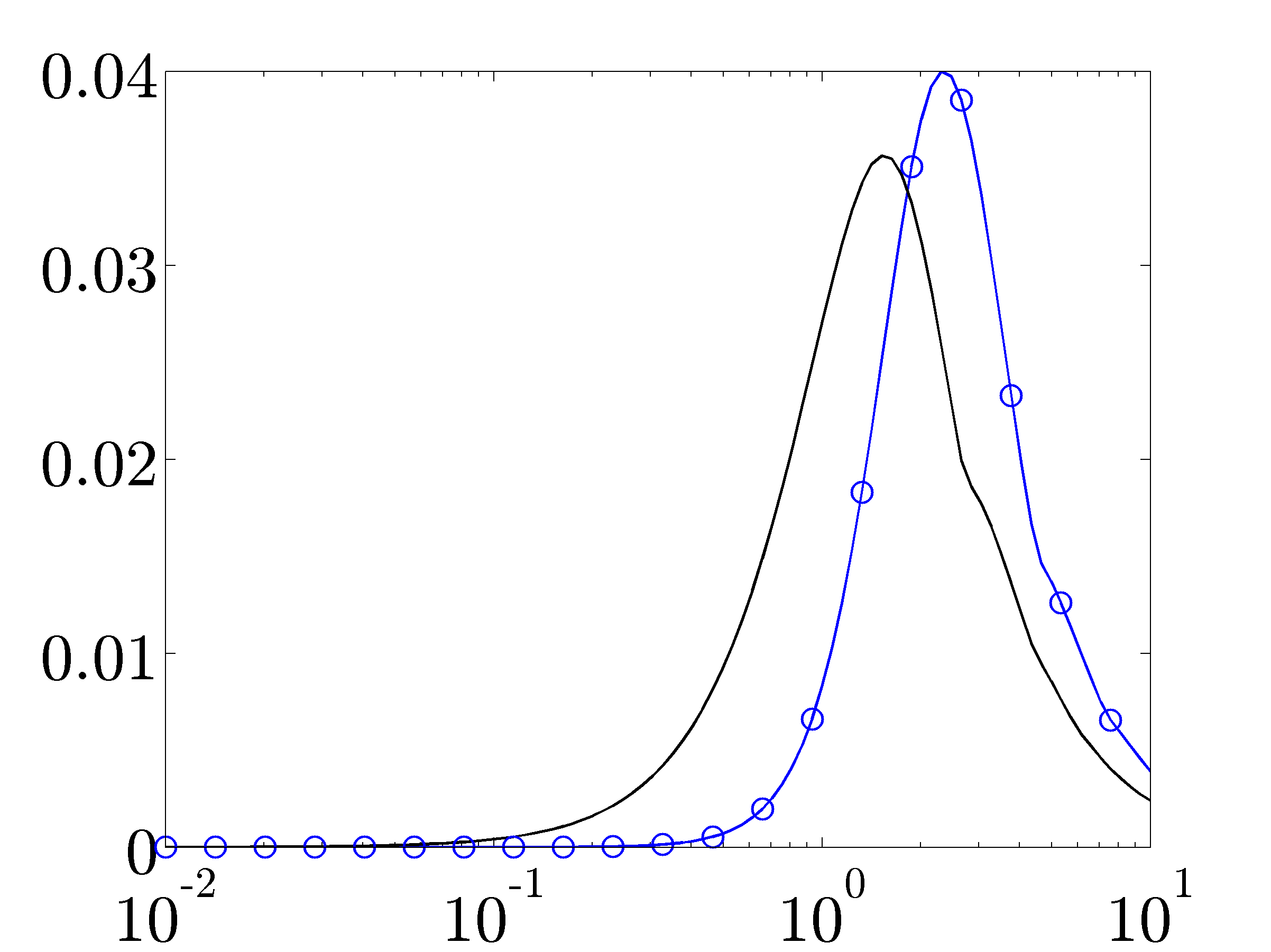}
				}
				\\
				{\large $k_z$}
				\\
				\subfigure[]
				{
				\label{fig.g12-g13-kx0}
				}
			\end{tabular}
			&
			\hspace{-0.7cm}
			\begin{tabular}{c}
				{\large $g_{22}(k_z)$, $g_{23}(k_z)$}
				\\
				{
				\includegraphics[width=0.45\columnwidth]
				{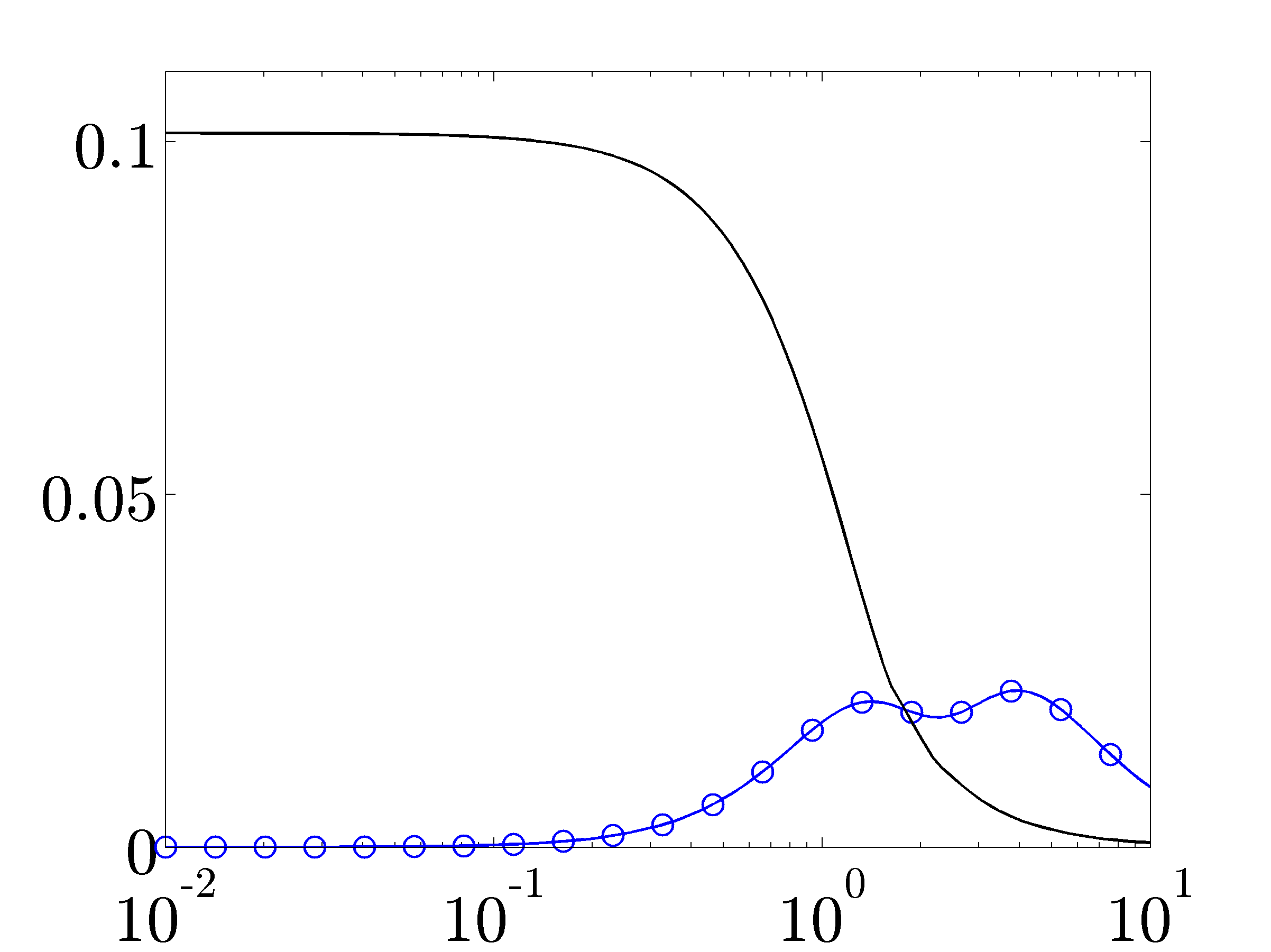}
				}
				\\
				{\large $k_z$}
				\\
				\subfigure[]
				{
				\label{fig.g22-g23-kx0}
				}
			\end{tabular}
		\end{tabular}
	\end{center}
	\caption{Spanwise frequency responses from $d_{j}$ to the polymer stress fluctuations $\tau_{22}$ ($g_{1 j}$), $\tau_{23}$ ($g_{2 j}$), and $\tau_{33}$ ($g_{3 j}$) with $j = \{ 2, 3\}$. The function $g_{12} =  g_{32}$ and the function $g_{13} = g_{33}$: (a) $g_{12}$, $g_{32}$ ($\circ$) and $g_{13}$, $g_{33}$ (solid); and (b) $g_{22}$ ($\circ$) and $g_{23}$ (solid).}
    	\label{fig.g-tau1-23-kx0}
 	\end{figure}

Figure~\ref{fig.g-tau1-23-kx0} shows the functions $g_{\ell j}$ with $\{ \ell = 1, 2, 3; j = 2, 3\}$ that quantify the spanwise wavenumber dependence of the respective frequency response operators. We note that $g_{32} = g_{12}$  and $g_{33} = g_{13}$. From figure~\ref{fig.g12-g13-kx0}, we see that $g_{12}$ and $g_{13}$ decay to zero at both low and high wavenumbers with the maximum values occurring at $k_z \approx 2.3$ and $k_z \approx 1.5$, respectively. Similarly, function $g_{22}$ decays to zero at both low and high values of $k_z$ and it achieves two peaks at $k_z \approx 1.4$ and $k_z \approx 4.0$. On the other hand, $g_{23}$ displays low-pass behavior and the maximum value of this frequency response is approximately three times larger than the maximum values of other responses in figure~\ref{fig.g-tau1-23-kx0}.
	
	\begin{figure}
   	\begin{center}
       		\begin{tabular}{ccc}
			\hspace{-0.3cm}
			\begin{tabular}{c}
				$\sigma^{2}_{\max} \left( \bG_{4 1}\right)$
				\\
				{
				\includegraphics[width=0.33\columnwidth]
				{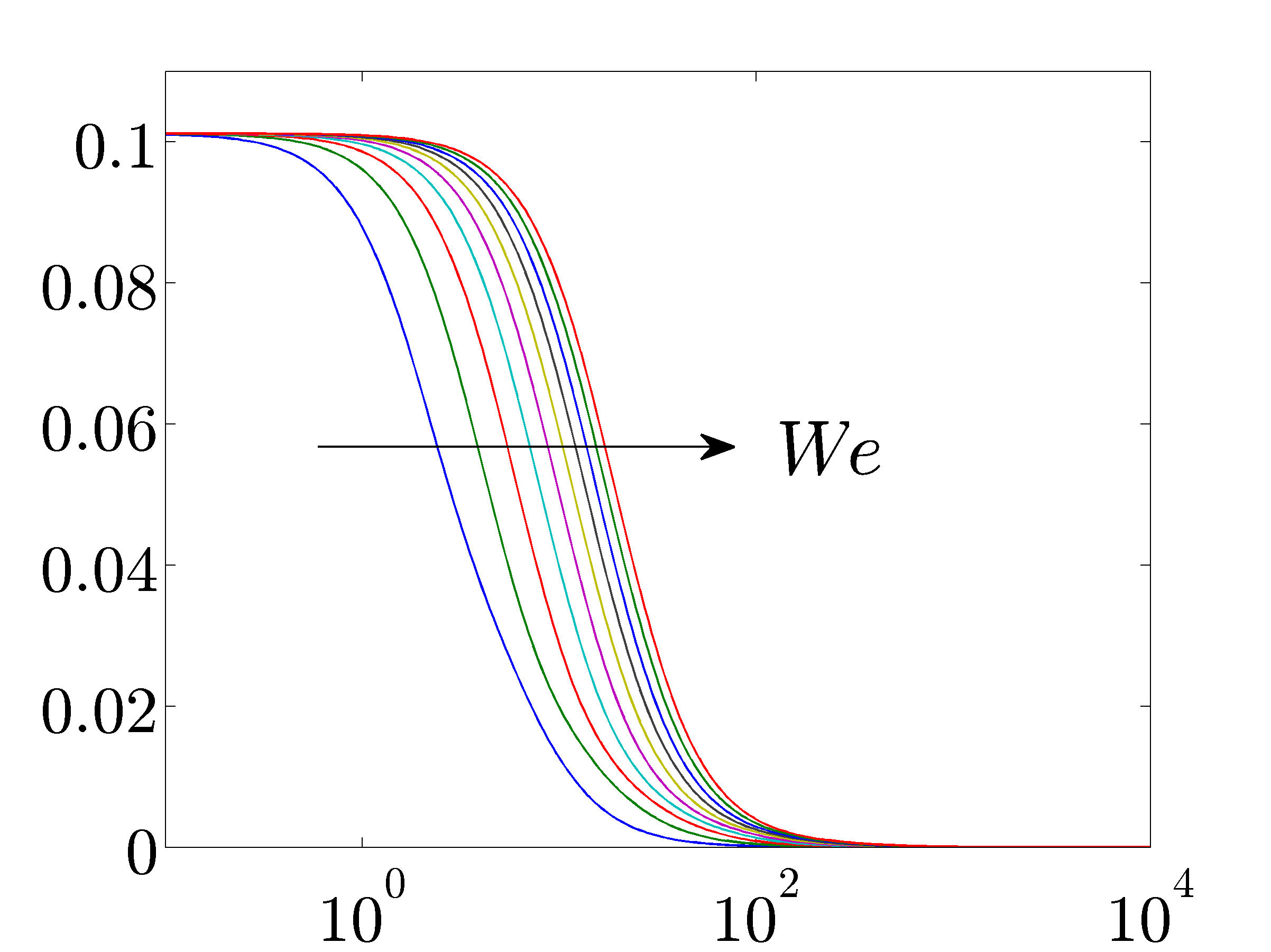}
				}
				\\
				{\large $\omega$}
				\\
				\subfigure[]
				{
				\label{fig.Smax-tau13-d1-kx0-L10}
				}
			\end{tabular}
			&
			\hspace{-0.7cm}
			\begin{tabular}{c}
				$\sigma^{2}_{\max} \left( \bG_{4 2}\right)$
				\\
				{
				\includegraphics[width=0.33\columnwidth]
				{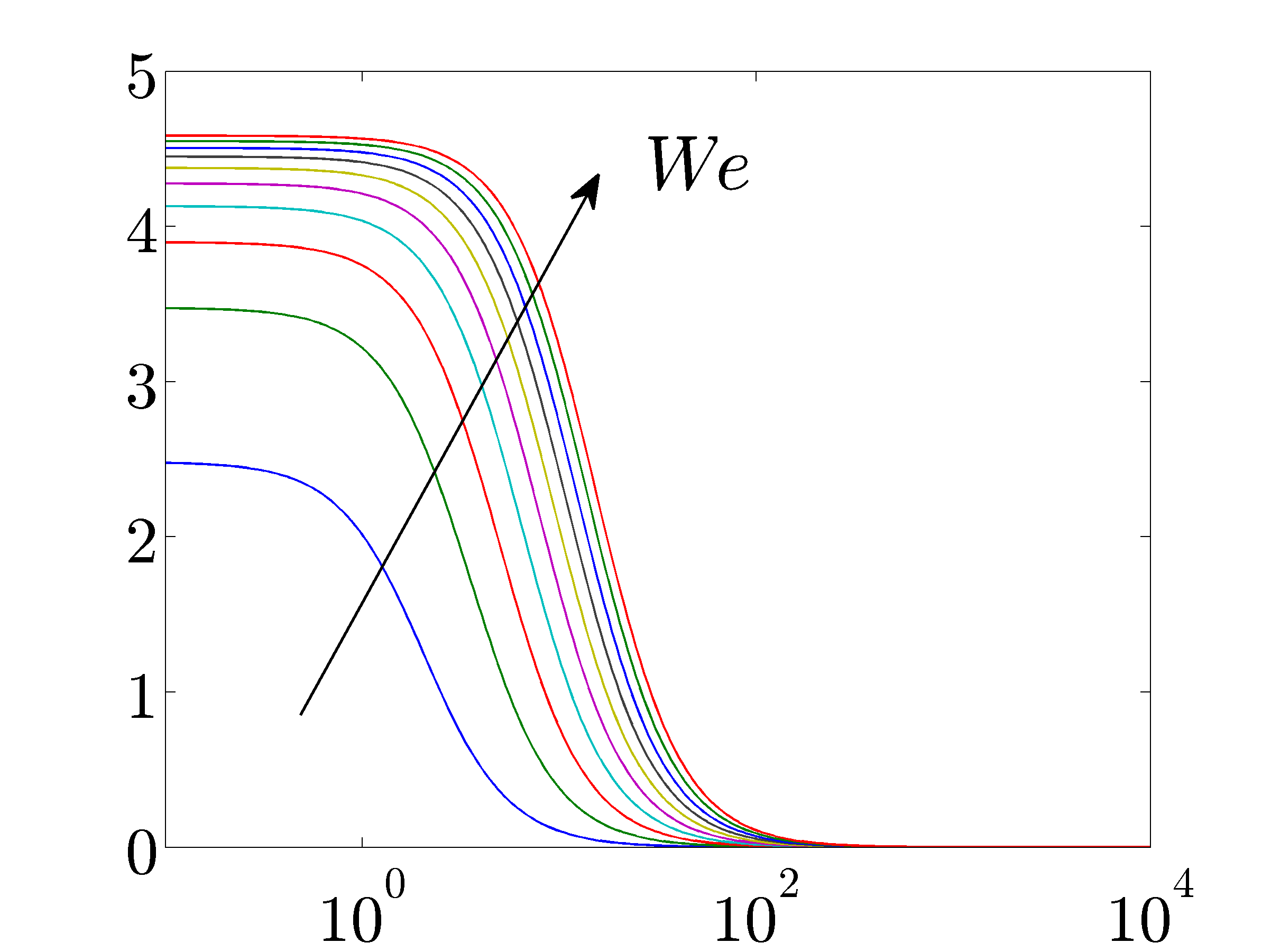}
				}
				\\
				{\large $\omega$}
				\\
				\subfigure[]
				{
				\label{fig.Smax-tau13-d2-kx0-L10}
				}
			\end{tabular}
			&
			\hspace{-0.7cm}
			\begin{tabular}{c}
				$\sigma^{2}_{\max} \left( \bG_{4 3}\right)$
				\\
				{
				\includegraphics[width=0.33\columnwidth]
				{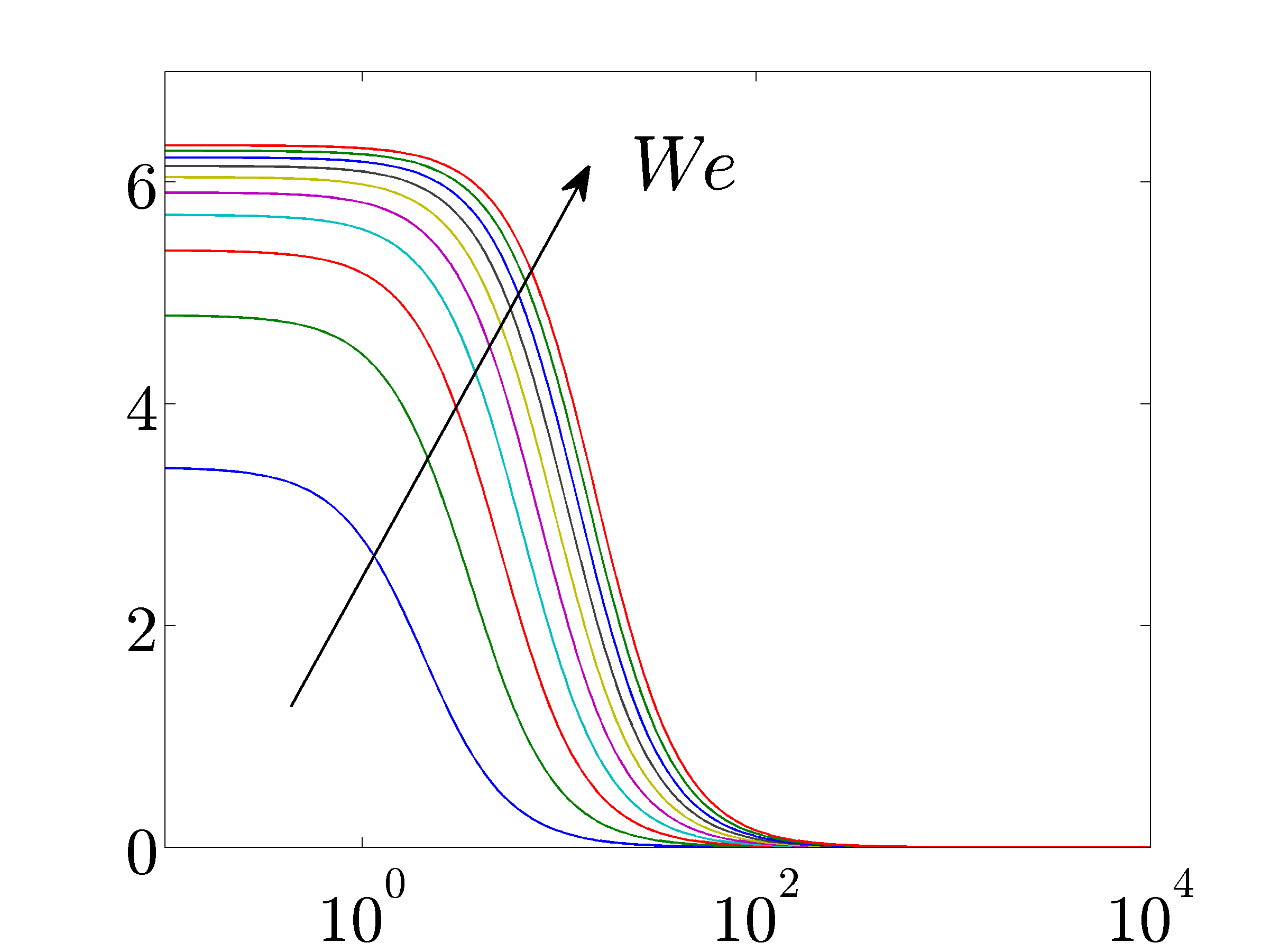}
				}
				\\
				{\large $\omega$}
				\\
				\subfigure[]
				{
				\label{fig.Smax-tau13-d3-kx0-L10}
				}
			\end{tabular}
			\\[2.5cm]
			\hspace{-0.3cm}
			\begin{tabular}{c}
				$\sigma^{2}_{\max} \left( \bG_{5 1}\right)$
				\\
				{
				\includegraphics[width=0.33\columnwidth]
				{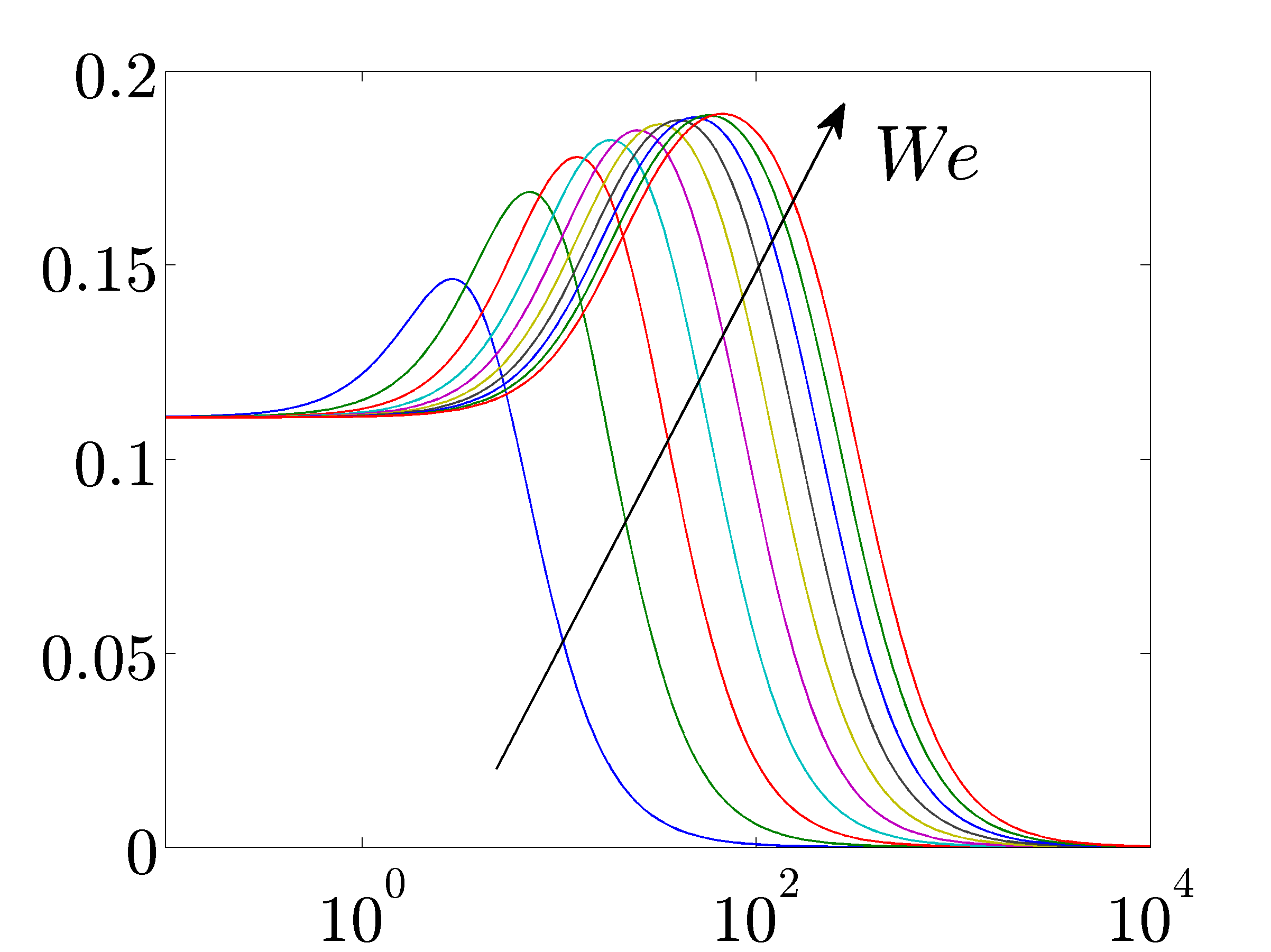}
				}
				\\
				{\large $\omega$}
				\\
				\subfigure[]
				{
				\label{fig.Smax-tau12-d1-kx0-L10}
				}
			\end{tabular}
			&
			\hspace{-0.7cm}
			\begin{tabular}{c}
				$\sigma^{2}_{\max} \left( \bG_{5 2}\right)$
				\\
				{
				\includegraphics[width=0.33\columnwidth]
				{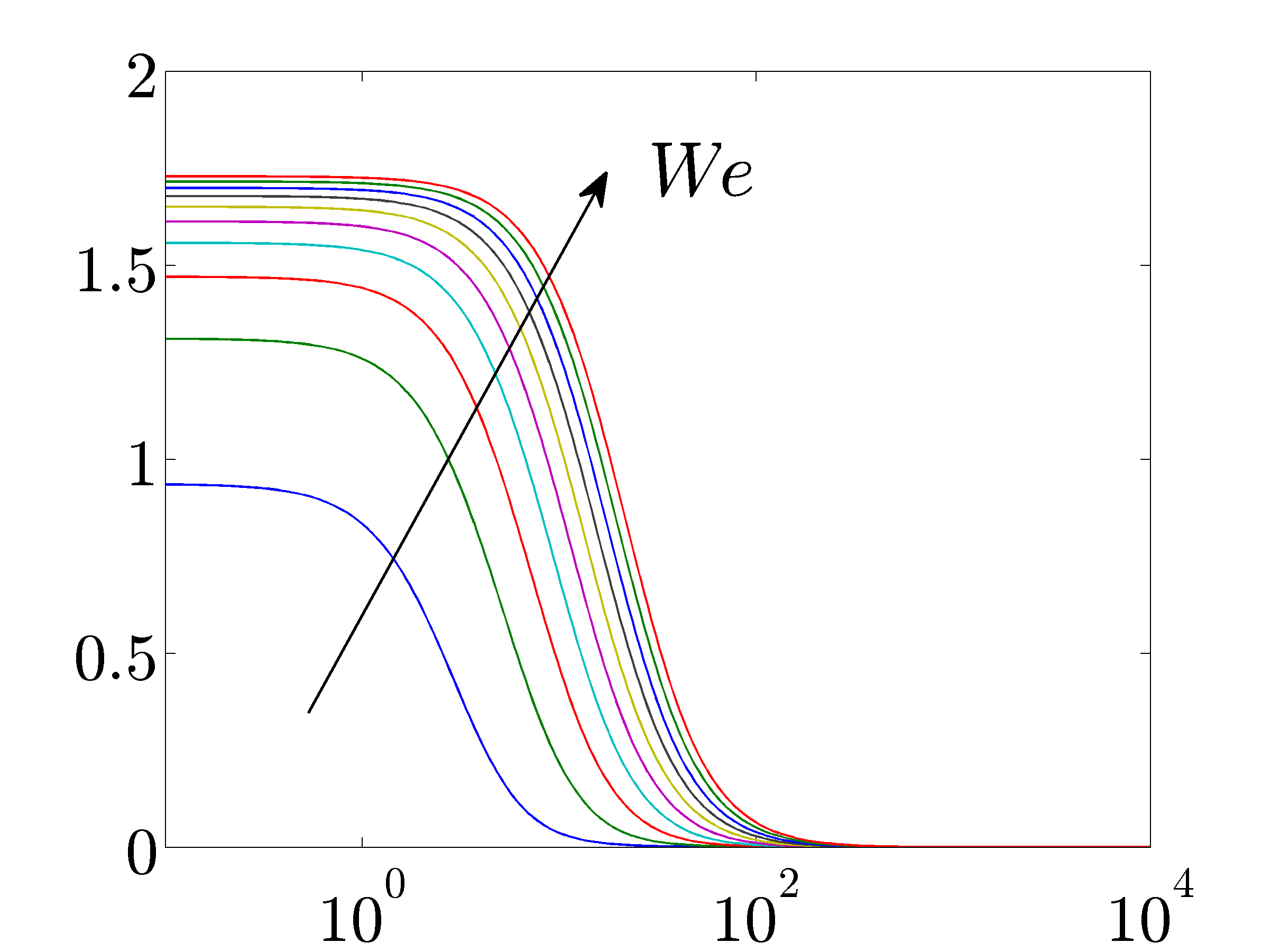}
				}
				\\
				{\large $\omega$}
				\\
				\subfigure[]
				{
				\label{fig.Smax-tau12-d2-kx0-L10}
				}
			\end{tabular}
			&
			\hspace{-0.7cm}
			\begin{tabular}{c}
				$\sigma^{2}_{\max} \left( \bG_{5 3}\right)$
				\\
				{
				\includegraphics[width=0.33\columnwidth]
				{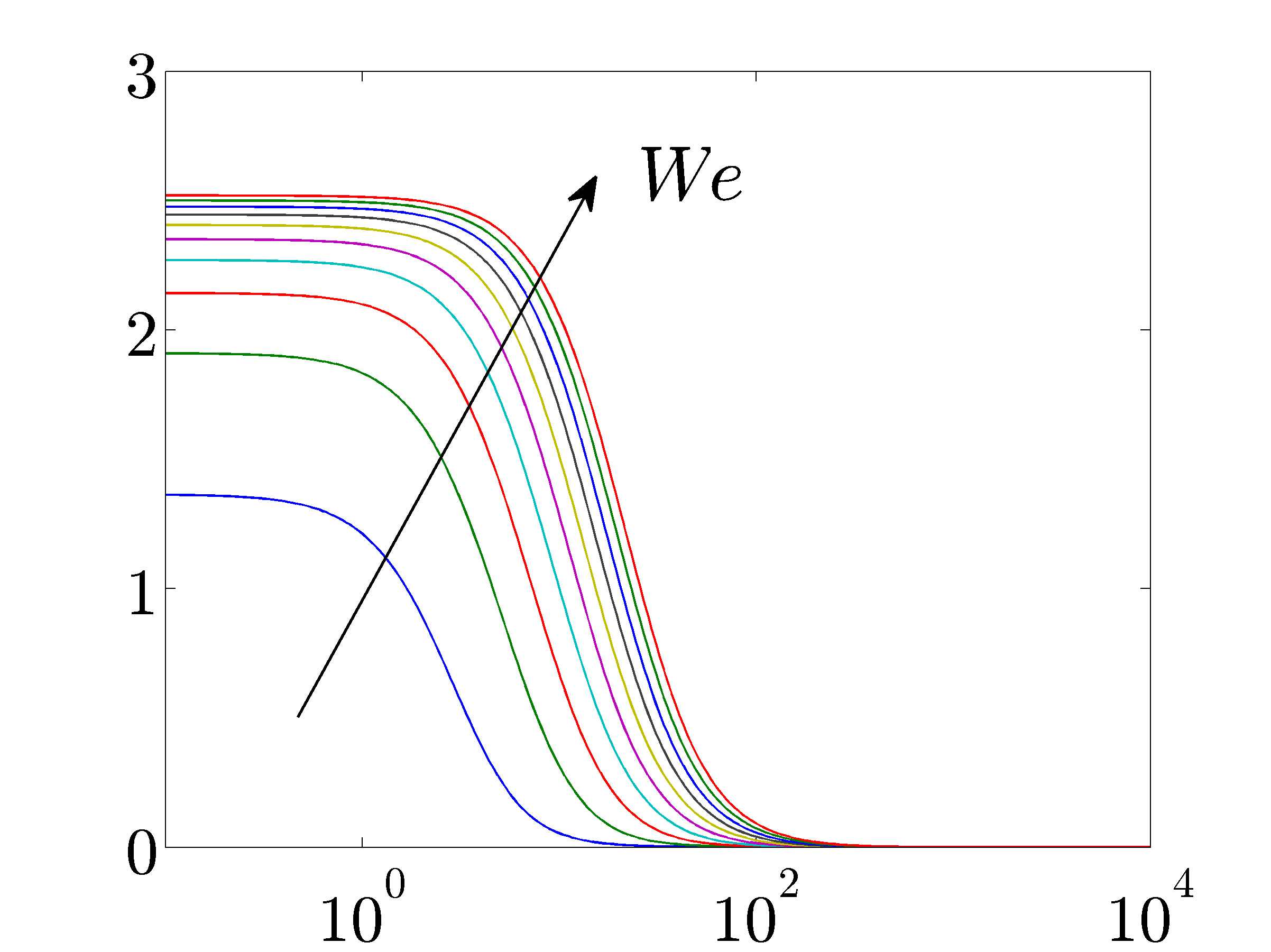}
				}
				\\
				{\large $\omega$}
				\\
				\subfigure[]
				{
				\label{fig.Smax-tau12-d3-kx0-L10}
				}
			\end{tabular}
			\\[2.5cm]
			\hspace{-0.3cm}
			\begin{tabular}{c}
				$\sigma^{2}_{\max} \left( \bG_{6 1}\right)$
				\\
				{
				\includegraphics[width=0.33\columnwidth]
				{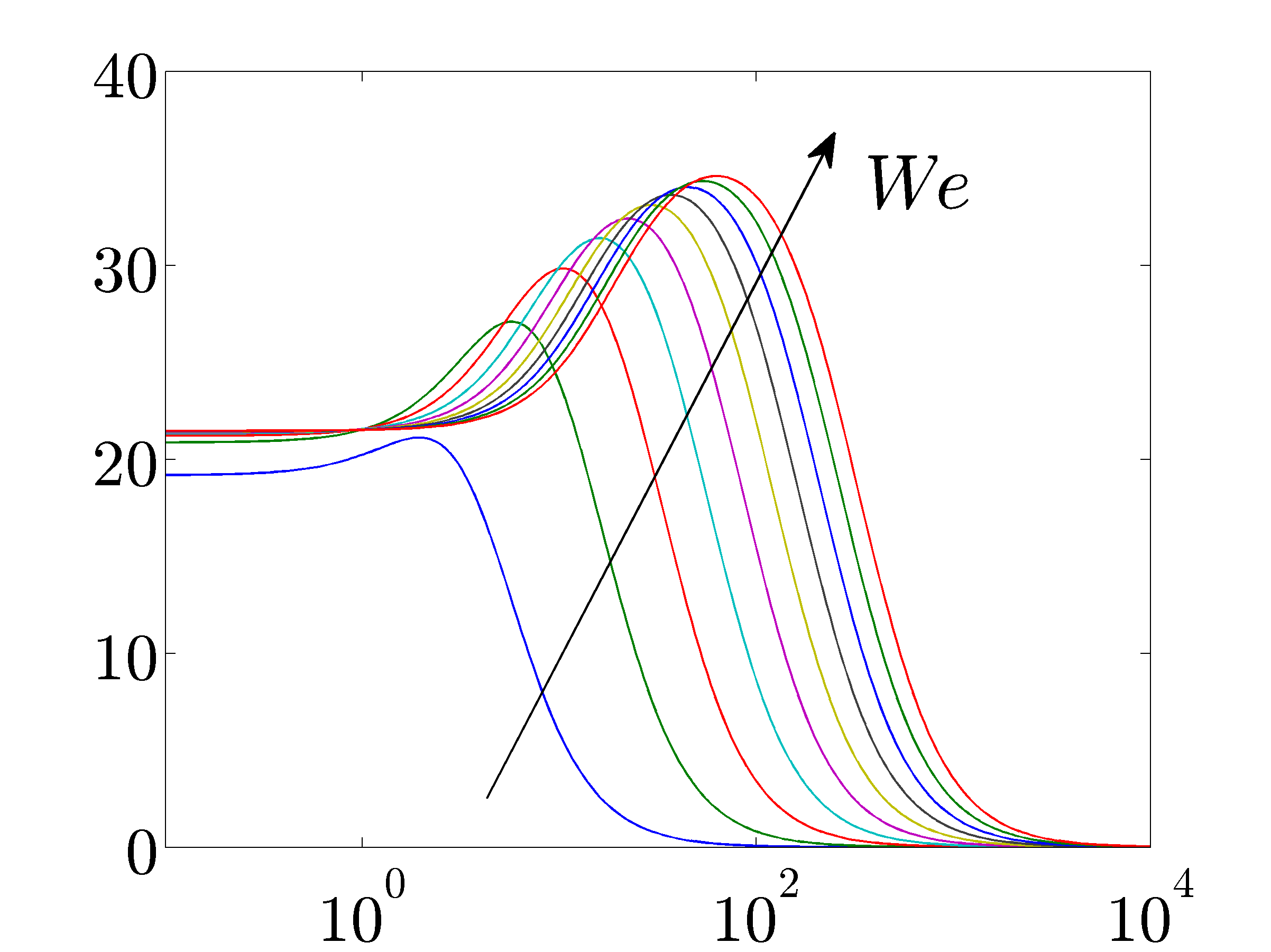}
				}
				\\
				{\large $\omega$}
				\\
				\subfigure[]
				{
				\label{fig.Smax-tau11-d1-kx0-L10}
				}
			\end{tabular}
			&
			\hspace{-0.7cm}
			\begin{tabular}{c}
				$\sigma^{2}_{\max} \left( \bG_{6 2}\right)$
				\\
				{
				\includegraphics[width=0.33\columnwidth]
				{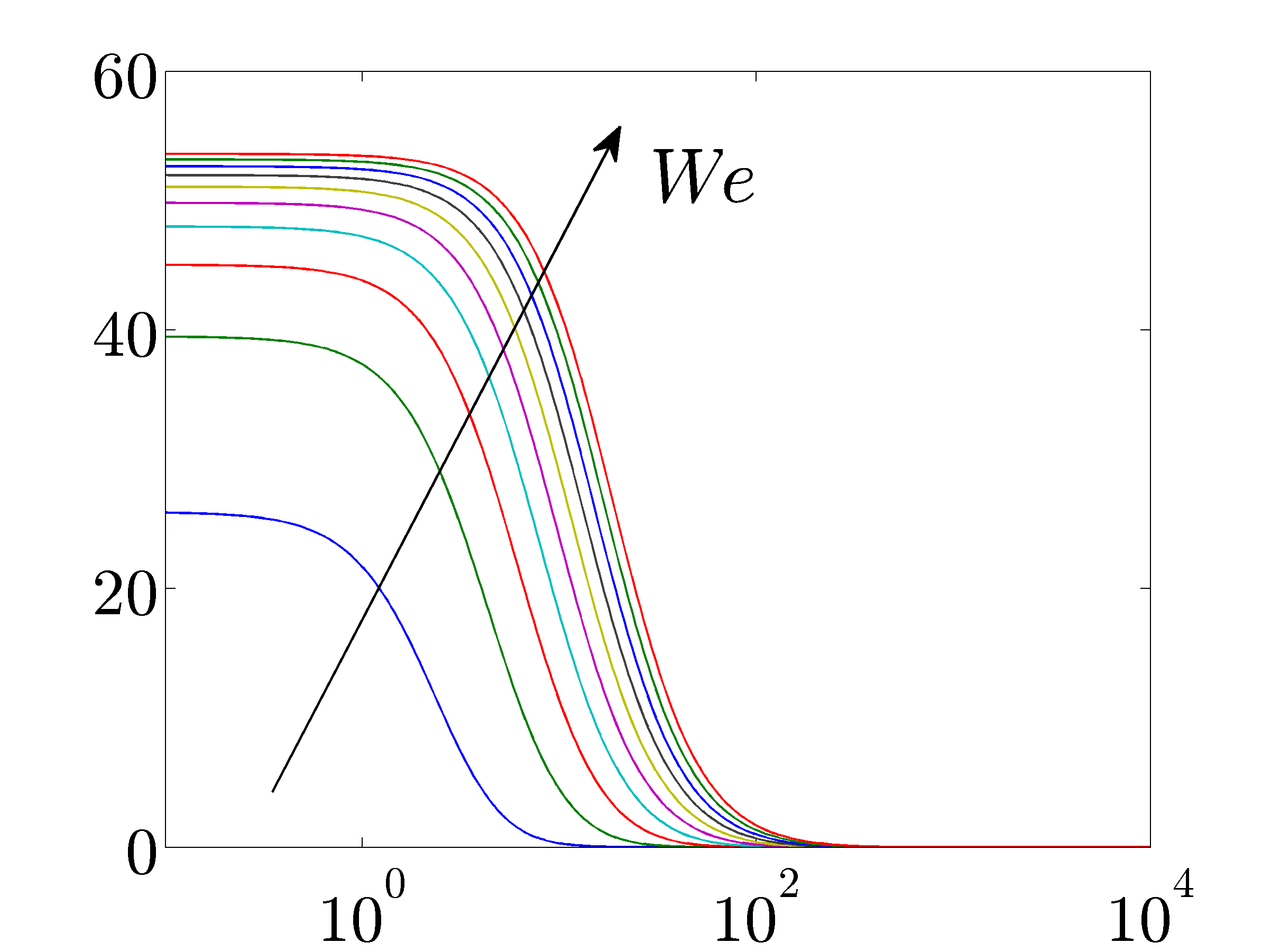}
				}
				\\
				{\large $\omega$}
				\\
				\subfigure[]
				{
				\label{fig.Smax-tau11-d2-kx0-L10}
				}
			\end{tabular}
			&
			\hspace{-0.7cm}
			\begin{tabular}{c}
				$\sigma^{2}_{\max} \left( \bG_{6 3}\right)$
				\\
				{
				\includegraphics[width=0.33\columnwidth]
				{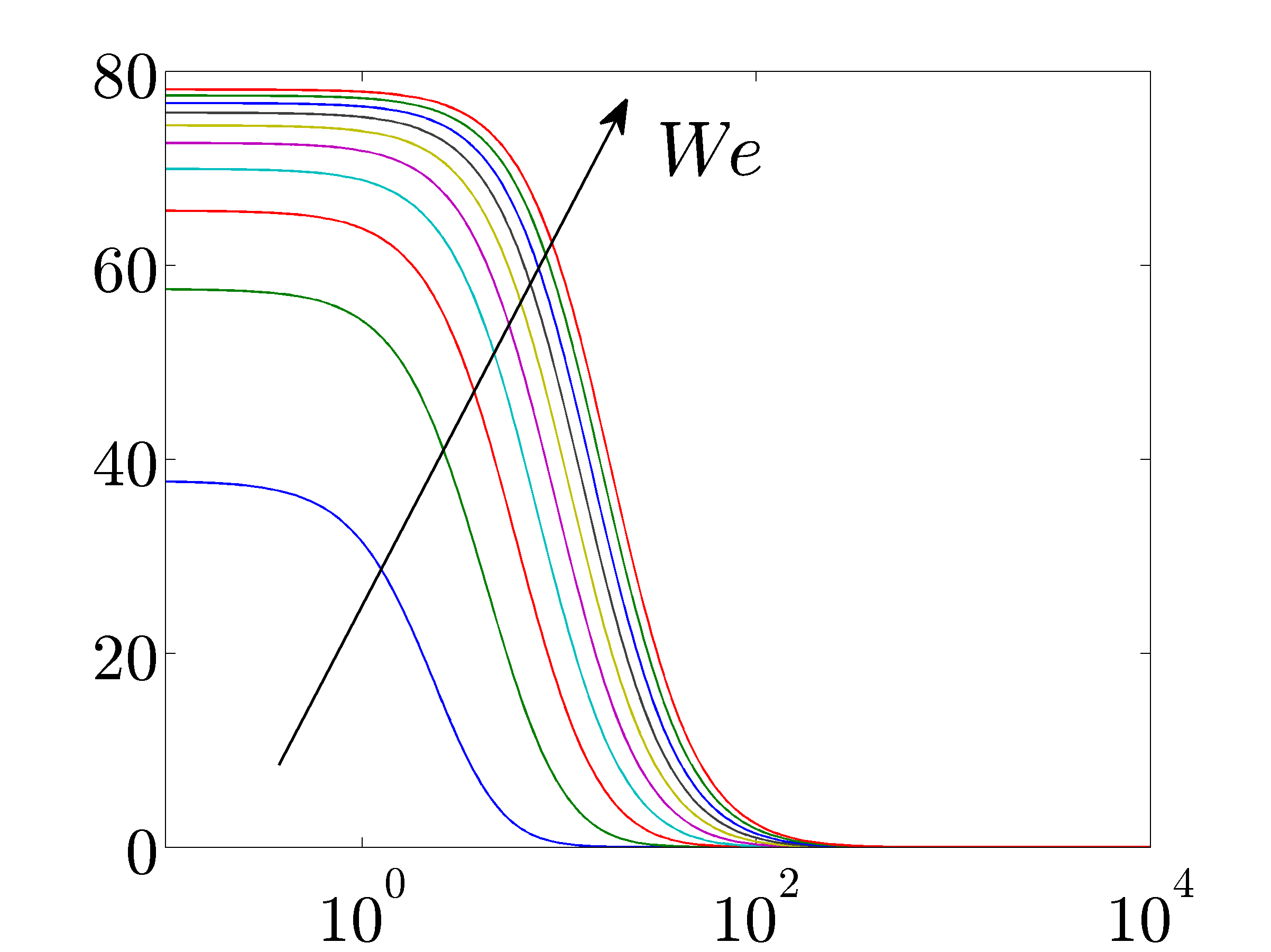}
				}
				\\
				{\large $\omega$}
				\\
				\subfigure[]
				{
				\label{fig.Smax-tau11-d3-kx0-L10}
				}
			\end{tabular}
		\end{tabular}
	\end{center}
	\caption{Maximum singular values of the streamwise-constant frequency responses operator from $d_j$ to $\tau_{13}$ ($\bG_{4j}$), $\tau_{12}$ ($\bG_{5j}$), and $\tau_{11}$ ($\bG_{6j}$) as a function of $\omega$ for $j = \{ 2, 3 \}$, $k_z = 1.5$, $\beta = 0.5$, $L = 10$ and $\We = [ 10, 100 ]$.}
    	\label{fig.Smax-tau2-d123-kx0}
 	\end{figure}
	
	We next analyze the temporal frequency responses of the operators $\bG_{\ell j}$ with $\{ \ell = 4, 5, 6; j = 1, 2, 3\}$ that map different forcing components to fluctuations in $\tau_{13}$, $\tau_{12}$, and $\tau_{11}$. Figure~\ref{fig.Smax-tau2-d123-kx0} shows $\omega$-dependence of $\sigma_{\max} \left( \bG_{\ell j} \right)$ for $k_z = 1.5$, $\beta = 0.5$, and $L = 10$. Figures~\ref{fig.Smax-tau13-d1-kx0-L10}~--~\ref{fig.Smax-tau13-d3-kx0-L10} show that $\sigma_{\max} \left( \bG_{4j} \right)$ with $j = \{ 1, 2, 3 \}$ achieve their respective peaks at $\omega = 0$. Furthermore, while the peak value of $\sigma_{\max} \left( \bG_{41} \right)$ does not depend on the Weissenberg number, the peak values of $\sigma_{\max} \left( \bG_{42} \right)$ and $\sigma_{\max} \left( \bG_{43} \right)$ increase with $\We$.
	
	The frequency responses $\bG_{\ell j}$ with $\{ \ell = 5, 6; j = 2, 3 \}$ that quantify amplification from $d_{2}$ and $d_{3}$ to $\tau_{12}$ and $\tau_{11}$ exhibit similar low-pass characteristics. On the other hand, figures~\ref{fig.Smax-tau12-d1-kx0-L10} and~\ref{fig.Smax-tau11-d1-kx0-L10} show that $\sigma_{\max} \left( \bG_{51} \right)$ and $\sigma_{\max} \left( \bG_{61} \right)$ achieve their peak values at non-zero temporal frequencies and that these values increase as $\We$ increases. We see that the forcing components in the wall-normal and spanwise directions induce larger amplification of polymer stress fluctuations compared to the streamwise forcing. Furthermore, the streamwise component of the polymer stress tensor $\tau_{11}$ experiences the largest amplification.
 		
	Following a series of algebraic manipulations, it can be shown that the worst-case amplification from $d_2$ and $d_3$ to $\tau_{13}$, $\tau_{12}$, and $\tau_{11}$ takes place at $\omega = 0$ and is given by
	\begin{equation}
	\label{eq.G456j}
	\begin{array}{rcl}
		G_{4 j}(k_z; \beta, \We, L)
		& \!\! = \!\! &
		\left( \bar{N}_{1} / 2 \right) g_{4j}(k_z; \beta),
		\\[0.2cm]
		G_{5 j}(k_z; \beta, \We, L)
		& \!\! = \!\! &
		\left( \bar{N}_{1} / 2 \right) \left( 2 + \beta \right)^2 \, g_{5j}(k_z),
		\\[0.15cm]
		G_{6 j}(k_z; \beta, \We, L)
		& \!\! = \!\! &
		\cfrac{\bar{N}_{1}^2}{\left( 1 \, + \, \bar{N}_1/\bar{L}^2 \right)^2}
        \;
        \left( 1 + 2 \beta \right)^2 g_{6j}(k_z),
		\;\;
		j = \{ 2, 3 \}.
	\end{array}
	\end{equation}
	The functions $g_{\ell j}$ with $\{ \ell = 4, 5, 6; j = 2, 3\}$ represent the $\We$- and $L$-independent spanwise frequency responses from the wall-normal and spanwise forces to $\tau_{13}$, $\tau_{12}$, and $\tau_{11}$. Equation~\eqref{eq.G456j} shows that the worst-case amplification of $\tau_{13}$ and $\tau_{12}$ is proportional to $\bar{N}_{1}$. On the other hand, the worst-case amplification of $\tau_{11}$ scales as $\bar{N}_{1}^2 / \left(1 + \bar{N}_{1}/\bar{L}^2 \right)^2$.
	
\begin{figure}
   	\begin{center}
       		\begin{tabular}{cc}
			\hspace{-0.3cm}
			\begin{tabular}{c}
				{\large $g_{42}(k_z; 0.5)$, $g_{43}(k_z; 0.5)$}
				\\
				{
				\includegraphics[width=0.45\columnwidth]
				{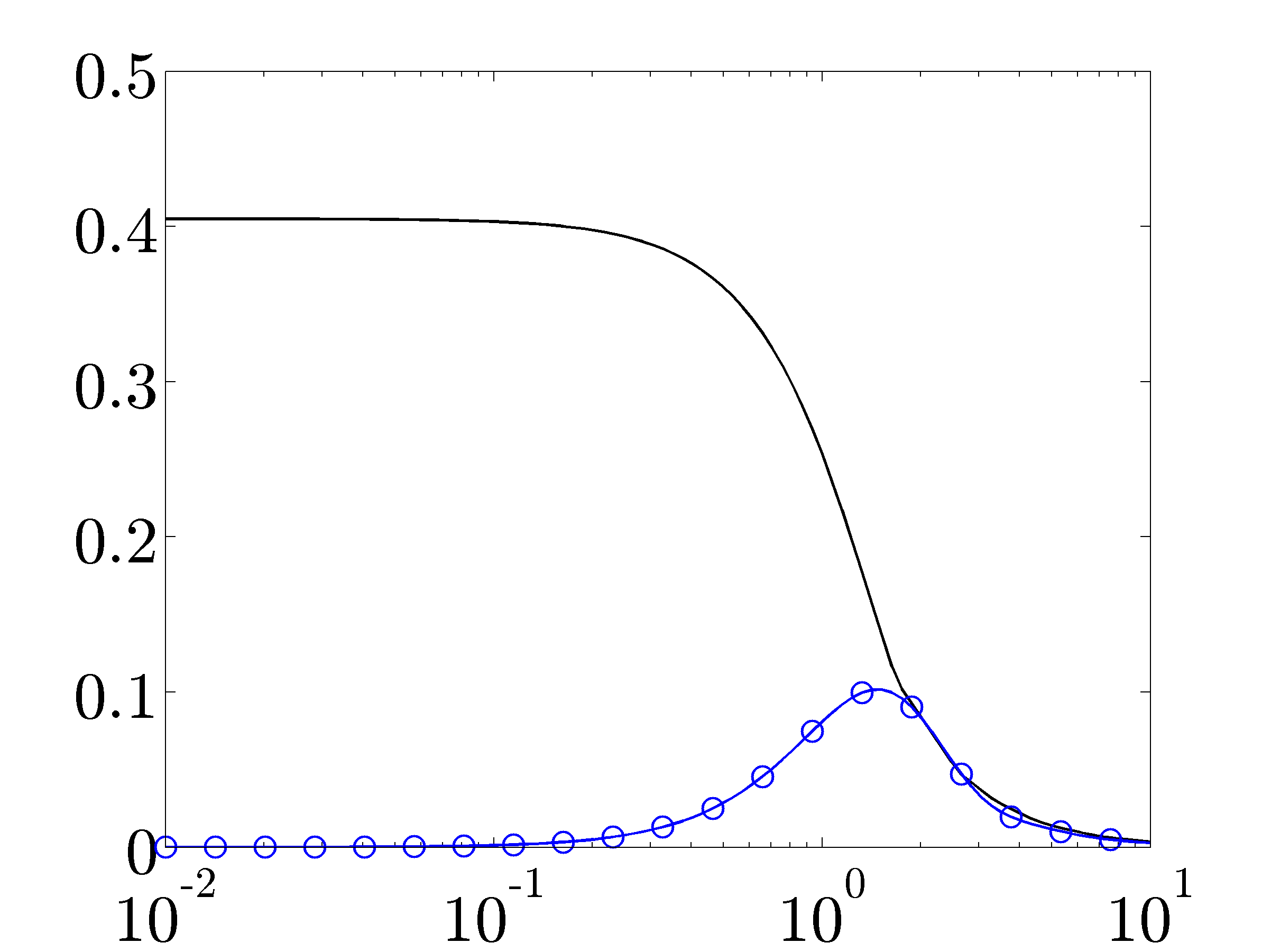}
				}
				\\
				{\large $k_z$}
				\\
				\subfigure[]
				{
				\label{fig.g42-g43-kx0}
				}
			\end{tabular}
			&
			\hspace{-0.7cm}
			\begin{tabular}{c}
				{\large $g_{52}(k_z)$, $g_{53}(k_z)$, $g_{62}(k_z)$, $g_{63}(k_z)$}
				\\
				{
				\includegraphics[width=0.45\columnwidth]
				{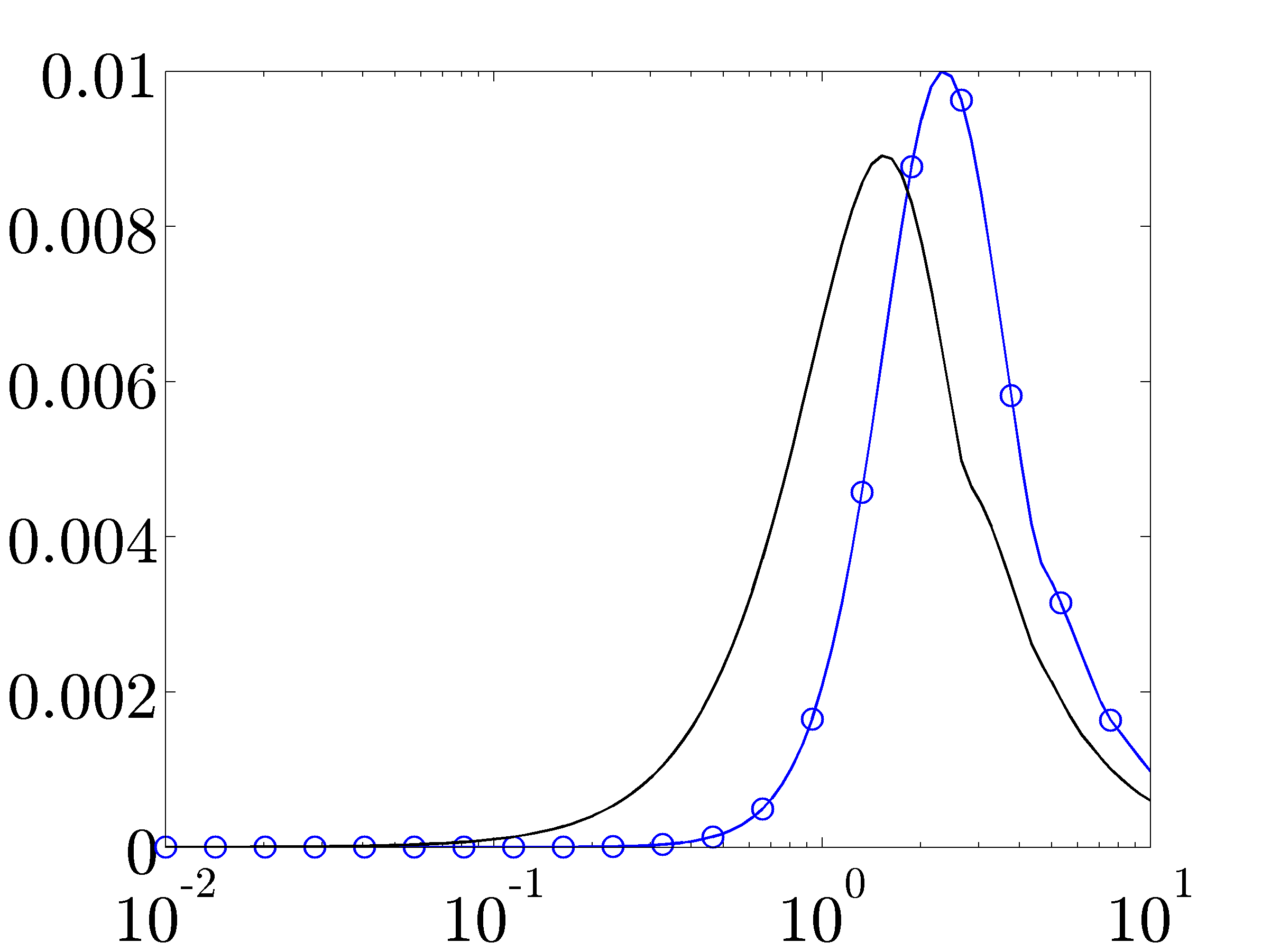}
				}
				\\
				{\large $k_z$}
				\\
				\subfigure[]
				{
				\label{fig.g62-g63-kx0}
				}
			\end{tabular}
		\end{tabular}
	\end{center}
	\caption{Spanwise frequency responses from $d_{j}$ to the polymer stress fluctuations $\tau_{1 3}$ ($g_{4 j}$), $\tau_{12}$ ($g_{5 j}$), and $\tau_{1 1}$ ($g_{6 j}$) for $j = \{ 2, 3\}$. The function $g_{5 2} = g_{6 2}$ and the function $g_{5 3} = g_{6 3}$: (a) $g_{4 2}$ ($\circ$) and $g_{4 3}$ (solid) for $\beta = 0.5$; and (b) $g_{5 2}$, $g_{6 2}$ ($\circ$) and $g_{53}$, $g_{6 3}$ (solid).}
    	\label{fig.g-tau2-23-kx0}
 	\end{figure}

	Figure~\ref{fig.g-tau2-23-kx0} shows the $k_z$-dependence of the functions $g_{\ell j}$ for $\{\ell = 4, 5, 6; j = 2, 3\}$. We note that $g_{52} = g_{62}$ and $g_{53} = g_{63}$. Function $g_{43}$ has low-pass shape and it peaks at $k_z = 0$. We also notice band-pass features of $g_{52}$, $g_{53}$, and $g_{42}$, with the peak values occurring at $k_z \approx 2.4$, $k_z \approx 1.6$, and $k_z \approx 1.4$, respectively.
	
	Table~\ref{table.G-tau2-limit} summarizes the worst-case amplification of $\tau_{13}$, $\tau_{12}$, and $\tau_{11}$ arising from the wall-normal and spanwise forces in the limit of infinitely large $L$ (or infinitely large $\We$). For Oldroyd-B fluids (i.e., as $L \rightarrow \infty$), both $G_{4j}$ and $G_{5j}$ scale quadratically with the Weissenberg number, and $G_{6j}$ scales quarticly with $\We$. On the other hand, as $\We \rightarrow \infty$, both $G_{4j}$ and $G_{5j}$ scale quadratically with $L$ and $G_{6j}$ scales quarticly with $L$. This demonstrates profound influence of $d_{2}$ and $d_{3}$ on $\tau_{11}$ in strongly elastic shear flows of viscoelastic fluids. Thus, even in the absence of inertia viscoelastic shear flows with large polymer relaxation times and large extensibility of polymer molecules exhibit high sensitivity to disturbances and low robustness to modeling imperfections.
	
	\begin{table}
	\centering
	\begin{tabular}{c | c | c}
		$j = \{ 2, 3 \}$ & $L \longrightarrow \infty$ & $\We \longrightarrow \infty$ \\ \hline
		\hspace{0.4cm} $G_{4j}\left( k_z; \beta, \cdot , \cdot \right)$ \hspace{0.45cm}
		&
		\hspace{0.45cm} $\We^2 \, g_{4j}( k_z; \beta )$ \hspace{0.45cm}
		&
		\hspace{0.45cm} $0.5 \, \bar{L}^2 \, g_{4j}( k_z; \beta )$ \hspace{0.4cm}
		\\[0.1cm] \hline
		\hspace{0.4cm} $G_{5j}\left( k_z; \beta, \cdot , \cdot \right)$ \hspace{0.45cm}
		&
		\hspace{0.45cm} $\We^2 \left( 2 + \beta \right)^2 g_{5j}( k_z )$ \hspace{0.45cm}
		&
		\hspace{0.45cm} $0.5 \, \bar{L}^2 \left( 2 + \beta \right)^2 g_{5j}( k_z )$ \hspace{0.4cm}
		\\[0.1cm] \hline
		\hspace{0.4cm} $G_{6j}\left( k_z; \beta, \cdot , \cdot \right)$ \hspace{0.45cm}
		&
		\hspace{0.45cm} $4 \, \We^4 \left( 1 + 2 \beta \right)^2 g_{6j}( k_z )$ \hspace{0.45cm}
		&
		\hspace{0.45cm} $0.25 \, \bar{L}^4 \left( 1 + 2 \beta \right)^2 g_{6j}( k_z )$ \hspace{0.4cm}
		\\[0.1cm]
	\end{tabular}
	 \caption{Worst-case amplification of $\tau_{13}$, $\tau_{12}$, and $\tau_{11}$ arising from $d_{2}$ and $d_3$ in the limit of infinitely large maximum extensibility or infinitely large Weissenberg number.}
      		\label{table.G-tau2-limit}
	\end{table}

    \vspace*{-4ex}
\section{Concluding remarks}
\label{sec.conclusion}

	In this study, we have examined non-modal amplification of disturbances in inertialess Couette flow of viscoelastic fluids using the FENE-CR model. The amplification is quantified by the maximal singular values of the frequency response operators that map sources of excitations (body forces) to the quantities of interest (velocity and polymer stress fluctuations). Spatio-temporal body forcing fluctuations are assumed to be purely harmonic in the horizontal directions and time, and deterministic in the wall-normal direction. Our three-dimensional component-wise frequency response analysis of the FENE-CR model sets the current paper apart from prior works which study the transient growth of velocity and polymer stress fluctuations in inertialess flows~\citep{jovkumPOF10} and non-modal amplification of stochastic disturbances in elasticity-dominated flows with non-zero inertia~\citep{hodjovkumJFM08,hodjovkumJFM09,jovkumJNNFM11} using the Oldroyd-B model. We have shown that streamwise-elongated flow structures are most amplified by disturbances. Furthermore, the component-wise frequency responses reveal that the wall-normal and spanwise forces have the strongest impact on the flow fluctuations, and that the influence of these forces is largest on streamwise components of velocity and polymer stress fluctuations.
	
	For streamwise-constant fluctuations, we have established analytically that the largest amplification of the streamwise velocity and streamwise component of the polymer stress tensor is proportional to the first normal stress difference of the nominal flow $\bar{N}_{1}$ and to $\bar{N}_{1}^2 / \left(1 + \bar{N}_{1}/\bar{L}^2 \right)^2$, respectively. This largest amplification is caused by wall-normal and spanwise forcing fluctuations and it takes place at low temporal frequencies and $\cO (1)$ spanwise wavenumbers. Using our analytical developments we have also shown that this worst-case amplification of $u$ and $\tau_{11}$ respectively scales as (i) ${\cal O}(\We^{2})$ and ${\cal O}(\We^{4})$ in the Oldroyd-B limit (i.e., as $L \rightarrow \infty$); and (ii) ${\cal O}(L^{2})$ and  ${\cal O}(L^{4})$ in the limit of infinitely large Weissenberg number. We thus conclude that in the presence of large polymer relaxation times and large extensibility of the polymer molecules, the velocity and polymer stress fluctuations can experience significant amplification even when inertial effects are completely absent. The underlying physical mechanism involves interactions of polymer stress fluctuations with a background shear, which induces a viscoelastic analog of the vortex tilting mechanism that is responsible for large amplification in inertial flows of Newtonian fluids.
	
	It is worth noting that, in the limit of infinitely large $\We$, the worst-case amplification of both velocity and polymer stress fluctuations is bounded by the maximum extensibility of the polymer molecules. This is in contrast to Oldroyd-B fluids where infinite extensibility allows the amplification of disturbances to grow unboundedly with $\We$~\citep{jovkumJNNFM11}. Our new observations demonstrate that
high sensitivity to disturbances and low robustness to modeling imperfections are reduced by finite extensibility of nonlinear dumbbells. Thus, both large polymer relaxation times and large extensibility of the polymer molecules are needed to achieve large amplification of velocity and polymer stress fluctuations in inertialess channel flows of viscoelastic fluids.
		
	The present work extends recent efforts~\citep{hodjovkumJFM08,hodjovkumJFM09,jovkumPOF10,jovkumJNNFM11} that examine possible mechanisms for triggering transition to elastic turbulence in channel flows of viscoelastic fluids. In addition to providing insight into worst-case amplification of velocity and polymer stress fluctuations in inertialess flows, we also demonstrate the importance of uncertainty quantification in flows of viscoelastic fluids. Our analysis shows high sensitivity of inertialess flows of viscoelastic fluids to external disturbances. Unfavorable scaling of the worst-case amplification of flow fluctuations with $\We$ and $L$ indicates that small-in-norm modeling imperfections can destabilize nominally stable flows. Hence, stability margins of inertialess channel flows of viscoelastic fluids decrease significantly with an increase in the Weissenberg number and the maximum extensibility of the polymer chains. This uncertainty may arise from inevitable imperfections in the laboratory environment or from the approximate nature of the constitutive equations. Our observations regarding model robustness also have important implications for numerical simulations, where numerical and/or roundoff errors may cause the simulated dynamics to differ from the actual dynamics.
		
	The present findings suggest a plausible mechanism for transition to elastic turbulence in channel flows of viscoelastic fluids. Large amplification of disturbances induces formation of streamwise streaks whose growth can put the flow into a regime where nonlinear interactions are no longer negligible. These nonlinear interactions can then induce secondary amplification~\citep{schhus02} or secondary instability~\citep{wal97} of streamwise streaks, their breakdown, and transition to a time-dependent disordered flow and elastic turbulence. To understand possible routes for the transition to elastic turbulence, it is essential to track later stages of disturbance development by considering nonlinearities in the constitutive equations and their interplay with streak development and high flow sensitivity. Our ongoing efforts are directed toward examining sensitivity of the streaks to three-dimensional disturbances. We also intend to study the presence of a self-sustaining mechanism (proposed for Newtonian fluids by~\cite{wal97}) and to numerically track later stages of disturbance development in strongly elastic channel flows of viscoelastic fluids.
	
    \vspace*{-4ex}
\section*{Acknowledgements}

	This work was supported in part by the National Science Foundation under CAREER Award CMMI-06-44793 (to M.R.J.), by the Department of Energy under Award DE-FG02-07ER46415 (to S.K.), by the University of Minnesota Digital Technology Center's 2010 Digital Technology Initiative Seed Grant (to M.R.J. and S.K.), and by the University of Minnesota Doctoral Dissertation Fellowship (to B.K.L.). The University of Minnesota Supercomputing Institute is acknowledged for providing computing resources.

    \vspace*{-2ex}
\appendix
\section{The underlying operators in 3D shear-driven channel flow of FENE-CR fluids}
\label{sec.app-operators}

	In this appendix, we define the underlying operators appearing in~\eqref{eq.veta}--\eqref{eq.ss} for a shear-driven channel flow. Operators $\{\bC, \bD\}$ in~(\ref{eq.veta}) are given by
	\begin{equation*}
		\begin{array}{rclrcl}
			\bC_{v}
			& \!\! = \!\! &
			\cfrac{(1 - \beta)}{\beta}
			\left[
			\begin{array}{cc}
				\bC_{v 1} & \bC_{v 2}
			\end{array}
			\right],
			&
			\bD_{v}
			& \!\! = \!\! &
			\cfrac{1}{\beta}
			\left[
			\begin{array}{ccc}
				\bD_{v 1} & \bD_{v 2} & \bD_{v 3}
			\end{array}
			\right],
			\\[0.25cm]
			\bC_{\eta}
			& \!\! = \!\! &
			\cfrac{(1 - \beta)}{\beta}
			\left[
			\begin{array}{cc}
				\bC_{\eta 1} & \bC_{\eta 2}
			\end{array}
			\right],
			&
			\bD_{\eta}
			& \!\! = \!\! &
			\cfrac{1}{\beta}
			\left[
			\begin{array}{ccc}
				\bD_{\eta 1} & \bD_{\eta 2} & \bD_{\eta 3}
			\end{array}
			\right],
		\end{array}
	\end{equation*}
	where
	\begin{equation*}
		\begin{array}{rcl}
			\bC_{v1}
			& \!\! = \!\! &
			\Delta^{-2}
			\left[
			\begin{array}{ccc}
				\bC_{v1,1} & \bC_{v1,2} & \bC_{v1,3}
			\end{array}
			\right],
			\;\;\;
			\bC_{v2}
			\; = \;
			\Delta^{-2}
			\left[
			\begin{array}{ccc}
				\bC_{v2,1} & \bC_{v2,2} & \bC_{v2,3}
			\end{array}
			\right],
			\\[0.3cm]
			\bC_{\eta 1}
			& \!\! = \!\! &
			\Delta^{-1}
			\left[
			\begin{array}{ccc}
				\bC_{\eta 1,1} & \bC_{\eta 1,2} & \bC_{\eta 1,3}
			\end{array}
			\right],
			\;\;\;
			\bC_{\eta 2}
			\; = \;
			\Delta^{-1}
			\left[
			\begin{array}{ccc}
				\bC_{\eta 2,1} & \bC_{\eta 2,2} & \bC_{\eta 2,3}
			\end{array}
			\right],
			\\[0.3cm]
			\bC_{v 1, 1}
			& \!\! = \!\! &
			\left( \cfrac{k^2 \, \bar{f}}{\We} \, - \, \cfrac{2 \, \We \, k_x^2}{\bar{L}^2}\right) \py
			\, + \,
			\cfrac{\mri k_x \, \bar{f}}{\bar{L}^2} \left( \pyy \, + \, k^2 \right),
			\;\;\;
			\bC_{v1, 2}
			\; = \;
			\cfrac{\mri k_z \, \bar{f}}{\We} \; \left( \pyy \, + \, k_z^2 \right),
			\\[0.6cm]
			\bC_{v1,3}
			& \!\! = \!\! &
			-\left( \cfrac{k_z^2 \, \bar{f}}{\We} \, + \, \cfrac{2 \, \We \, k_x^2}{\bar{L}^2}\right) \py
			\, + \,
			\cfrac{\mri k_x \, \bar{f}}{\bar{L}^2} \left( \pyy \, + \, k^2 \right),
			\;\;\;
			\bC_{v2,1}
			\; = \;
			-\cfrac{2 \, k_z k_x \, \bar{f}}{\We} \; \py,
			\\[0.5cm]
			\bC_{v 2, 2}
			& \!\! = \!\! &
			\cfrac{\mri k_x \, \bar{f}}{\We} \; \left( \pyy + k^2 \right),
			\;\;\;
			\bC_{v 2, 3}
			\; = \;
			-\left( \cfrac{k_x^2 \, \bar{f}}{\We} \, + \, \cfrac{2 \, \We \, k_x^2}{\bar{L}^2}\right) \py
			\, + \,
			\cfrac{\mri k_x \, \bar{f}}{\bar{L}^2} \left( \pyy \, + \, k^2 \right),
			\\[0.5cm]
			\bC_{\eta 1, 1}
			& \!\! = \!\! &
			\left( 2 \, \We / \bar{L}^2 \right) k_x \, k_z
			\, - \,
			\left( \mri k_z \, \bar{f} / \bar{L}^2 \right) \py,
			\;\;\;
			\bC_{\eta 1, 2}
			\; = \;
			\left( \mri k_x \, \bar{f} / \We \right) \py,
			\\[0.3cm]
			\bC_{\eta 1, 3}
			& \!\! = \!\! &
			\left(
			\cfrac{2 \, \We}{\bar{L}^2}
			\, - \,
			\cfrac{\bar{f}}{\We}
			\right)
			k_x \, k_z
			\, - \,
			\cfrac{\mri k_z \, \bar{f}}{\bar{L}^2} \; \py,
			\;\;\;
			\bC_{\eta 2, 1}
			\; = \;
			\cfrac{\bar{f}}{\We} \; \left( k_z^2 \, - \, k_x^2\right),
			\\[0.5cm]
			\bC_{v 2, 2}
			& \!\! = \!\! &
			- \cfrac{\mri k_z \, \bar{f}}{\We} \; \py,
			\;\;\;
			\bC_{v 2, 3}
			\; = \;
			\left(
			\cfrac{2 \, \We}{\bar{L}^2}
			\, + \,
			\cfrac{\bar{f}}{\We}
			\right)
			k_x \, k_z
			\, - \,
			\cfrac{\mri k_z \, \bar{f}}{\bar{L}^2} \; \py,
			\\[0.6cm]
			\bD_{v 1} & = & \mri k_x \Delta^{-2} \py,
			\,\,
			\bD_{v 2} \, = \, k^2 \Delta^{-2},
			\,\,
			\bD_{v 3} \, = \, \mri k_z \Delta^{-2} \py,
			\\[0.3cm]
			\bD_{\eta 1} & = & -\mri k_z \Delta^{-1},
			\,\,
			\bD_{\eta 2} \, = \, 0,
			\,\,
			\bD_{\eta 3} \, = \, \mri k_x \Delta^{-1}.
		\end{array}
	\end{equation*}
	Here, $k^2 = k_x^2 + k_z^2$, $\mri = \sqrt{-1}$, $\Delta = \pyy - k^2$ with Dirichlet boundary conditions, $\Delta^2 = \pyyyy - 2 k^2 \pyy + k^4$ with both Dirichlet and Neumann boundary conditions.
	
	The $\bF$-operators appearing in~(\ref{eq.psi}) are determined by
	\begin{equation*}
	\begin{array}{rcl}
		\bF_{1 1}
		& \!\! = \!\! &
		\left[
		\begin{array}{rrr}
			-D & 0 & 0 \\[0.3em]
			0 & -D & 0 \\[0.3em]
			0 & 0 & -D \\[0.3em]
		\end{array}
		\right],
		\,\,
		\bF_{2 2}
		\, = \,
		\left[
		\begin{array}{ccc}
			-D & 0 & 0 \\[0.1cm]
			0 & -D & -\We \bar{f} / \bar{L}^2 \\[0.15cm]
			0 & 2 \, \We & - \left( D \, + \, 2 \, \We^2 / \bar{L}^2 \right)
		\end{array}
		\right],
		\\[0.8cm]
		\bF_{2 1}
		& \!\! = \!\! &
		\left[
		\begin{array}{ccc}
			0 & \We & 0 \\[0.1cm]
			\We \left( 1 \, - \, \bar{f} / \bar{L}^2 \right) & 0 & -\We \bar{f} / \bar{L}^2 \\[0.15cm]
			-2 \, \We^2 / \bar{L}^2 & 0 & -2 \, \We^2 / \bar{L}^2
		\end{array}
		\right],
		\,\,
		\;\;
		D
		\; = \;
		\left( \bar{f} \, + \, \We \, \mri k_x \, U \right),
	\end{array}
	\end{equation*}
	\begin{equation*}
	\begin{array}{rcl}
		\bF_{1 v}
		& \!\! = \!\! &
		\cfrac{\We}{k^2}
		\left[
		\begin{array}{c}
			2 \, k^2 \left( \py \, + \, \mri k_x \, \bar{R}_{12} \right) \\[0.5em]
			\mri k_z \left( \pyy \, + \, k^2 \right) \, - \, \We \, k_x \, k_z \, \bar{R}_{12} \, \py  \\[0.5em]
			-2 \, k_z^2 \, \py
		\end{array}
		\right],
		\;\;\;
		\bF_{2 v}
		\; = \;
		\cfrac{\We}{k^2}
		\left[
		\begin{array}{c}
			\bF_{2 v}^{1} \\[0.4em]
			\bF_{2 v}^{2} \\[0.4em]
			\bF_{2 v}^{3}
		\end{array}
		\right],
		\\[0.7cm]
		\bF_{1 \eta}
		& \!\! = \!\! &
		\cfrac{\We}{k^2}
		\left[
		\begin{array}{c}
			0 \\[0.3em]
			\mri k_x \, \py \, - \, k_x^2 \, \bar{R}_{12}  \\[0.3em]
			-2 \, k_x \, k_z \\[0.3em]
		\end{array}
		\right],
		\,\,
		\bF_{2 \eta}
		\; = \;
		\cfrac{\We}{k^2}
		\left[
		\begin{array}{c}
			k_z^2 + \mri k_x \, \bar{R}_{12} \, \py \, - \, k_x^2 \, \bar{R}_{11} \\[0.3em]
			k_x \, k_z \, \bar{R}_{12} \, - \, \mri k_z \, \py \\[0.3em]
			2 \left( k_x \, k_z \, \bar{R}_{11} \, - \, \mri k_z \, \bar{R}_{12} \, \py \right)
		\end{array}
		\right],
		\\[0.8cm]
		\bF_{2v}^{1}
		& \!\! = \!\! &
		\mri k_z \, \bar{R}_{12} \, \pyy \, - \, k_x k_z \left( 1 + \bar{R}_{11} \right) \py,
		\\[0.3cm]
		\bF_{2 v}^{2}
		& \!\! = \!\! &
		\mri k_x \, k^2 \, \bar{R}_{11} \, + \, k_z^2 \, \bar{R}_{12} \, \py \, + \, \mri k_x \, \pyy,
		\;\;\;
		\bF_{2 v}^{3}
		\; = \;
		2 \left( \mri k_x \, \bar{R}_{12} \, \pyy  \, - \, k_x^2 \, \bar{R}_{11} \, \py \right).			
	\end{array}
	\end{equation*}
	The operators appearing in the evolution equations~\eqref{eq.ss} are given by
	\begin{equation*}
	\begin{array}{rcl}
		\bA
		& \!\! = \!\! &
		\left[
		\begin{array}{cc}
			\bF_{11} \, + \, \bF_{1v} \bC_{v1} \, + \, \bF_{1\eta} \bC_{\eta 1}
			&
			\bF_{1v} \bC_{v2} \, + \, \bF_{1\eta} \bC_{\eta 2}
			\\[0.3em]
			\bF_{21} \, + \, \bF_{2v} \bC_{v1} \, + \, \bF_{2 \eta} \bC_{\eta 1}
			&
			\bF_{22} \, + \, \bF_{2v} \bC_{v2} \, + \, \bF_{2 \eta} \bC_{\eta 2}
		\end{array}
		\right],
		\\[0.7cm]
		\bC
		& \!\! = \!\! &
		\left[
		\begin{array}{cc}
			\bC_{uv} \bC_{v1} \, + \, \bC_{u\eta} \bC_{\eta1}
			&
			\bC_{uv} \bC_{v2} \, + \, \bC_{u\eta} \bC_{\eta2}
			\\[0.3em]
			\bC_{v1}
			&
			\bC_{v2}
			\\[0.3em]
			\bC_{wv} \bC_{v1} \, + \, \bC_{w\eta} \bC_{\eta1}
			&
			\bC_{wv} \bC_{v2} \, + \, \bC_{w\eta} \bC_{\eta2}
		\end{array}
		\right],
		\\[0.7cm]
		\bB
		& \!\! = \!\! &
		\left[
		\begin{array}{c}
			\bF_{1v} \bD_{v} \, + \, \bF_{1\eta} \bD_{\eta} \\[0.3em]
			\bF_{2v} \bD_{v} \, + \, \bF_{2\eta} \bD_{\eta}
		\end{array}
		\right],
		\;\;\;
		\bD
		\; = \;
		\left[
		\begin{array}{c}
			\bC_{uv} \bD_{v} + \bC_{u\eta} \bD_{\eta} \\[0.3em]
			\bD_{v} \\[0.3em]
			\bC_{wv} \bD_{v} + \bC_{w\eta} \bD_{\eta}
		\end{array}
		\right],
		\\[0.8cm]
		\bC_{u v}
		& \!\! = \!\! &
		\left( \mri k_x / k^2 \right) \py,
		\;\;
		\bC_{u \eta}
		\; = \;
		-\mri k_z / k^2,
		\;\;
		\bC_{w v}
		\; = \;
		\left( \mri k_z / k^2 \right) \py,
		\;\;
		\bC_{w \eta}
		\; = \;
		\mri k_x / k^2.
		\end{array}
	\end{equation*}
	
		\vspace*{-2ex}
\section{Explicit scaling of worst-case amplification of streamwise-constant velocity fluctuations}
\label{sec.app-derivation}

	In this section, we discuss how to obtain explicit expressions for the worst-case amplification from the wall-normal and spanwise forces to the streamwise velocity fluctuations in the limit of infinitely large $\We$ or infinitely large $L$. We note that derivation of the analytical expressions for $G_{u2}$ and $G_{u3}$ is more challenging because the worst-case amplification of $u$ arising from $d_2$ and $d_3$ depends on both $\We$ and $L$. However, since our computations presented in~\S~\ref{sec.streamwise-constant} demonstrate that the worst-case amplification from $d_2$ and $d_3$ to $u$ takes place at low temporal frequencies, the essential features can be captured by analyzing the corresponding frequency responses at $\omega = 0$; see figures~\ref{fig.Smax-ud2-kx0-L10} and~\ref{fig.Smax-ud3-kx0-L10}. For flows without temporal variations, we can obtain the following static-in-time expressions that relate the conformation tensor fluctuations with the streamwise velocity $u$ and $(y,z)$-plane streamfunction $\psi$ (i.e., $v = \mri k_z \psi$, $w = - \py \psi$)
	\begin{subequations}
	\label{eq.r-conformation-omega0}
	\begin{align}
		\label{eq.r22-r23-r33}
		r_{22}
		\; = \;\; &
		\cfrac{2 \We}{\bar{f}} \; \mri k_z \, \py \psi,
		\;\;\;
		r_{33}
		\; = \;
		-\cfrac{2 \We}{\bar{f}} \; \mri k_z \, \py \psi,
		\;\;\;
		r_{23}
		\; = \;
		-\cfrac{\We}{\bar{f}} \left( \pyy + k_z^2 \right) \psi,
		\\[0.2cm]
		\label{eq.r13}
		r_{13}
		\; = \;\; &
		\left( \We / \bar{f} \right)
		\left(
		U' r_{23}
		\, - \,
		\left( \We / \bar{f} \right) \pyy \psi
		\, + \,
		\mri k_z u
		\right),
		\\[0.1cm]
		\label{eq.r12}
		r_{12}
		\; = \;\; &
		\left( \We / \bar{f} \right)
		\left(
		U' r_{22}
		\, - \,
		\left( \bar{f} / \bar{L}^2 \right) r_{11}
		\, + \,
		\left( \We / \bar{f} \right) \mri k_z \, \py \psi
		\, + \,
		\py u
		\right),
		\\[0.1cm]
		\label{eq.r11}
		r_{11}
		\; = \;\; &
		\left( 2 \, \We^2 / \zeta_0 \right)
		\left(
		\left( 3 \, \We / \bar{f} \right) \mri k_z \, \py \psi
		\, + \,
		2 \, \py u
		\right).
	\end{align}
	\end{subequations}
	The above expression~\eqref{eq.r-conformation-omega0} is obtained by taking the temporal Fourier transforms of~\eqref{eq.psiA}--\eqref{eq.psiB} and replacing the wall-normal velocity and vorticity with
	\[
		v \; = \; \mri k_z \, \psi,
		\;\;\;
		\eta \; = \; \mri k_z \, u.
	\]
	In the absence of streamwise forcing and streamwise variations, the static-in-time momentum equation (in the streamwise direction) provides a relation between the streamwise velocity and the streamwise components of the conformation tensor
	\begin{equation}
	\label{eq.u-momentum}
		\Delta u
		\; = \;
		-\cfrac{1 - \beta}{\beta}
		\left(
		\cfrac{\bar{f}}{\We}
		\;
		\py \, r_{12}
		\, + \,
		\cfrac{\bar{f}}{\We}
		\;
		\mri k_z \, r_{13}
		\, + \,
		\cfrac{\bar{f}}{\bar{L}^2}
		\;
		\py \, r_{11}
		\right).
	\end{equation}
	Substituting~\eqref{eq.r13} and~\eqref{eq.r12} into~\eqref{eq.u-momentum} yields a relation between streamwise velocity and the conformation tensor fluctuations in the wall-normal/spanwise plane,
	\begin{equation}
	\label{eq.u-r22-r23}
		\Delta u
		\; = \;
		- \left( 1 - \beta \right)
		\left(
		\py( U' \, r_{22}) \, + \, \mri k_z (U' r_{23})
		\right).
	\end{equation}
	Furthermore, we can obtain an expression relating $u$ and $\psi$ by substituting~\eqref{eq.r22-r23-r33} into~\eqref{eq.u-r22-r23} which yields
	\begin{equation}
	\begin{array}{rcl}
		\Delta u
		& \!\! = \!\! &
		- \,
        		\left( \We / \bar{f} \right)
		\,
		\left(1 - \beta \right)
		\mri k_z
		\,
		\Delta
		\psi.
	\end{array}
	\label{eq.u-psi}
	\end{equation}
	It can be shown that $\psi$ is induced by the action of the wall-normal and spanwise forcing
	\begin{equation}
	\label{eq.psi-d2d3-app}
		\psi
		\; = \;
		\Delta^{-2}
		\left[
		\begin{array}{cc}
			-\mri k_z & \py
		\end{array}
		\right]
		\left[
		\begin{array}{c}
			d_{2} \\[0.1cm]
			d_{3}
		\end{array}
		\right],
	\end{equation}

	Finally, the frequency response from $d_{2}$ and $d_{3}$ to $u$ at $\omega = 0$ is obtained by substituting~\eqref{eq.psi-d2d3-app} into~\eqref{eq.u-psi} which yields
	\begin{equation*}
		\bH_{uj}(k_z, 0; \beta, \We, L)
		\; = \;
		-\left( \We / \bar{f} \right)
		\left( 1 \, - \, \beta \right)
		\bD_{v j},
		\;\;\;
		j \; = \; \{ 2, 3\},
	\end{equation*}
	where
	\begin{equation*}
		\bD_{v 2}
		\; = \;
		k_z^2 \Delta^{-2},
		\;\;\;
		\bD_{v 3}
		\; = \;
		\mri k_z \, \Delta^{-2} \py.
	\end{equation*}
	The worst-case amplification of $u$ caused by $d_2$ and $d_3$ can be reliably approximated by
	\begin{equation*}
	\begin{array}{rcl}
		G_{u j} \left( k_z; \beta, \We, L \right)
		& \!\! \approx \!\! &
            \sigma^2_{\max} \left( \bH_{u j}(k_z,0; \beta, \We, L) \right)
            \\[0.2cm]
		& \!\! = \!\! &
		{\ds
		\left( \We / \bar{f} \right)^2
            \,
		\left( 1 - \beta \right)^2
		\sigma_{\max}^{2} \left( \bD_{v j} \right)
		}
		\\[0.2cm]
		& \!\! = \!\! &
		{\ds
		\left(
		\bar{N}_{1} / 2
		\right)
		\left( 1 - \beta \right)^2
		g_{u j}(k_z),
		\;\;\;
		j \; = \; \{ 2, 3 \},
		}
	\end{array}
	\end{equation*}
	where the functions $g_{u j}$ with $j = \{ 2, 3 \}$ quantify the spanwise frequency responses from $d_{2}$ and $d_{3}$ to $u$. In the Oldroyd-B limit (i.e., as $L \rightarrow \infty$), the first normal stress difference $\bar{N}_{1} \rightarrow 2 \, \We^2$ and the function $G_{u j}$ is given by
	\begin{equation*}
		\lim_{L \rightarrow \infty}
		G_{u j}(k_z; \beta, \We, L)
		\; = \;
		\We^2 \left( 1 - \beta \right)^2 \, g_{u j}(k_z).
	\end{equation*}
	On the other hand, in the limit of infinitely large $\We$, the first normal stress difference $\bar{N}_{1} \rightarrow \bar{L}^2$ and the function $G_{u j}$ is given by
	\begin{equation*}
		\lim_{\We \rightarrow \infty}
		G_{u j}(k_z; \beta, \We, L)
		\; = \;
		\left( \bar{L}^4 / 2 \right) \left( 1 - \beta \right)^2 \, g_{u j}(k_z).
	\end{equation*}

	\vspace*{-2ex}
\section{The frequency response operators from body forces to polymer stresses in streamwise-constant Couette flow of FENE-CR fluids}
\label{sec.app-Gop-2d3c}

	The streamwise-constant frequency response operators in~\eqref{eq.hinf-tau1-d23-2d3c} that map different forcing components to the polymer stress fluctuations are given by
	\begin{equation*}
	\begin{array}{rcl}
		\bG_{1j}(k_z, \omega; \beta, \We, L)
		& \!\! = \!\! &
		\cfrac{2 \, \bar{f}}{\mri \omega \beta + \bar{f}} \,\, \py \, \bD_{v j},
		\;\;
		\bG_{3j}(k_z, \omega; \beta, \We, L)
		\; = \;
		-\cfrac{2 \, \bar{f}}{\mri \omega \beta + \bar{f}} \,\, \py \, \bD_{v j},
		\\[0.4cm]
		\bG_{2j}(k_z, \omega; \beta, \We, L)
		& \!\! = \!\! &
		\left( \bar{f} / \left( \mri \omega \beta + \bar{f} \right) \right) (\mri / k_z) \left( \pyy + k_z^2 \right) \, \bD_{v j},
		\\[0.3cm]
		\bG_{41}(k_z, \omega; \beta, \We, L)
		& \!\! = \!\! &
		\left( \mri k_z \bar{f} / \left( \mri \omega \beta + \bar{f} \right) \right) \bE_{uu}^{-1} \, \bC_{u \eta} \, \bD_{\eta 1},
		\\[0.2cm]
		\bG_{4j}(k_z, \omega; \beta, \We, L)
		& \!\! = \!\! &
		\cfrac{\We \bar{f}}{(\mri \omega \beta + \bar{f})(\mri \omega + \bar{f})} \,\, (\mri / k_z) \left( \pyy + k_z^2 \right) \, \bD_{v j} \, + \,
		\\[0.3cm]
		&&
		\cfrac{\We}{(\mri \omega \beta + \bar{f})} \,\, (\mri / k_z) \pyy \, \bD_{v j}
		\, + \,
		\cfrac{\mri k_z \bar{f}}{(\mri \omega + \bar{f})} \,\, \bE_{uu}^{-1} \, \bE_{uv} \, \bD_{v j},
		\\[0.4cm]
		\bG_{51}(k_z, \omega; \beta, \We, L)
		& \!\! = \!\! &
		\cfrac{2 \We^2}{\bar{L}^2} \, \cfrac{2 \mri \omega \bar{f} - \omega^2}{\left( \mri \omega \beta + \bar{f} \right) \left( \zeta_0 \, - \, \omega^2 \, + \, \mri \omega \zeta_1 \right)} \,\, \py \, \bE_{uu}^{-1} \, \bC_{u \eta} \, \bD_{\eta 1} \, + \,
		\\[0.4cm]
		&&
		\cfrac{\bar{f}}{\mri \omega \beta + \bar{f}} \,\, \py \, \bE_{uu}^{-1} \, \bC_{u \eta} \, \bD_{\eta 1},
		\\[0.5cm]
		\bG_{5j}(k_z, \omega; \beta, \We, L)
		& \!\! = \!\! &
		\cfrac{2 \We^2}{\bar{L}^2} \, \cfrac{2 \mri \omega \bar{f} - \omega^2}{\left( \mri \omega + \bar{f} \right) \left( \zeta_0 \, - \, \omega^2 \, + \, \mri \omega \zeta_1 \right)} \,\, \py \, \bE_{uu}^{-1} \, \bE_{uv}\, \bD_{v j} \, + \,
		\\[0.4cm]
		&&
		\cfrac{2 \We^3}{\bar{L}^2} \, \cfrac{3 \mri \omega \bar{f} - \omega^2}{\left( \mri \omega + \bar{f} \right) \left( \mri \omega \beta + \bar{f} \right) \left( \zeta_0 \, - \, \omega^2 \, + \, \mri \omega \zeta_1 \right)} \,\, \py \, \bD_{v j} \, + \,
		\\[0.4cm]
		&&
		\cfrac{\We \left( \mri \omega + 3 \bar{f} \right)}{\left( \mri \omega + \bar{f} \right) \left( \mri \omega \beta + \bar{f} \right)} \,\, \py \, \bD_{v j}
		\, + \,
		\cfrac{\bar{f}}{\mri \omega + \bar{f}} \,\, \py \, \bE_{uu}^{-1} \, \bE_{uv} \, \bD_{vj},
		\\[0.4cm]
		\bG_{61}(k_z, \omega; \beta, \We, L)
		& \!\! = \!\! &
		\left( \cfrac{\bar{f}}{\We} \, + \, \cfrac{2 \We}{\bar{L}^2}\right) \cfrac{2 \We^2 (\mri \omega + 2 \bar{f})}{\bar{f} \left( \zeta_0 \, - \, \omega^2 \, + \, \mri \omega \zeta_1 \right)} \,\, \cfrac{\mri \omega + \bar{f}}{\mri \omega \beta + \bar{f}}\,\, \py \, \bE_{uu}^{-1} \, \bC_{u \eta} \, \bD_{\eta 1},
		\\[0.5cm]
		\bG_{6j}(k_z, \omega; \beta, \We, L)
		& \!\! = \!\! &
		\left( \cfrac{\bar{f}}{\We} \, + \, \cfrac{2 \We}{\bar{L}^2}\right) \left( \cfrac{2 \We^2 (\mri \omega + 2 \bar{f})}{\bar{f} \left( \zeta_0 \, - \, \omega^2 \, + \, \mri \omega \zeta_1 \right)} \,\, \py \, \bE_{uu}^{-1} \, \bE_{uv}\, \bD_{v j} \, + \, \right.
		\\[0.4cm]
		&&
		\left. \cfrac{2 \We^3 \left( \mri \omega + 3 \bar{f} \right)}{\bar{f} \left( \mri \omega \beta + \bar{f} \right) \left( \zeta_0 \, - \, \omega^2 \, + \, \mri \omega \zeta_1 \right)} \,\, \py \bD_{v j} \right),
		\;\;
		j = \{ 2, 3 \}.
	\end{array}
	\end{equation*}
		
		\vspace*{-2ex}

\end{document}

%% file: commands.tex
\newcommand{\enma}[1]   {\ensuremath{#1}}

\newcommand{\beq}{\begin{equation}}
\newcommand{\eeq}{\end{equation}}
\newcommand{\bseq}{\begin{subequations}}
\newcommand{\eseq}{\end{subequations}}
\newcommand{\beqn}{\begin{eqnarray}}
\newcommand{\eeqn}{\end{eqnarray}}
\newcommand{\ba}{\begin{array}}
\newcommand{\ea}{\end{array}}
\newcommand{\bct}{\begin{center}}
\newcommand{\ect}{\end{center}}
\newcommand{\btab}{\begin{tabular}}
\newcommand{\etab}{\end{tabular}}
\newcommand{\btmz}{\begin{itemize}}
\newcommand{\etmz}{\end{itemize}}
\newcommand{\benum}{\begin{enumerate}}
\newcommand{\eenum}{\end{enumerate}}

\newcommand{\trace}     {\enma{\mathrm{trace}}}

\newcommand{\bv}{{\bf v}}

\newcommand{\matbegin}{
        \left[
}
\newcommand{\matend}{
        \right]
}

\newcommand{\tbo}[2]{
  \matbegin \begin{array}{c}
       #1 \\ #2
       \end{array} \matend }

\newcommand{\obt}[2]{
  \matbegin \begin{array}{cc}
       #1 & #2
       \end{array} \matend }

\newcommand{\tbt}[4]{
  \matbegin \begin{array}{cc}
       #1 & #2 \\ #3 & #4
       \end{array} \matend }

\newcommand{\be}{\begin{equation}}
\newcommand{\ee}{\end{equation}}

\newcommand{\cplxs}{ C\kern -.35em \rule{0.03 em}{.7 ex}~   }

\def\complex{\hbox{C\kern -.45em \rule{0.03 em}{1.5 ex}}~}

\newcommand{\bi}{\begin{itemize}}
\newcommand{\ei}{\end{itemize}}



%% file: simpleNEW.tex
\begin{picture}(7.5,1)(0,0)

    \put(0,0.5){\vector(1,0){0.75}}
    \put(0.375,0.7){\makebox(0,0)[b]{$d$}}
    \put(0.75,0){\framebox(1.75,1){$\dfrac{1}{s \, + \, \lambda_1}$}}

    \put(2.5,0.5){\vector(1,0){0.75}}
    \put(2.875,0.7){\makebox(0,0)[b]{$\phi_1$}}
    \put(3.25,0){\framebox(1,1){\textcolor{red}{$R$}}}
    \put(4.25,0.5){\vector(1,0){0.75}}
    \put(5,0){\framebox(1.75,1){$\dfrac{1}{s \, + \, \lambda_2}$}}
    \put(6.75,0.5){\vector(1,0){0.75}}
    \put(7.125,0.7){\makebox(0,0)[b]{$ \varphi $}}

	\vspace{0.4cm}

\end{picture}

%% file: simpleGnew.tex
\begin{picture}(7.55,2.25)(0,0)

    \put(0,1.75){\vector(1,0){0.75}}
    \put(0.375,1.95){\makebox(0,0)[b]{$d$}}
    \put(0.75,1.25){\framebox(1.75,1){$\dfrac{1}{s \, + \, \lambda_1}$}}

    \put(2.5,1.75){\vector(1,0){0.75}}
    \put(2.875,1.95){\makebox(0,0)[b]{$\phi_1$}}
    \put(3.25,1.25){\framebox(1,1){\textcolor{red}{$R$}}}
    \put(4.25,1.75){\vector(1,0){0.75}}
    \put(5,1.25){\framebox(1.75,1){$\dfrac{1}{s \, + \, \lambda_2}$}}
    \put(6.75,1.75){\line(1,0){0.4}}
    \put(7.15,1.75){\circle*{0.08}}
    \put(7.15,1.75){\vector(1,0){0.4}}
    \put(7.15,1.95){\makebox(0,0)[b]{$ \varphi $}}
    
    \put(7.15,1.75){\line(0,-1){0.075}} \multiput(7.15,1.675)(0,-0.2){5}{\line(0,-1){0.1}} \put(7.15,0.675){\line(0,-1){0.175}}
    \put(7.15,0.5){\line(-1,0){0.03}} \multiput(7.12,0.5)(-0.2,0){13}{\line(-1,0){0.1}} \put(4.52,0.5){\vector(-1,0){0.27}}
    \put(3.25,0){\dashbox{0.1}(1,1){$\textcolor{blue}{\Gamma}$}}
    \put(3.25,0.5){\line(-1,0){0.075}} \multiput(3.175,0.5)(-0.2,0){15}{\line(-1,0){0.1}} \put(0.175,0.5){\line(-1,0){0.175}}
    \put(0,0.5){\line(0,1){0.075}} \multiput(0,0.575)(0,0.2){5}{\line(0,1){0.1}} \put(0,1.575){\line(0,1){0.175}}

	\vspace{0.4cm}
    
\end{picture}

%% file: fbkGamma.tex
\begin{picture}(7,2.75)(0,0)

    \put(0,2){\vector(1,0){1.}}
    \put(0.5,2.2){\makebox(0,0)[b]{$\bd$}}
    \put(1,1.25){\framebox(5,1.5){\btab{c} \mbox{\tc{black}{\bf Nominal}} \\[0.15cm] \mbox{\tc{black}{\bf Linearized Dynamics}} \etab}}

    \put(6,2){\line(1,0){0.5}}
    \put(6.5,2){\circle*{0.08}}
    \put(6.5,2){\vector(1,0){0.5}}
    \put(6.5,2.2){\makebox(0,0)[b]{$ \bvarphi $}}
    
    \put(6.5,2){\line(0,-1){0}} \multiput(6.5,2)(0,-0.2){7}{\line(0,-1){0.1}} \put(6.5,0.6){\line(0,-1){0.1}}
    \put(6.5,0.5){\line(-1,0){0.03}} \multiput(6.47,0.5)(-0.2,0){11}{\line(-1,0){0.1}} \put(4.27,0.5){\vector(-1,0){0.27}}
    \put(3,0){\dashbox{0.1}(1,1){$\tc{blue}{\Gamma}$}}
    \put(3.5,-0.2){\makebox(0,0)[t]{\btab{c} \tc{blue}{\bf modeling uncertainty} \\ \tc{black}{(can be nonlinear or time-varying)} \etab}}
    \put(3,0.5){\line(-1,0){0.05}} \multiput(2.95,0.5)(-0.2,0){14}{\line(-1,0){0.1}} \put(0.15,0.5){\line(-1,0){0.15}}
    \put(0,0.5){\line(0,1){0}} \multiput(0,0.5)(0,0.2){7}{\line(0,1){0.1}} \put(0,1.9){\line(0,1){0.1}}

\end{picture}